\documentclass[11pt,a4paper]{article}
\pdfoutput=1

\usepackage{jheppub}\usepackage{graphicx,placeins,amsmath,amssymb,subfigure,color,ulem}
\usepackage[applemac]{inputenc}
\usepackage{cancel}

\def\lsim{\raise0.3ex\hbox{$<$\kern-0.75em\raise-1.1ex\hbox{$\sim$}}}
\def\gsim{\raise0.3ex\hbox{$>$\kern-0.75em\raise-1.1ex\hbox{$\sim$}}}

\makeatletter
\def\gl@align#1#2{\lower.6ex\vbox{\baselineskip\z@skip\lineskip\z@
     \ialign{$\m@th#1\hfill##\hfil$\crcr#2\crcr\sim\crcr}}}
\makeatother

\newcommand{\meff}{m^{\rm eff}}

\newcommand{\meffB}{m^{\rm eff}_{B}}
\newcommand{\meffN}{m^{\rm eff}_{N}}
\newcommand{\meffXi}{m^{\rm eff}_{\Xi}}

\newcommand{\EeffBB}{E^{\rm eff}_{BB}}
\newcommand{\EeffNN}{E^{\rm eff}_{NN}}
\newcommand{\EeffXiXi}{E^{\rm eff}_{\Xi\Xi}}
\newcommand{\EeffTri}{E^{\rm eff}_{^3{\rm He}}}
\newcommand{\EeffHe}{E^{\rm eff}_{^4{\rm He}}}

\newcommand{\DelEeffBB}{\Delta E^{\rm eff}_{BB}}
\newcommand{\DelEeffNN}{\Delta E^{\rm eff}_{NN}}
\newcommand{\DelEeffXiXi}{\Delta E^{\rm eff}_{\Xi\Xi}}
\newcommand{\DelEeffTri}{\Delta E^{\rm eff}_{^3{\rm He}}}
\newcommand{\DelEeffHe}{\Delta E^{\rm eff}_{^4{\rm He}}}

\newcommand{\DelEBB}{\Delta E_{BB}}
\newcommand{\DelENN}{\Delta E_{NN}}
\newcommand{\DelEXiXi}{\Delta E_{\Xi\Xi}}
\newcommand{\DelETri}{\Delta E_{^3{\rm He}}}
\newcommand{\DelEHe}{\Delta E_{^4{\rm He}}}

\newcommand{\tmax}{t_{\rm max}}
\newcommand{\tmin}{t_{\rm min}}

\newcommand{\rcrit}{r_{\rm crit}}

\begin{document}

\title{
Mirage in Temporal Correlation functions for Baryon-Baryon Interactions in Lattice QCD
}

\author[a]{T.~Iritani,}
\author[b]{T.~Doi,}
\author[c,d]{S.~Aoki,}
\author[e]{S.~Gongyo}
\author[b,f]{T.~Hatsuda,}
\author[b,g]{Y.~Ikeda,}
\author[h]{T.~Inoue,}
\author[g]{N.~Ishii,}
\author[g]{K.~Murano,}
\author[d]{H.~Nemura,}
\author[d,i]{K.~Sasaki,}
\author{(HAL QCD Collaboration)}

\affiliation[a]{Department of Physics and Astronomy, Stony Brook University, NY 11794-3800, USA}
\affiliation[b]{Theoretical Research Division, Nishina Center, RIKEN, Wako 351-0198, Japan}
\affiliation[c]{Center for Gravitational Physics, Yukawa Institute for Theoretical Physics, Kyoto University, Kitashirakawa Oiwakecho, Sakyo-ku, Kyoto 606-8502, Japan}
\affiliation[d]{Center for Computational Sciences, University of Tsukuba, Tsukuba 305-8577, Japan}
\affiliation[e]{CNRS, Laboratoire de Math\'ematiques et Physique Th\'eorique, Universit\'ede Tours, 37200 France}
\affiliation[f]{iTHES Research Group, RIKEN, Wako 351-0198, Japan}
\affiliation[g]{Research Center for Nuclear Physics (RCNP), Osaka University, Osaka 567-0047, Japan}
\affiliation[h]{Nihon University, College of Bioresource Sciences, Kanagawa 252-0880, Japan}
\affiliation[i]{Yukawa Institute for Theoretical Physics, Kyoto University, Kitashirakawa Oiwakecho, Sakyo-ku, Kyoto 606-8502, Japan}

\emailAdd{takumi.iritani@stonybrook.edu}
\emailAdd{doi@ribf.riken.jp}
\emailAdd{saoki@yukawa.kyoto-u.ac.jp}
\emailAdd{shinya.gongyo@yukawa.kyoto-u.ac.jp}
\emailAdd{thatsuda@riken.jp}
\emailAdd{yikeda@riken.jp}
\emailAdd{inoue.takashi@nihon-u.ac.jp}
\emailAdd{ishii@ribf.riken.jp}
\emailAdd{murano@rcnp.osaka-u.ac.jp}
\emailAdd{nemura@riken.jp}
\emailAdd{kenjis@het.ph.tsukuba.ac.jp}

\preprint{RIKEN-QHP-244, YITP-16-91}
\abstract{Single state saturation of the temporal
  correlation function is  a key condition to extract physical observables such as 
   energies and matrix elements  of hadrons from lattice QCD simulations.
  A method  commonly employed to check the saturation is 
    to seek for a plateau of the observables  for large  Euclidean time. 
  Identifying  the plateau in the cases having nearby states, however,
  is non-trivial and one may even be misled by a fake plateau.
  Such a situation takes place typically for a system with two or more baryons.
  In this study, we demonstrate explicitly the danger from a possible fake plateau in the
   temporal correlation functions mainly for two baryons ($\Xi\Xi$ and $NN$),
   and  three and four baryons ($^3{\rm He}$ and $^4{\rm He})$ as well,
    employing (2+1)-flavor lattice QCD at $m_{\pi}=0.51$ GeV on four lattice volumes  with 
    $L=$ 2.9, 3.6, 4.3 and 5.8 fm.
    Caution is required when drawing conclusions about the bound $NN$, $3N$ and $4N$ systems
    based only on the standard plateau fitting
    of the temporal correlation functions.
   }

\keywords{lattice QCD, baryon interactions, ground state saturation,  plateau of
the effective energy}

\maketitle

\section{Introduction} 

In lattice QCD, observables such as the energies and the matrix elements of hadrons are
commonly extracted from temporal correlation functions at large Euclidean time where 
 ground state  saturation is expected to be realized. For
example, a two point correlation function $C(t)$ for the operator $O_{1,2}(t, \vec x)$ 
is related to physical quantities as
\begin{eqnarray}
C(t)&\equiv&\sum_{\vec x} \langle 0 \vert O_1(t, \vec x) O_2(0, \vec 0) \vert 0 \rangle 
=\sum_{\vec x} \langle 0 \vert O_1(t, \vec x)\sum_{k=1}^\infty \vert k\rangle\langle k \vert  \ O_2(0, \vec 0) \vert 0 \rangle 
= \sum_{n=1}^\infty Z_n e^{-m_n t} +\cdots\nonumber \\
\end{eqnarray}
where $ \vert n \rangle$ is the $n$-th one-particle (zero momentum) eigenstate of QCD with mass $m_n$
which couples to the operator $O_{1,2}$, and $Z_n$ is the corresponding pole residue,  
$Z_n =  \langle 0 \vert O_1(0,\vec 0) \vert n \rangle\langle n  \vert O_2 (0,\vec 0) \vert 0 \rangle$. The ellipsis represents contributions from two or more particle states.  Assuming the ordering that  $0 < m_1 < m_2 < m_3 \cdots $, we can extract the mass and the matrix element for the lowest energy state from the large $t$ behavior of $C(t)$ as
\begin{eqnarray}
  C(t) &\simeq & Z_1 e^{-m_1 t} + \mathcal{O}(e^{-m_2 t} ), \qquad t\rightarrow \infty , 
\end{eqnarray}
where  contributions from two or more particle states are suppressed  for $t\rightarrow \infty$.

In practice, we take large but finite $t$, so that $e^{-(m_2-m_1)t}$ becomes
negligibly small. If $m_2 - m_1 = \mathcal{O}(\Lambda_{\rm QCD})$, which is generally
true for single hadron states in QCD, it requires $t \ge \mathcal{O}(1)$~fm.
Therefore,  one can safely extract  single-hadron masses as long as $C(t)$ 
is accurate enough at $t\sim \mathcal{O}( 1 )$~fm.
To check whether $C(t)$ is dominated by the ground state, the effective mass, defined by
\begin{eqnarray}
\meff  (t) &=& -\frac{1}{a}\log\left( \frac{C(t+a)}{C(t)} \right),
\end{eqnarray}
is often employed,  where $a$ is the lattice spacing.  If $\meff(t)$ becomes almost independent of $t$ at $t \ge \tmin$ (``the plateau''),
$C(t)$ is considered to be dominated by the ground state and the mass is extracted  from $C(t)$ by using the data at
 $t \ge \tmin$.  

For multi-hadrons,  the energy shift of the whole system on the lattice relative to the threshold defined by the 
sum of each hadron masses is of interest, since it has information on the binding energy and the scattering phase shift \cite{Luscher:1990ux}.  
For the energy shift of  the two-baryon system, $\DelEBB \equiv E_{BB} - 2m_{B}$, where
 $E_{BB}$ is the lowest energy of the two-baryon system and $m_{B}$ 
is the baryon mass, one introduces the effective energy shift defined by
\begin{eqnarray}
\DelEeffBB (t) &\equiv & \EeffBB(t) - 2 \meffB(t) = - \frac{1}{a}\log\left( \frac{R_{BB}(t+a)}{ R_{BB}(t)}\right), 
\end{eqnarray}
where $R_{BB}$ is the two-baryon propagator 
$C_{BB}(t)$ divided by the one-baryon propagator $C_B(t)$  
squared as 
\begin{eqnarray}
R_{BB}(t) &\equiv & \frac{ C_{BB}(t) }{ C_B(t)^2}
\end{eqnarray}
 with
 \begin{eqnarray}
C_{BB} (t) &\equiv & \langle B(t)^2 \bar B(0)^2\rangle, \qquad
C_{B}(t) \equiv  \langle B(t)\bar B(0)\rangle ,
\end{eqnarray}
and the effective energy of two-baryon system $E_{BB}^\mathrm{eff}(t)$,
which is defined by
\begin{equation}
  E_{BB}^\mathrm{eff}(t) = - \frac{1}{a}\log
  \left( \frac{C_{BB}(t+a)}{C_{BB}(t)} \right).
\end{equation}
In actual numerical simulations, it is often observed that  the statistical error for $\DelEeffBB (t)$
is substantially reduced from the individual errors for
$\EeffBB(t)$ and $2\meffB(t)$ due to their mutual correlations.
In addition, $\DelEeffBB (t)$  shows a plateau-like behavior at relatively earlier time $t$ than 
 it is supposed to be, so that one may be tempted to extract physical information from such
 a behavior.

In this paper, we address the issue whether the plateau-like behavior observed for the effective energy shift of the multi-baryons system is reliable or not.  Indeed,
it was previously claimed, by fitting the plateau-like behavior of the effective energy shifts,  that 
 dineutron, deuteron, ${}^3{\rm He}$ and ${}^4 {\rm He}$  are all bound for heavy  pion masses,
  $m_\pi \simeq 510$ MeV~\cite{Yamazaki:2012hi} and $m_\pi\simeq 300$ MeV~\cite{Yamazaki:2015asa}.
 For making detailed comparisons with such previous results, we employ the same lattice setup  as  Ref.~\cite{Yamazaki:2012hi}.
 We perform  more measurements of baryon correlation functions than those in previous studies to investigate the reliability of the plateau-like behavior from the point of view of statistics, while we take two different source operators (the smeared source used in~\cite{Yamazaki:2012hi}  
and the wall source%
\footnote{The wall source has been adopted in
 the HAL QCD method~\cite{Ishii:2006ec,Aoki:2008hh,Aoki:2009ji,Aoki:2011ep,Aoki:2012tk}, which utilizes the
 space-time correlation functions instead of just the temporal correlation to study multi-hadrons. 
 In this method,
 bound states for dineutron and deuteron 
 are not found at similar values of pion masses~\cite{HALQCD:2012aa,Inoue:2010es,Inoue:2011ai}.  
 Detailed comparison between the HAL QCD approach 
 and the approach discussed in this paper by using the same lattice data will
 be given in  independent publications under preparation
  and will not be discussed in the present paper.})
 as well as two different single-baryon operators (the non-relativistic type used in~\cite{Yamazaki:2012hi}  and the
 relativistic one)  to study the reliability of the plateau-like behavior from the 
 point of view of systematics.

 This paper is organized as follows.
In Sec.~\ref{sec:general}, we give general considerations on the plateau identification in multi-baryon system, and explicitly demonstrate the danger of the fake plateau using the mock-up data.
In Sec.~\ref{sec:params}, lattice simulation parameters used in this paper are summarized. In this paper, we consider the effective energy shift for $\Xi\Xi$ as well as $NN, 3N, 4N$ systems. In  Sec.~\ref{sec:XiXi}, we study the $\Xi\Xi$ systems in detail,
since signal to noise ratio ($S/N$) in lattice QCD is better for $\Xi$ than $N$. This is 
due to the fact that $\Xi$ contains two heavier strange quarks, while $N$ consists of lighter up and down quarks only.   
As demonstrated in Sec.~\ref{sec:general}, 
we observe plateau-like behaviors in the effective energy shift around $t\sim 1$ fm, which however disagree between the smeared source and the wall source.
We then discuss that  it is difficult to judge which plateau (or neither) is true only 
from the information of time correlation functions.
In Sec.~\ref{sec:NN} and \ref{sec:He}, we analyze the $NN,3N,4N$ systems in a similar manner. Although 
statistical errors are  larger, we observe similar disagreements between two sources as in the case of $\Xi\Xi$ systems.
In Sec.~\ref{sec:conclusion},
conclusions in this paper are given and some discussions follow.
In appendix \ref{sec:app:sink-dep}, we present the study on the sink operator dependence 
for effective energy shifts.
Disagreements are observed 
among plateau values from different sink operators for the smeared source, 
but not for the wall source.
In appendix \ref{sec:app:Eeff}, 
we collect the figures for effective energy shifts on various volumes.

\section{General considerations}
\label{sec:general}

\subsection{Difficulties in multi-baryon systems}
\label{sec:difficulty}

Even though the plateau method works in principle, and indeed in practice for, e.g., the ground state meson masses,
the method sometimes suffers from difficulties, in particular, in the case of multi-baryon systems.

First of all, 
we note that the requirement of the ground state dominance encounters a fundamentally new challenge
when one studies multi-hadron systems instead of single-hadron systems.
In fact, $\tmin$ required for the ground state dominance becomes much larger for multi-hadron states,
since $\delta E\equiv E_2 - E_1$ is much smaller where $E_1$ is the ground state energy while $E_2$ is the lowest excited state energy.
For example,  in the case of bound states, $\delta E$ is a few MeV for deuteron and a few tens of MeV for $^4$He.
With the absence of bound states as is the case for dineutron or diproton, 
there exist only continuum states and thus no intrinsic energy gap exists.
In lattice calculations,
the energy spectrum is discretized in a finite box with the spatial extension $L$,
leading to  $\delta E\simeq (2\pi)^2/(L^2 m_N)$, which 
 is also small as, for instance,  $\delta E \lesssim
 25$ MeV
 at large enough $L \gtrsim 8$ fm
 for two baryons at the physical quark masses.
These small splittings are in sharp contrast to the single-hadron systems, where 
$\delta E \sim {\cal O}(\Lambda_{\rm QCD})$.

The requirement of taking large $t$ causes a serious difficulty in lattice QCD,
since the data at larger $t$ are in general accompanied with much worse $S/N$.
The situation is severe in particular for multi-baryon systems at large $t$, for which we have~\cite{Parisi:1983ae,Lepage:1989hd}
\begin{eqnarray}
\frac{S_A(t)}{N_A(t)} &\sim& \exp\left[ -A \left(m_B-\frac{3m_M}{2}\right) t \right],  
\label{eq:S2N}
\end{eqnarray}
where $m_B$ and $m_M$ are the ground state baryon mass and the meson mass coupled to
 the $B\bar{B}$ annihilation channel, respectively. 
 The signal $S_A(t)$ is given by a propagator for an $A$-baryon system, 
 schematically denoted as
\begin{eqnarray}
  S_A(t) &=& \langle [B(t)]^A    [\bar{B}(0)]^A \rangle 
\end{eqnarray}
with (zero momentum) baryon creation and annihilation operators $\bar B(t)$ and $B(t)$, while the noise $N_A(t)$ is given by
\begin{eqnarray}
N_A(t)^2 &=& \langle \left\vert [B(t)]^A    [\bar{B}(0)]^A \right\vert^2 \rangle - \vert S_A(t) \vert^2 .
\end{eqnarray}
The asymptotic formula Eq.~(\ref{eq:S2N}) says that $S/N$ 
becomes worse for bigger $t$ as well as larger numbers 
of baryons and/or smaller quark mass ({\it i.e.} lighter meson).
This may prevent us from taking sufficiently large $t$ to guarantee
the $t$ independence of 
$E_A^{\rm eff}(t)$, so that
we can not reliably control  systematic errors from excited state contaminations. 

In order to demonstrate the danger of such excited state contaminations, we
consider the mock-up data given by
\begin{eqnarray}
  R(t)
  = b_1 e^{-\Delta E_{BB}t} + b_2 e^{-(\delta E_{\rm el}+\Delta E_{BB}) t}+c_1 e^{-(\delta E_{\rm inel} +\Delta E_{BB}) t},
\end{eqnarray}
where $\Delta E_{BB} = E_{BB} - 2 m_B$ with the ground state energy $E_{BB}$, while
$\delta E_{\rm el} = E^*_{BB} - E_{BB}$ and $\delta E_{\rm inel} = E_{\rm inel} - E_{BB}$
with the first excited elastic state energy  $E^*_{BB}$ and the lowest inelastic state energy $E_{\rm inel}$, respectively. 
Thus the effective energy shift becomes
\begin{eqnarray}
  \DelEeffBB(t)
  &\equiv& -\frac{1}{a}\log \left(\frac{R (t+a)}{R (t)}\right) \nonumber \\
  &=&  \DelEBB
  - \frac{1}{a}\log \left(\frac{1+(b_2/b_1)\cdot e^{-\delta
    E_{\rm el}(t+a)}+(c_1/b_1)\cdot e^{-\delta  E_{\rm inel}(t+a)}}
  {1+(b_2/b_1)\cdot e^{-\delta  E_{\rm el}  t}
+(c_1/b_1)\cdot e^{-\delta E_{\rm inel} t}}\right) ,
\end{eqnarray}
so that
$\DelEeffBB(t) - \DelEBB$ corresponds to the  deviation in $\DelEeffBB(t)$ from its true value.
Note that  both $b_2/b_1$ and $c_1/b_1$ can be negative if source and sink operators are different.
As an example,
we consider $\delta E_{\rm el} = 50$ MeV, which is the typical lowest excitation energy of elastic two-baryon scattering states in our numerical setup with $La = 4.3$ fm lattice
(see Sec.~\ref{sec:params}), while we take $\delta E_{\rm inel} = 500$ MeV, 
which is roughly the order of 
$m_\pi$  in our simulations.
In lattice QCD, one often tries to tune the interpolating operator 
so that excited state contaminations are suppressed.
Since the difference between inelastic states and the ground state is expected to be intrinsic in QCD,
one may ideally take a good operator for baryons which have small overlaps with inelastic states.
We therefore adopt a small value $c_1/b_1=0.01$
as the contamination from the inelastic state.
On the other hand, 
it is much more difficult to separate the ground state from the elastic excited state
by tuning the operator,
since the difference between these states do not originate from QCD, but from 
the use of a finite lattice box.
Accordingly, 
we take $b_2/b_1= \pm 0.1$ as well as $b_2/b_1=0$ for a comparison,
as the contamination of  the excited elastic state.

\begin{figure}[tb]
\centering
  \includegraphics[width=0.49\textwidth]{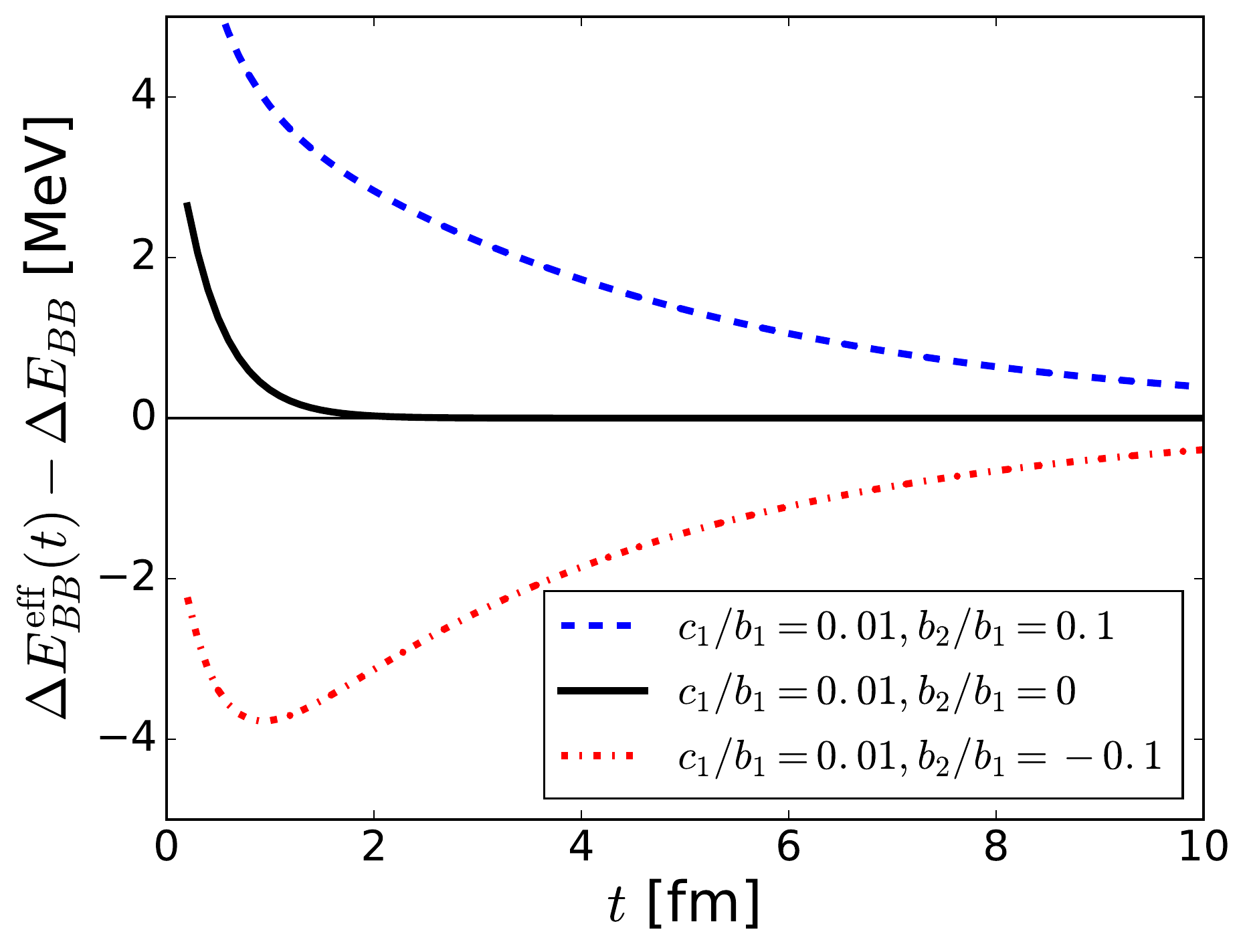}
  \includegraphics[width=0.49\textwidth]{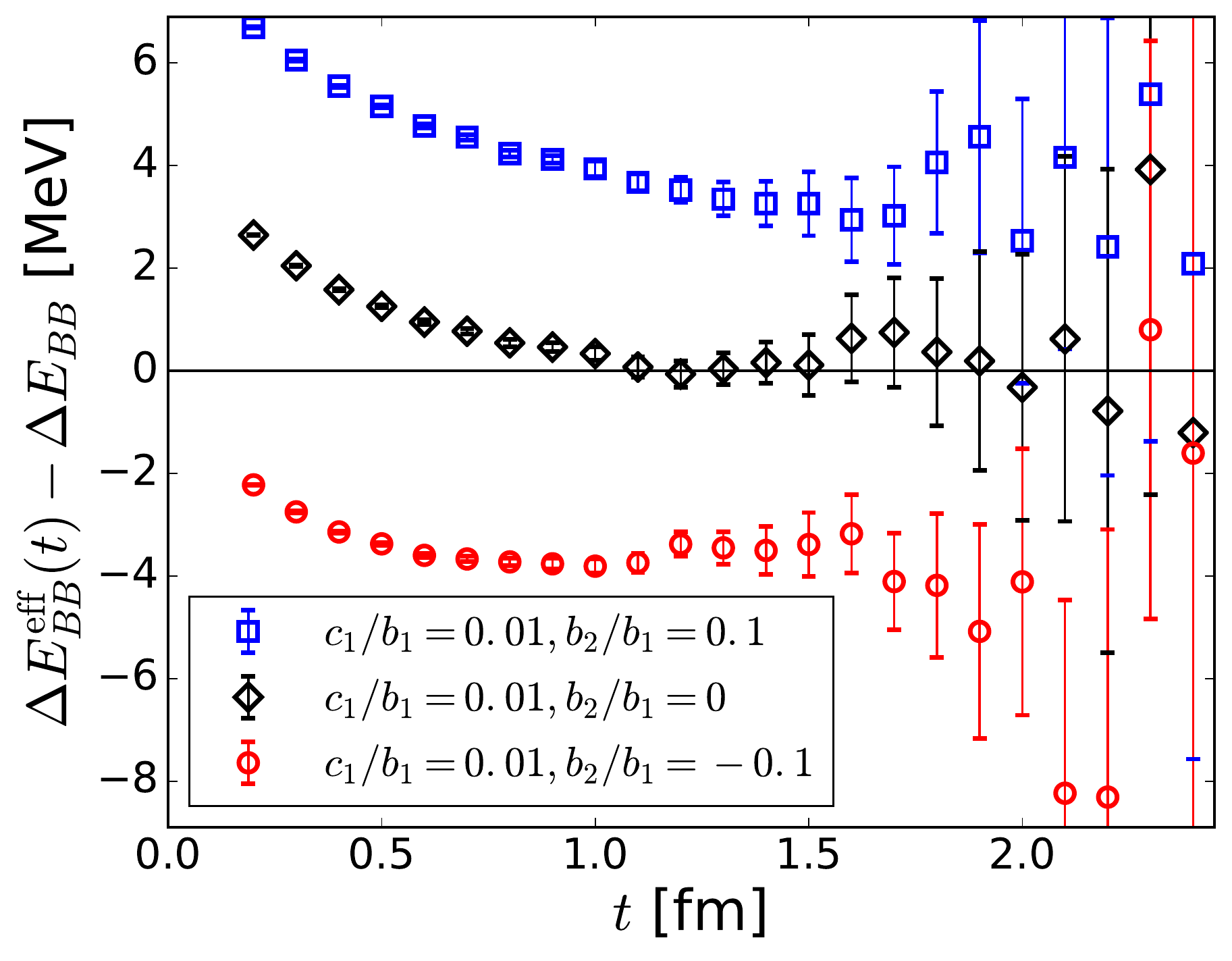}
 \caption{
   (Left) $\DelEeffBB(t) - \DelEBB$ as a function of 
   $t$ for $\delta  E_{\rm el} = 50$ MeV, $\delta E_{\rm inel} = 500$ MeV 
   and $c_1/b_1=0.01$ with $b_2/b_1=0$ (black solid line), 
   $b_2/b_1=-0.1$ (red dot-dashed line) 
   and $b_2/b_1= 0.1$ (blue dashed line). 
   (Right) Discrete data with fluctuations and errors for $t \le 2.5$ fm.
}
 \label{fig:demo}
\end{figure}

In Fig.~\ref{fig:demo} (Left), we plot $\DelEeffBB(t) - \DelEBB$ as a function of $t$  for the above choice of parameters.   
Let us consider the case with $b_2/b_1 =0$ (black line) first. In the absence of the excited elastic  state,
the effective energy shift $\DelEeffBB(t)$  smoothly approaches to the plateau 
(from above for the positive $c_1/b_1$) 
and $t \lesssim
 1$ fm is  sufficient to reduce the systematic error from the contamination to the level of accuracy we need for $\DelEBB$. 
 Unfortunately, this ideal situation cannot be realized in practice, and
for a more realistic cases with $\pm 10$ \%
contamination of the 1st excited elastic state at $t=0$, we need $t \gtrsim 8-10$~fm  to achieve  the level of accuracy we need, as shown  by red and  blue lines.
In practice, however, $\tmin\simeq 8-10$~fm is too large to have a good signal due to the exponentially  decreasing signal to noise ratio  for multi-baryons as mentioned before.
Shown in  Fig.~\ref{fig:demo} (Right)
are $\DelEeffBB(t)-\DelEBB$ 
as a function of the discrete time (integer $t/a$ with lattice spacing $a=0.1$ fm) for $ t \le 2.5$ fm,
which would appear in typical numerical simulations.
To obtain the data in this demonstration,
we  assign random fluctuations to $R(t)$ whose magnitude increases exponentially in time and
is comparable to that of our lattice data, and then calculate
the central value and statistical error of $\DelEeffBB(t)$ at each $t$.
This figure clearly demonstrates that it is almost impossible to have data with enough accuracy 
 at $t \simeq 8 - 10$ fm in current simulations.

Another point which is noteworthy in
Fig.~\ref{fig:demo} (Right) is that  
the plateau-like behaviors show up at $t\simeq 1 - 2$ fm.
Provided that $t \simeq 1-2$~fm is the region
  where statistical errors for two-baryon system
  may be controlled in present-day lattice simulations,
one may easily misidentify this plateau-like behaviors as a real plateau.
The estimate for $\DelEBB$ then contains 
the systematic error of $\pm 4$ MeV ($b_2/b_1= \pm 0.1$),
which is significant  to the typical value of $\DelEBB$, 10 MeV or less, 
for the two-baryon system.

The behaviors demonstrated in Fig.~\ref{fig:demo} certainly depend on parameters such as
$\delta  E_{\rm el}$, $\delta E_{\rm inel}$, $c_1/b_1,b_2/b_1$,
and a fake plateau may or may not appear in a specific lattice QCD simulation.
There exists a potential danger, however,
that a fake plateau appears during a search of a plateau at accessible $t$, by tuning, for example,
the interpolating operators. Thus
it is always necessary to find the explicit evidence that 
the obtained plateau-like structure is not fake.
Due to the 
exponentially increasing noise in time, 
this task is extremely difficult, and
becomes even impossible practically   at physical quark masses 
with a larger lattice box, since  $\delta  E_{\rm el}$ becomes much smaller 
as discussed before.

\subsection{Fitting range for temporal correlations}
\label{sec:procedure}

In the following sections, we will analyze the lattice data to show explicitly the 
 problem raised in Sec.~\ref{sec:difficulty}. 
   In the time correlation analysis of the actual lattice data for two baryons, 
   one can only utilize the data at the moment up to $t=\tmax \sim$ 2 fm in the temporal direction
 due to the exponential decrease of $S/N$ in $ E_{BB}^{\rm eff}(t)$ for large $t$.
  Also, the lower limit of $t=\tmin$ is constrained by the ground state saturation by 
  a single hadron in   $m_{B}^{\rm eff}(t)$.  Therefore, 
  a practical procedure adopted in many of the previous works are 
  to look for the plateau of $\Delta E_{BB}^{\rm eff}(t) = E_{BB}^{\rm eff}(t) - 2 m_{B}^{\rm eff}(t)$
  under the  expectation that some cancellation of systematic as well as statistical errors. 
  We will adopt  the same practical procedure below for choosing the fitting window in the temporal direction, and
   show that the procedure leads to inconsistent results as expected.

\section{Lattice parameters}
\label{sec:params}

In this paper, we employ the same gauge configurations in
Ref.~\cite{Yamazaki:2012hi},
i.e., 2+1 flavor QCD with the Iwasaki gauge action at $\beta=1.90$ and 
the nonperturbatively $\mathcal{O}(a)$-improved Wilson quark action  at $c_{\rm SW} = 1.715$~\cite{Aoki:2005et}. 
The lattice spacing determined from $m_\Omega = 1.6725$ GeV is
$a=0.08995(40)$ fm ($a^{-1} = 2.194(10)$ GeV).
While we take the physical value of the strange quark mass, we employ heavier degenerate up and down quark masses
with hopping parameters $(\kappa_{ud}, \kappa_s) = (0.1373316, 0.1367526)$,
which corresponds to $m_\pi = 0.51$ GeV, $m_N = 1.32$ GeV and $m_\Xi = 1.46$ GeV.
We use four lattice sizes as adopted in
Ref.~\cite{Yamazaki:2012hi}, $L^3\times T = (32^3, 40^3,48^3)\times 48$,  
and $64^3\times 64$, corresponding to $La=2.9, 3.6, 4.3$ and 5.8 fm, respectively.

For measurements of multi-baryon correlation functions, we employ two different
sources, one is the smeared quark source, the other is the wall quark source, to check
whether plateau-like behaviors agree between two sources.  For the
smeared source, we take exactly the same smearing function and parameters used in
Ref.~\cite{Yamazaki:2012hi}: Quark propagators are solved 
using the exponentially smeared source of the form that%
\footnote{
Smearing function in Eq.~(\ref{eq:smearing}) is slightly different 
from the one written in Eq.~(12) of Ref.~\cite{Yamazaki:2012hi}.
In reality, we were notified that the one in Eq.~(\ref{eq:smearing}) is the actual formula used in Ref.~\cite{Yamazaki:2012hi}.
We thank T.~Yamazaki for the information.
}
\begin{align}
  q_s(\vec x, t) &= \sum_{\vec{y}} f(\vert\vec{x}-\vec{y}\vert) q(\vec{y},t)
  \quad \text{with}
  \quad
  f(r) \equiv
\begin{cases}
  A e^{-Br} 
\quad \text{for} \ 0 < r < (L-1)/2,  \\
 1 \quad \text{for} \ r = 0,  \\ 
 0 \quad \text{for} \ (L-1)/2 \leq r, \\
 \end{cases} 
\label{eq:smearing}
\end{align}
after the Coulomb gauge fixing is applied to gauge configurations. 
For the wall source, 
we take
\begin{eqnarray}
q_w( t) &=& \sum_{\vec y} q(\vec y,t).
\label{eq:wall}
\end{eqnarray}

Relativistic interpolating operators for proton, neutron and $\Xi$ are given by
\begin{eqnarray}
p_\alpha &=& \epsilon_{abc} ( u^{a\,T} C\gamma_5 d^b) u^c_\alpha,\qquad
n_\alpha = \epsilon_{abc} ( u^{a\,T} C\gamma_5 d^b) d^c_\alpha,\nonumber \\
\Xi^0_\alpha &=&  \epsilon_{abc} ( s^{a\,T} C\gamma_5 u^b) s^c_\alpha, \qquad
\Xi^-_\alpha =  \epsilon_{abc} ( s^{a\,T} C\gamma_5 d^b) s^c_\alpha,
\label{eq:operator}
\end{eqnarray}
where $C=\gamma_4\gamma_2$ is the charge conjugation matrix, 
$\alpha$ and $a,b,c$ are the spinor index and color indices, respectively. 
Non-relativistic operator exclusively used in Ref.~\cite{Yamazaki:2012hi}
  is obtained by replacing $C\gamma_5$ in Eq.~(\ref{eq:operator})
by $C\gamma_5 (1 + \gamma_4)/2$.
We employ both non-relativistic and relativistic operators
in this paper to estimate the systematic errors from the different choices.

For the source operators, we insert $q_s$ or $q_w$  in each flavor of Eq.~(\ref{eq:operator}) or its non-relativistic variant.
In the case of the smeared source, we take the same $\vec x$ for all  quarks in Eq.~(\ref{eq:operator}), 
as is done in Ref.~\cite{Yamazaki:2012hi}.
For sink operators, on the other hand, 
each baryon operator is composed of point quark fields, 
and projected to zero spatial momentum by averaging over the spatial position.
For the choice of relativistic and non-relativistic baryon operator, 
we consider the same choice at both source and sink in this study.
In the case of $^4$He, however, 
the non-relativistic nucleon operator is used for the source regardless of the choice for the sink operator,
in order to reduce the numerical cost.
Altogether, we consider four different combinations for each multi-baryon system, 
two from wall and smeared quark sources times two from relativistic and non-relativistic baryon operators. 

Quark propagators are solved with the periodic boundary condition in all directions
using the quark source described above.
Correlation functions (with relativistic and non-relativistic baryon operators)
are then calculated accordingly,
where we use the unified contraction algorithm (UCA)~\cite{Doi:2012xd}.
UCA  significantly reduces the computational cost of correlation functions,
in particular for those of $^3$He and $^4$He.
(See also related works~\cite{Yamazaki:2009ua,Detmold:2012eu,Gunther:2013xj,Nemura:2015yha}.)

On each gauge configuration, we repeat the measurement of correlation functions 
for a number of smeared sources at different spatial point and time slices  and a number of wall sources at  different time slices.
For the $48^3 \times 48$ and $64^3 \times 64$ lattices, correlation functions are calculated not only in  one direction 
but also in other three as the time direction on each configuration
using the rotational symmetry.
In order to reduce the computational cost for the quark solver,
the following stopping conditions $|\rcrit|$ for the residual error are employed:
$|\rcrit| = 10^{-4} (10^{-12})$ for smeared (wall) source on the $32^3\times 48$ lattice, 
$|\rcrit| = 10^{-6} (10^{-4})$  for smeared (wall) source on the $40^3\times 48$ lattice, 
$|\rcrit| = 10^{-6} (10^{-4})$  for smeared (wall) source on the $48^3\times 48$ lattice,
$|\rcrit| = 10^{-6}$ for smeared source on the $64^3\times 64$ lattice and 
$|\rcrit| = 10^{-6}$ (for half of the total statistics) or $10^{-12}$ (for the other half) for wall source on the $64^3\times 64$ lattice.
In all cases, we check that systematic errors associated with the choice of the stopping condition is much smaller 
than the statistical fluctuations  in this study.
Nonetheless, we correct these errors
by using the all-mode-averaging (AMA) technique~\cite{Blum:2012uh,Shintani:2014vja}
with the translational invariance.%
\footnote{
Rigorously speaking, there exists a possible bias in our AMA corrections
associated with the numerical round-off errors~\cite{Blum:2012uh,Shintani:2014vja}.
Such bias, however, is expected to be negligible
since the magnitude of AMA correction themselves are already small 
in our relatively conservative choice for $|\rcrit|$.
}
Here, the AMA corrections for relaxed stopping conditions of $|\rcrit| = 10^{-4}$ or $10^{-6}$ data
are obtained by the corresponding computations with ``exact'' solver ($|\rcrit| = 10^{-12}$)
with the following measurements:
$1$                        source  for smeared        source on the $32^3\times 48$ lattice, 
$1$          (2)           sources for smeared (wall) source on the $40^3\times 48$ lattice, 
$4\times 1$  ($4\times 2$) sources for smeared (wall) source on the $48^3\times 48$ lattice and
$1\times 1$  ($4\times 1$) sources for smeared (wall) source on the $64^3\times 64$ lattice,
where the factor of 4 or 1 for the $48^3 \times 48$ and $64^3 \times 64$ lattices 
denotes the enhancement factor in statistics by the rotational symmetry.

\begin{table}[tb]
\begin{center}
\begin{tabular}{|cc|c|cc|c|}
\hline
size & $La$ & \# of conf & \# of smeared sources& $(A, B)$ & \# of wall sources \\
\hline
$32^3\times 48$ & 2.9 fm & 402 & 384 &  (1.0, 0.18) & 48 \\
$40^3\times 48$ & 3.6 fm & 207 & 512 &  (0.8, 0.22) & 48 \\
$48^3\times 48$ & 4.3 fm & 200 & $4\times 384$  & (0.8, 0.23) & $4\times 48$ \\
$64^3\times 64$ & 5.8 fm & 327 & $1 \times 256$ & (0.8, 0.23) & $4\times 64$ \\
\hline
\end{tabular}
\caption{Lattice size, \# of configurations,  
  \# of smeared sources and wall sources on each configuration, and smearing
  parameters $(A,B)$. The factor of 4 in \# of sources for $48^4$ and $64^4$ means
  that all 4 directions ($x,y,z,t$) are used as the time direction. 
}
\label{tab:param}
\end{center}
\end{table}

The lattice parameters, and 
the number of configurations as well as 
the number of smeared sources and wall sources
are tabulated in Table~\ref{tab:param}. 
As noted above, the number of measurements for the $48^3 \times 48$ and $64^3 \times 64$ lattices can be increased 
by exploiting the rotational symmetry,
and the factor of 4 in Table~\ref{tab:param} represents 
this enhancement.
In addition, 
for each measurement on any lattice volumes, we calculate correlation functions in forward and backward propagations ($t > 0$ and $t < 0$, respectively)
and take an average to improve the signal. The corresponding 
factor of 2 is not included in Table~\ref{tab:param}.
We note that the numbers 
of configurations and measurements 
for the smeared source in this work are much larger than those in~\cite{Yamazaki:2012hi}.
As [\# of conf. $\times$ \# of smeared sources] in Ref.~\cite{Yamazaki:2012hi} is 
$[200 \times 192]$, $[200 \times 192]$, $[200 \times 192]$ and $[190 \times256]$ 
on a lattice volume with $La = 2.9, 3.6, 4.3$ and $5.8$ fm, respectively,
the ratio of the number of measurements  in this work to Ref.~\cite{Yamazaki:2012hi} amounts to be
about 4.0, 2.8, 8.0 and 1.7 for each volume.

In our analyses, statistical errors are estimated by the jackknife method.
We find that the auto-correlation in terms of configuration trajectory is small 
by observing that the statistical errors are almost independent 
among the choices of bin-size of 2, 5, 10, 20 configurations.
Hereafter, we show the results obtained with the bin-size of 
10 configurations (100 trajectories), unless otherwise stated.

\section{$\Xi\Xi$ systems} 
\label{sec:XiXi} 

\subsection{$\Xi\Xi$ $(^1S_0, ^3S_1)$ with smeared source}
\label{subsec:XiXi_smeared}

Let us first consider the $\Xi\Xi$ system 
in the spin-singlet channel with zero orbital angular momentum, $\Xi\Xi({}^1S_0)$, where 
the interpolating operator is given by $ \Xi^Q_{1}\Xi^Q_{2} - \Xi^Q_{2}\Xi^Q_{1} $ with $Q=0,-$.
The reasons to choose this channel  are twofold:
Firstly, the signal to noise ratio for strange baryons  is better than non-strange baryons.
Secondly,  in the flavor SU(3) limit,  it belongs to the same  {\bf 27} multiplet as the $NN({}^1S_0)$, 
so that one may obtain some insights into  the bound dineutron suggested in previous works.
To make a firm connection to previous works, we 
 start our analyses with  the smeared source Eq.~(\ref{eq:smearing}) and later consider the 
 case with the wall source.

\begin{figure}[tb]
  \centering
  \includegraphics[width=0.47\textwidth]{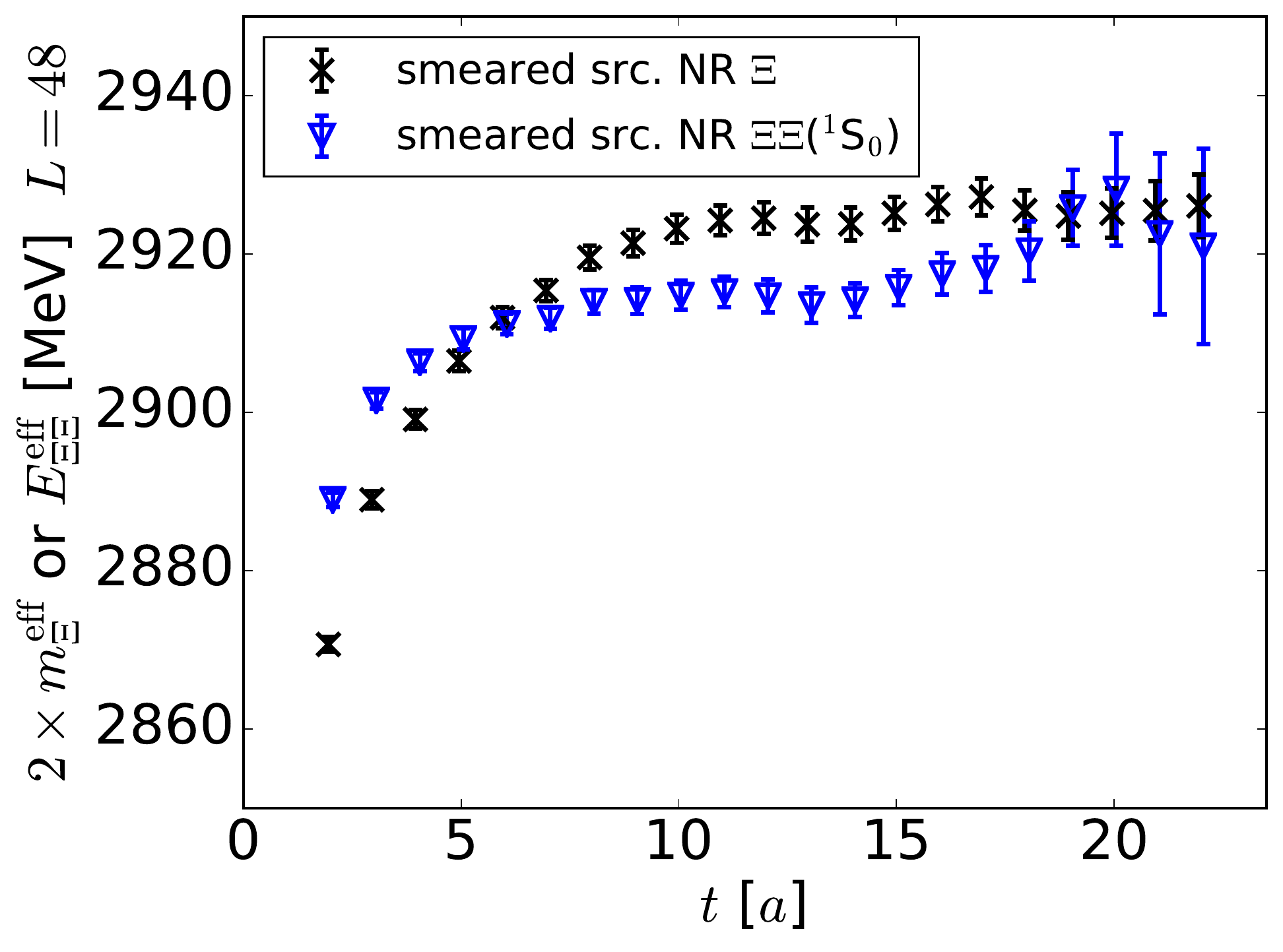}
  \includegraphics[width=0.47\textwidth]{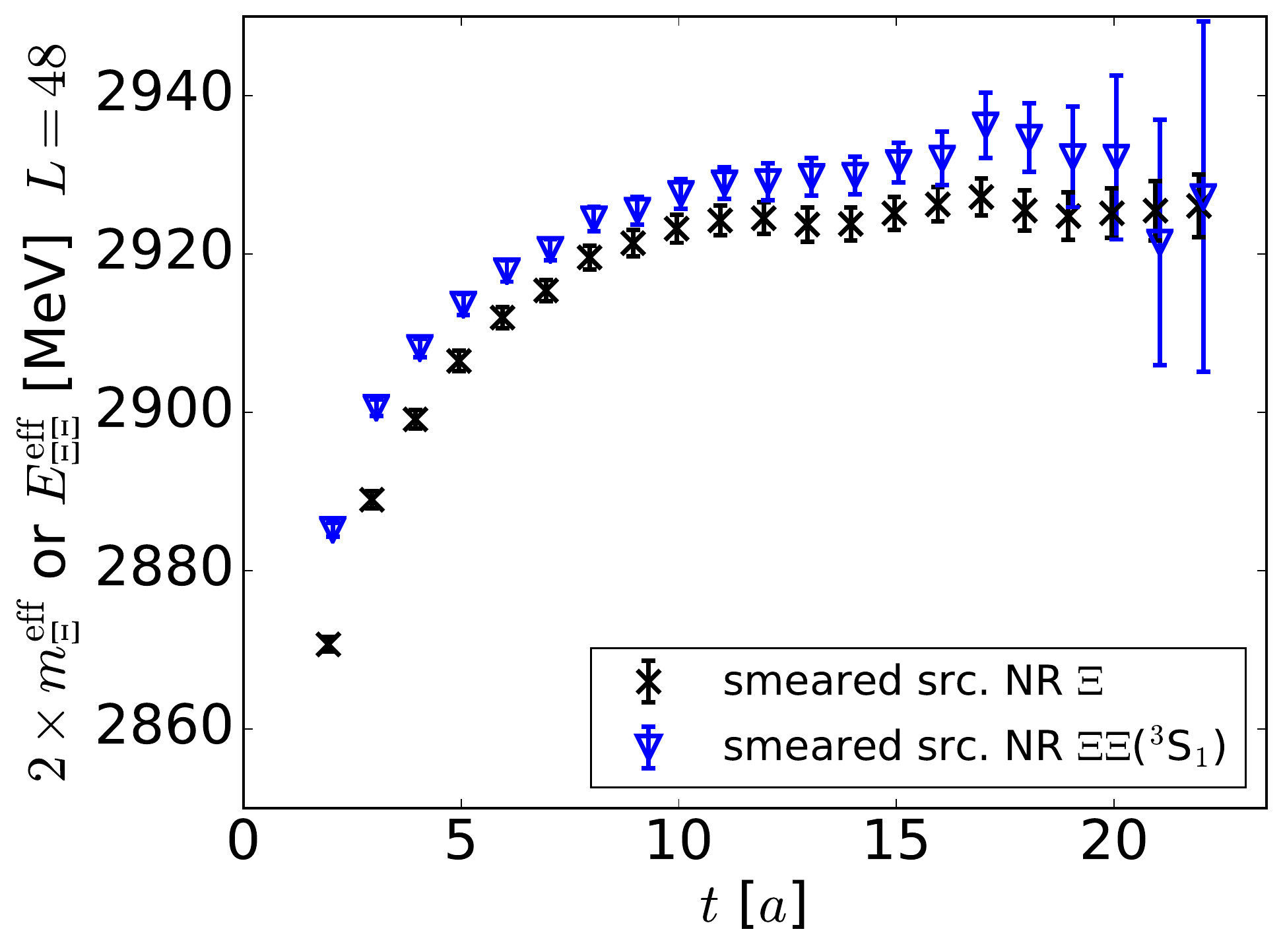}\\
  \includegraphics[width=0.47\textwidth]{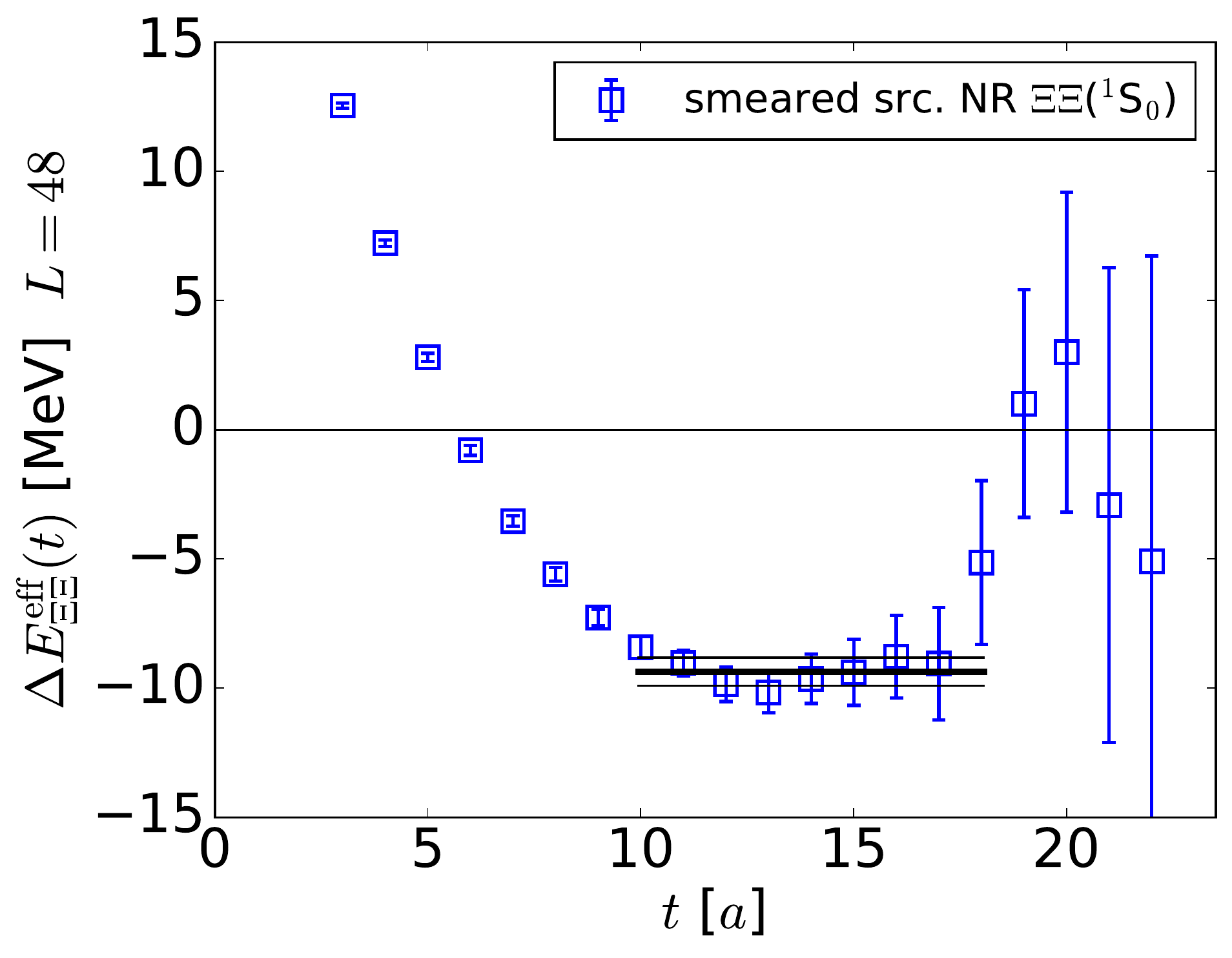}
  \includegraphics[width=0.47\textwidth]{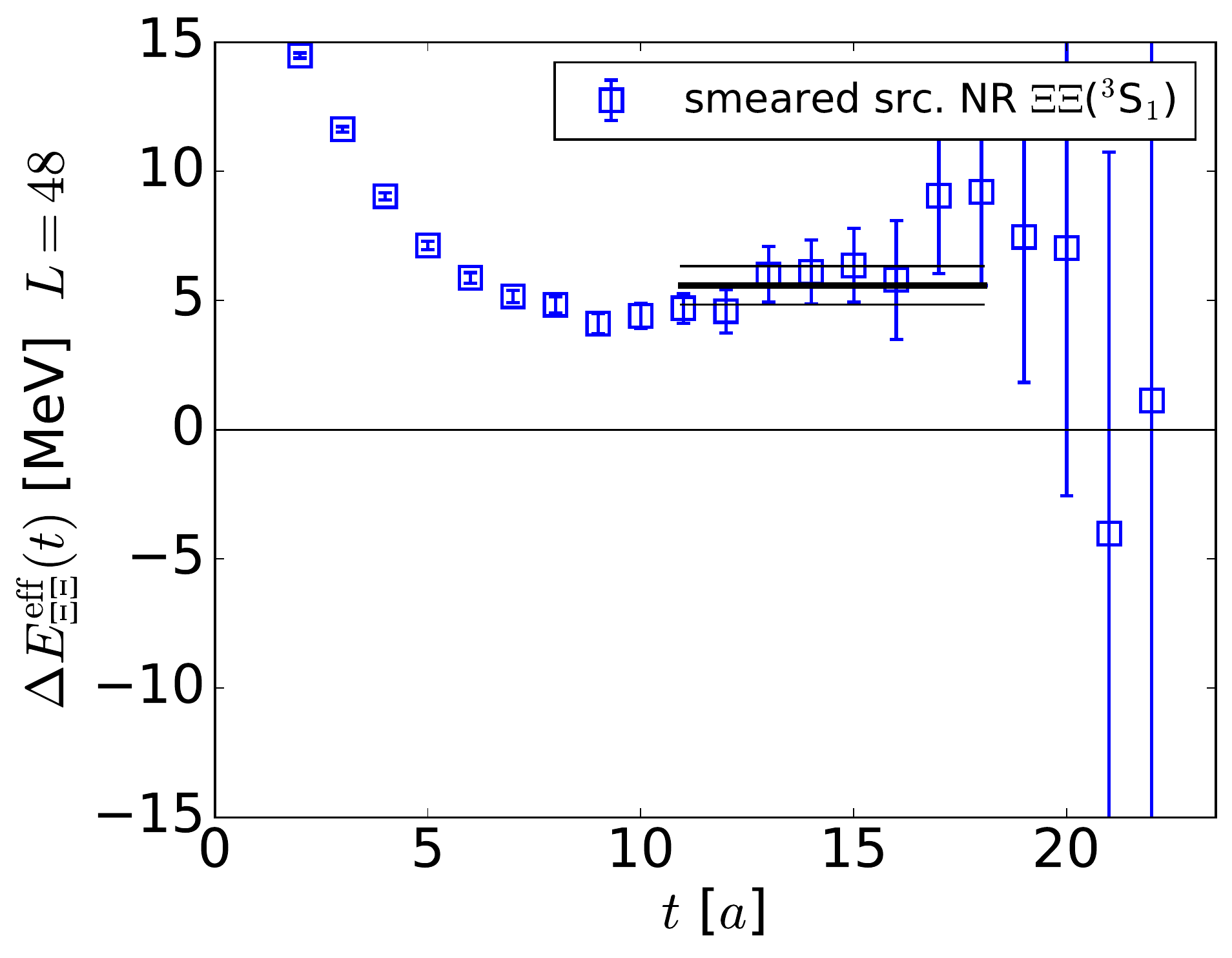}\\
  \includegraphics[width=0.47\textwidth]{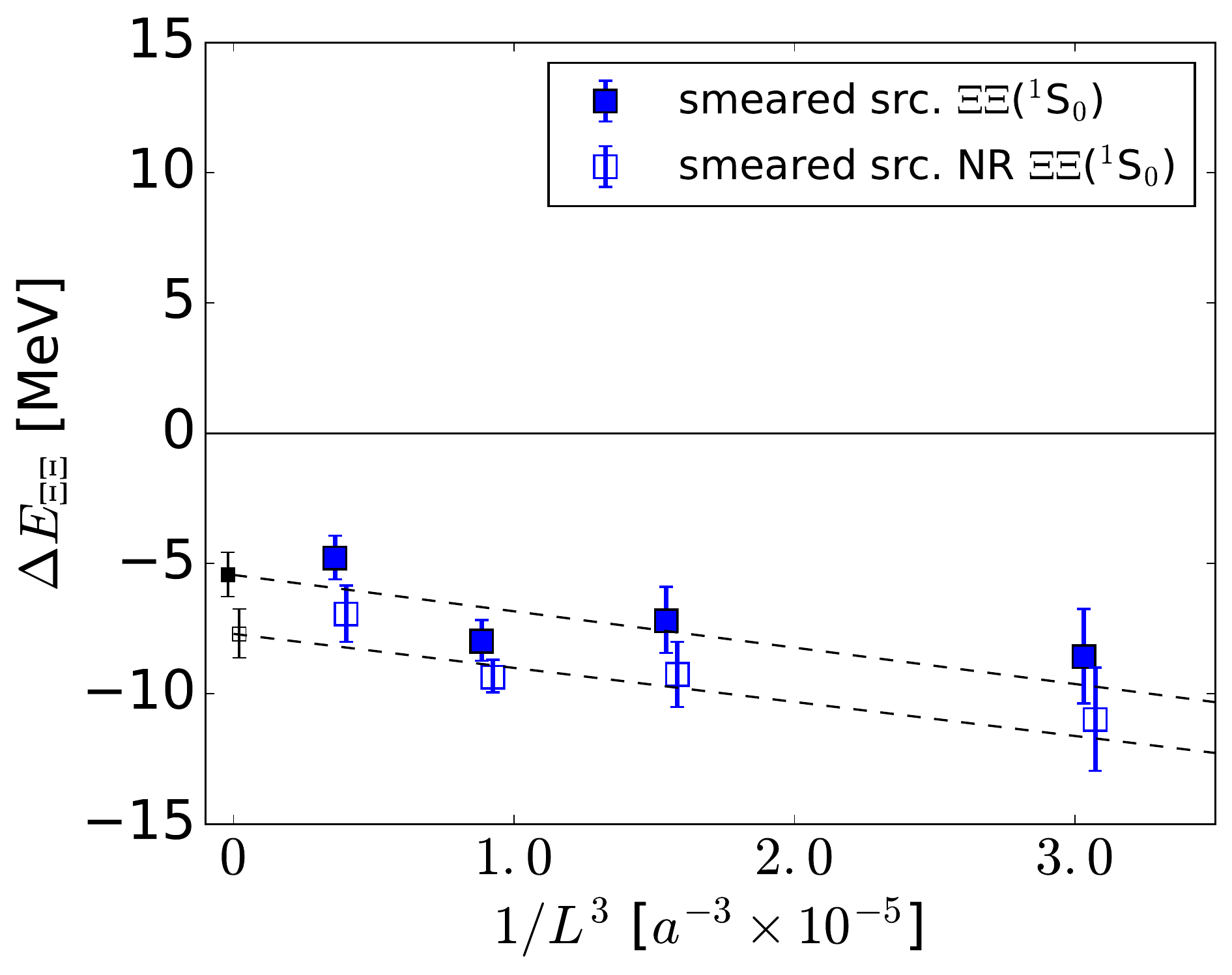}
  \includegraphics[width=0.47\textwidth]{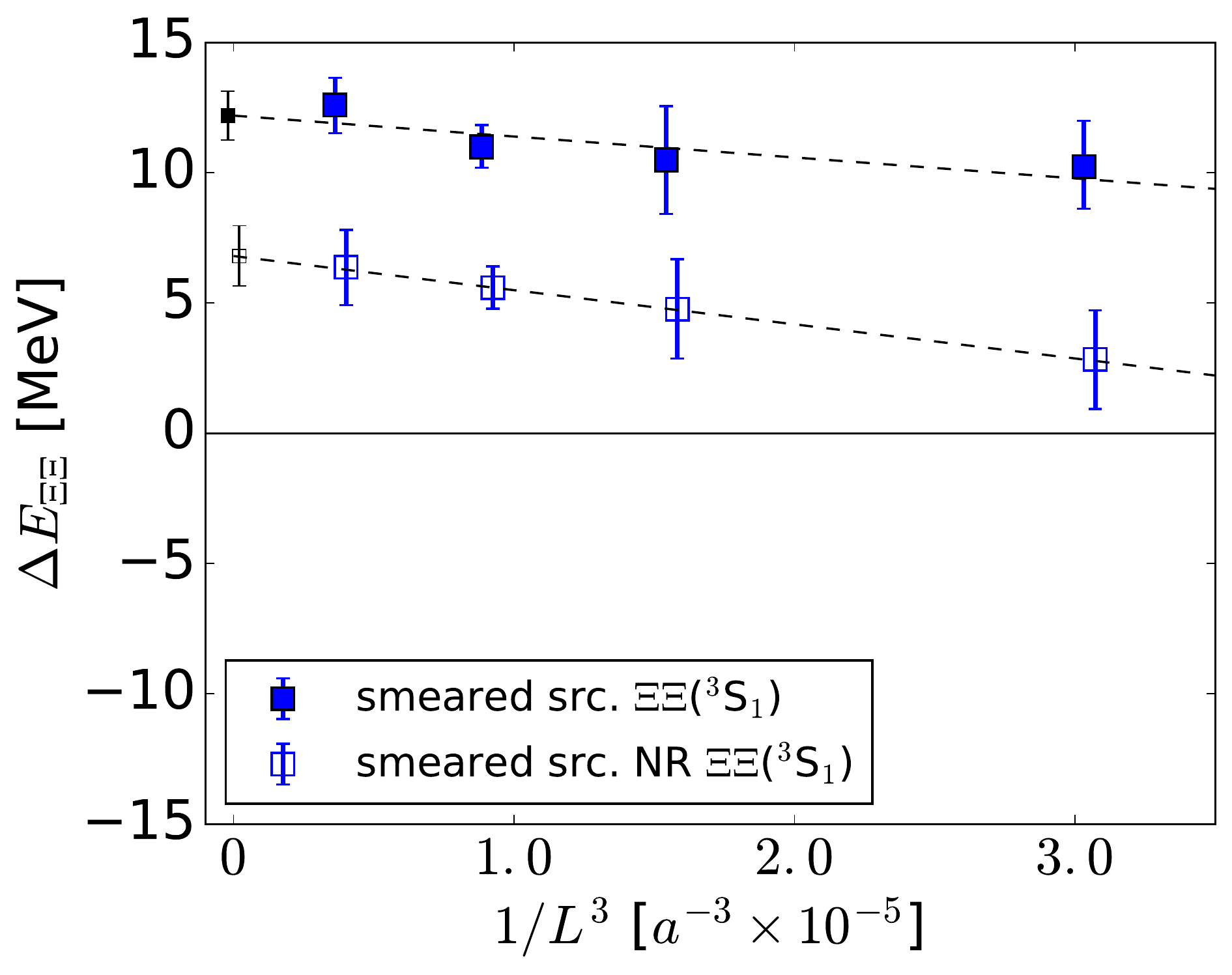}
  \caption{(Upper left) Effective mass $2\meffXi(t)$ (black cross) and effective energy $\EeffXiXi(t)$ (blue triangle)
  in the $\Xi\Xi({}^1S_0)$ channel as a function of $t/a$ on the $48^3\times 48$
  lattice  from the smeared  source with the non-relativistic operator.
  (Middle left) Effective energy shift $\DelEeffXiXi(t) \equiv \EeffXiXi(t) - 2 \meffXi (t)$,
  together with the fit (statistical only)  in the $\Xi\Xi({}^1S_0)$ channel.
  (Lower left)  The energy shift $\DelEXiXi$ in the $\Xi\Xi({}^1S_0)$ channel
   as a function of $1/L^3$
   from the smeared source with the non-relativistic operator (open square) 
   as well as the relativistic one (solid square), 
   together with their infinite volume extrapolations. 
   The errors are obtained from statistical and systematic errors added in quadrature.
 (Upper right, Middle right, Lower right) Same quantities  in the $\Xi\Xi({}^3S_1)$ channel. 
}
 \label{fig:Eff_Xi_smear}
\end{figure}

Fig.~\ref{fig:Eff_Xi_smear} (Upper left) shows $ 2 \meffXi(t)$ (black cross) and $\EeffXiXi(t)$ (blue triangle)  for non-relativistic interpolating operators
on the $48^3 \times 48$ lattice, while Fig.~\ref{fig:Eff_Xi_smear} (Middle left) shows the 
errors and fluctuations of the effective energy shift, $\DelEeffXiXi(t)=\EeffXiXi(t) - 2 \meffXi (t)$.

 One finds a plateau-like behavior in  Fig.~\ref{fig:Eff_Xi_smear} (Middle left) for 
 $10 \le t/a \le 18$
 before the explosion of the noise over the signal  for larger $t$.
 As we have argued in Sec.~\ref{sec:general} by the mock data,
 such a plateau is likely to be fake due to the contamination of  the higher scattering states. 
  Nevertheless, following the practical procedure taken by the previous works,
 let us try 
 an exponential fit of $R_{\Xi\Xi}(t)$ in this ``plateau" region
 and to take a large volume extrapolation.
 The  fitted result is shown by the horizontal bars (the thick line and the thin lines are the central value and the 1$\sigma$ statistical errors, respectively). 
 We perform similar analyses for other volumes and also 
 for relativistic interpolating operators, and
 results for $\DelEXiXi (^1S_0)$ are summarized in Table~\ref{tab:summary_XiXi}.
The numbers in the first parenthesis denote the statistical error, 
while the numbers in the second parenthesis denote systematic errors from the fit.
  Taking the same criterion adopted in Ref.~\cite{Yamazaki:2012hi},
we estimate the systematic errors by variations among the fitting window as [$t_\mathrm{min} \pm 1, t_\mathrm{max}\pm 1]$
\footnote{We have checked that the window $[t_\mathrm{min} \pm 2,
    t_\mathrm{max} \pm 2]$ gives 
almost the same results as far as the fit is stable.}.

 In Fig.~\ref{fig:Eff_Xi_smear} (Lower left), plotted as a function of $1/L^3$ is  
 $\DelEXiXi (^1S_0)$, with statistical and systematic errors added in quadrature, 
 together with the value at infinite volume obtained by linear extrapolation.
 At $L \rightarrow \infty$,
 we find  $\DelEXiXi= -[7.70(0.89)(^{+0.37}_{-0.20})]$ MeV (non-relativistic operator) and
$ -[5.44(0.82)(^{+0.28}_{-0.09})]$ MeV (relativistic operator).
They indicate  the existence of a  $\Xi\Xi$ bound state in the  $^1S_0$ channel, which 
is qualitatively ``consistent" with  previous studies finding dineutron bound state at this
quark mass. Obviously the big question is whether such conclusion is reliable or not 
as we have discussed in Sec.~\ref{sec:general}.

To answer the above question solely in terms of the lattice data, 
let us move on to analyze $\Xi\Xi (^3S_1)$  with the same fitting procedure.
In this case, the  interpolating operator is given by $\Xi_\alpha^0\Xi_\alpha^{-} -
\Xi_\alpha^{-}\Xi_\alpha^{0}$ with $\alpha=1,2$. In the flavor SU(3) limit,  
this channel is in the {\bf 10} multiplet where no $NN$ channels belong to. 

One finds again that a ``plateau''-like behavior in $11 \le t/a \le 18$
before the explosion of the 
noise over the signal in larger $t$ as shown in Fig.~\ref{fig:Eff_Xi_smear}
(Middle right) in the case of non-relativistic operator on the $48^3\times
48$ lattice.  We perform the same analysis in other
volumes and also for relativistic interpolating operators.  In
Fig.~\ref{fig:Eff_Xi_smear} (Lower right),  $\DelEXiXi (^3S_1)$, with
statistical and systematic errors added in quadrature, is plotted as a function
of $1/L^3$, together with the values at infinite volume obtained by linear
extrapolation.  The results are summarized in Table~\ref{tab:summary_XiXi} with
the infinite volume limit,
$\DelEXiXi (^3S_1) = 6.81(1.04)(^{+0.52}_{-0.48})$ MeV (non-relativistic operator) and 
$12.20(94)(^{+0.02}_{-0.12})$ MeV (relativistic operator).

These results in the $^3S_1$ channel clearly indicate that the procedure to analyze the data was wrong as expected.
If one could correctly identify the ground state energy of a two particle system on the finite lattice,
its infinite volume extrapolation must be either  zero (for the scattering state) or negative (for the bound state):
Positive definite $\DelEXiXi (^3S_1)$ as seen in Fig.~\ref{fig:Eff_Xi_smear} (Lower right)  cannot be allowed.  
Therefore, we conclude that
the plateaux seen in $\DelEeffXiXi (t)$ for spin-singlet and spin-triplet channels are fake and 
are likely to be the mirages of true plateaux located in much larger $t$ as we have discussed in Sec.~\ref{sec:general}.

\subsection{$\Xi\Xi$ $(^1S_0, ^3S_1)$ with wall source}
\label{subsec:XiXi_wall}

 \begin{figure}[tb]
  \centering
  \includegraphics[width=0.47\textwidth]{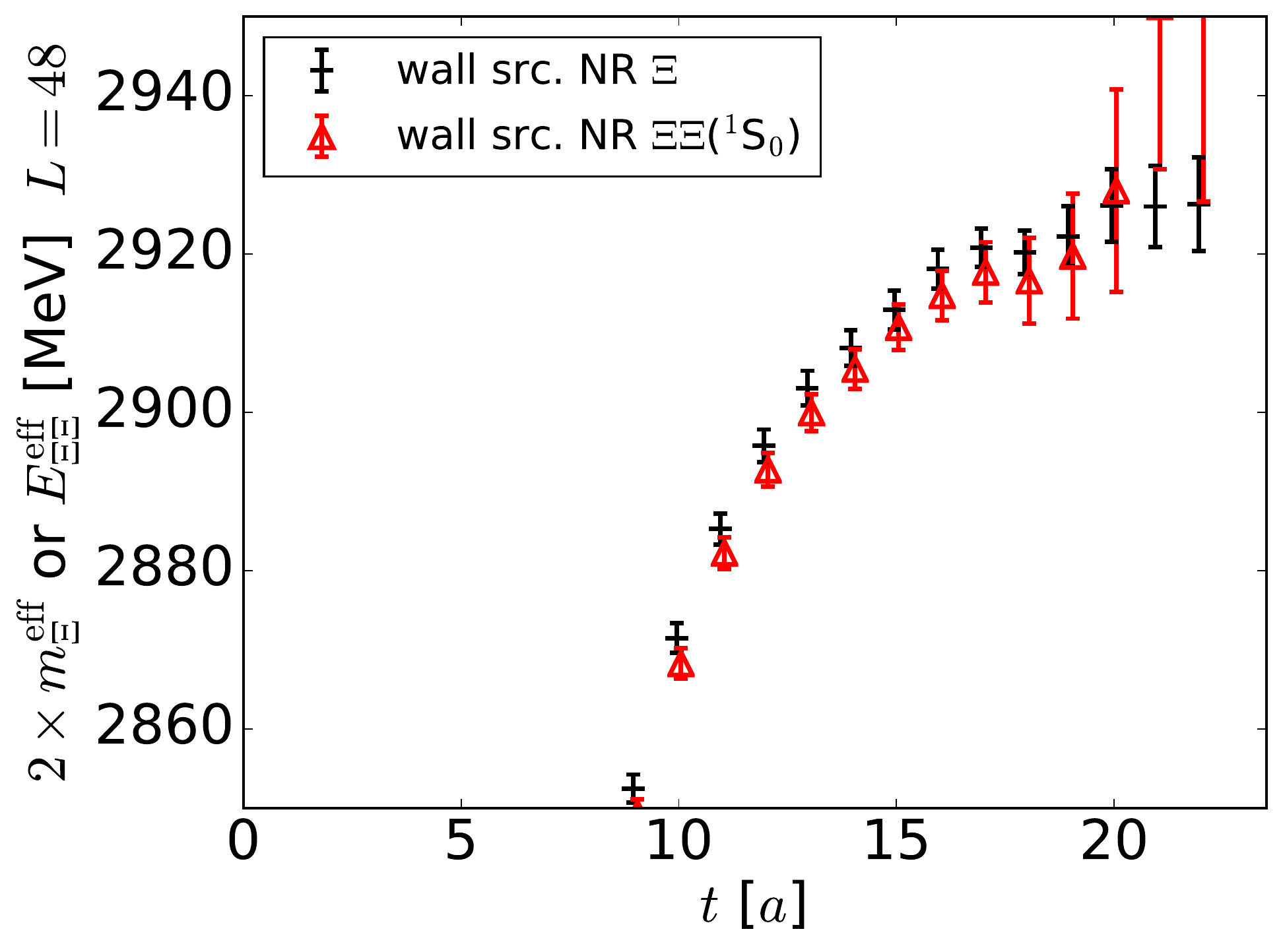}
  \includegraphics[width=0.47\textwidth]{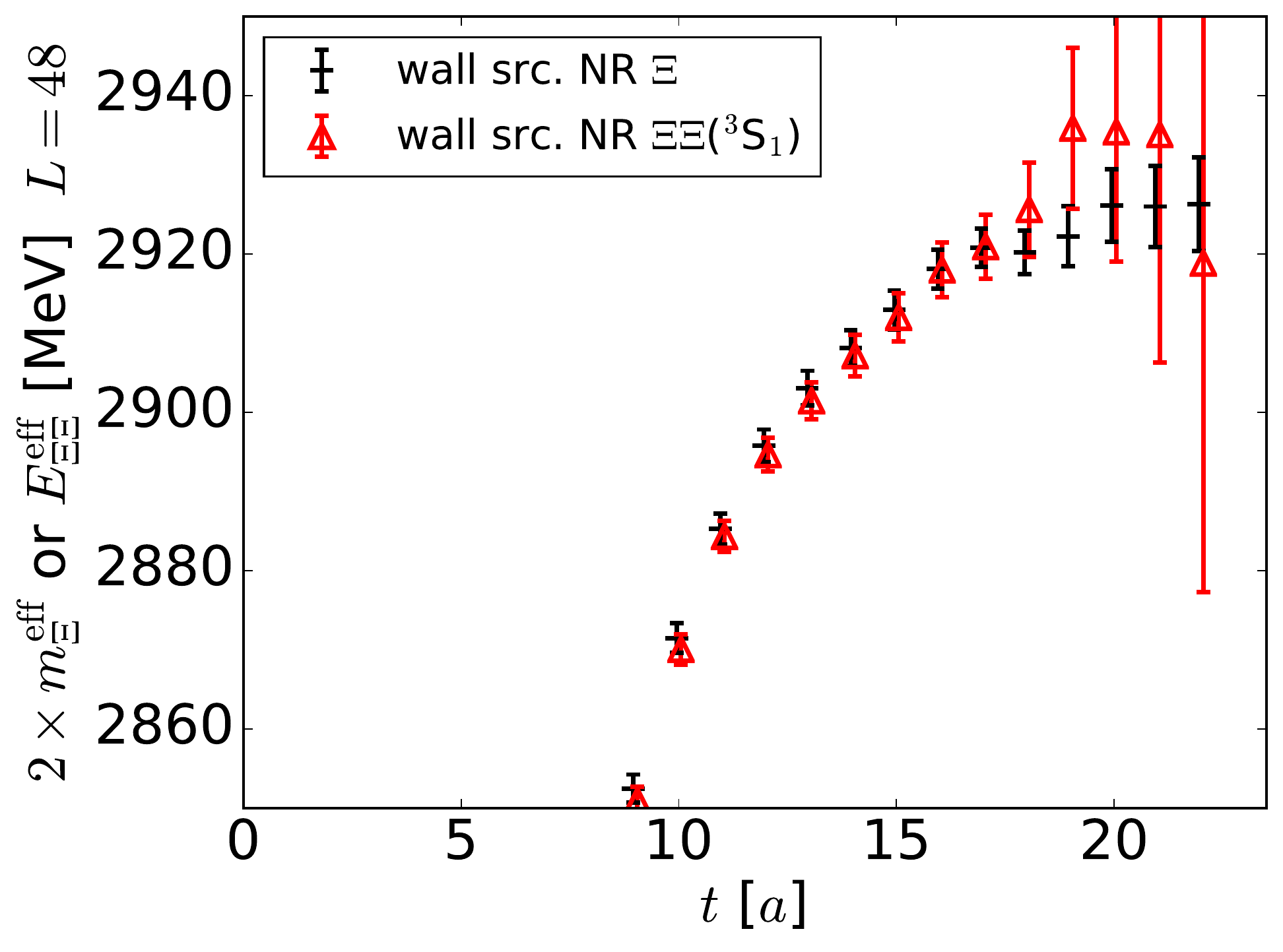}\\
  \includegraphics[width=0.47\textwidth]{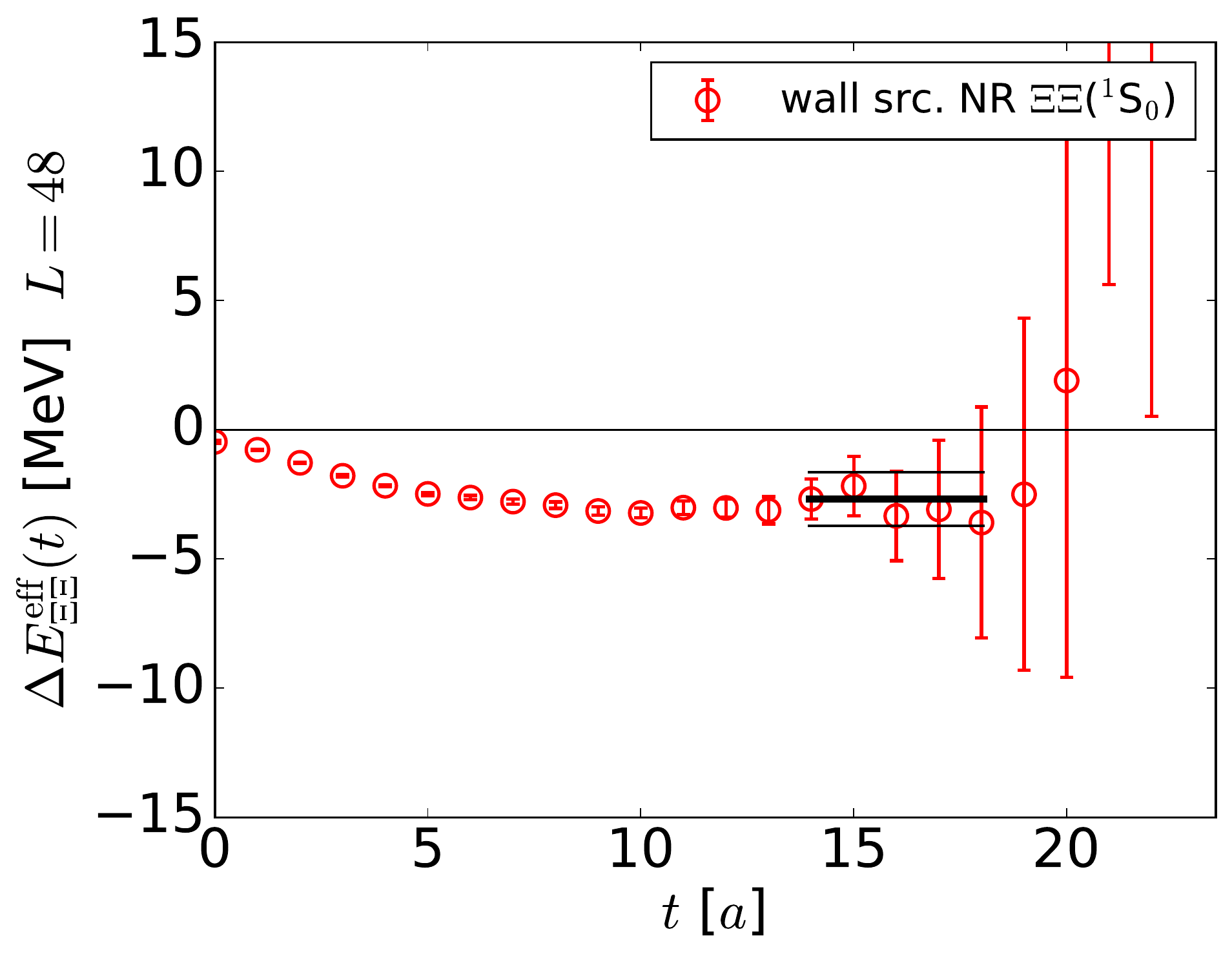}
  \includegraphics[width=0.47\textwidth]{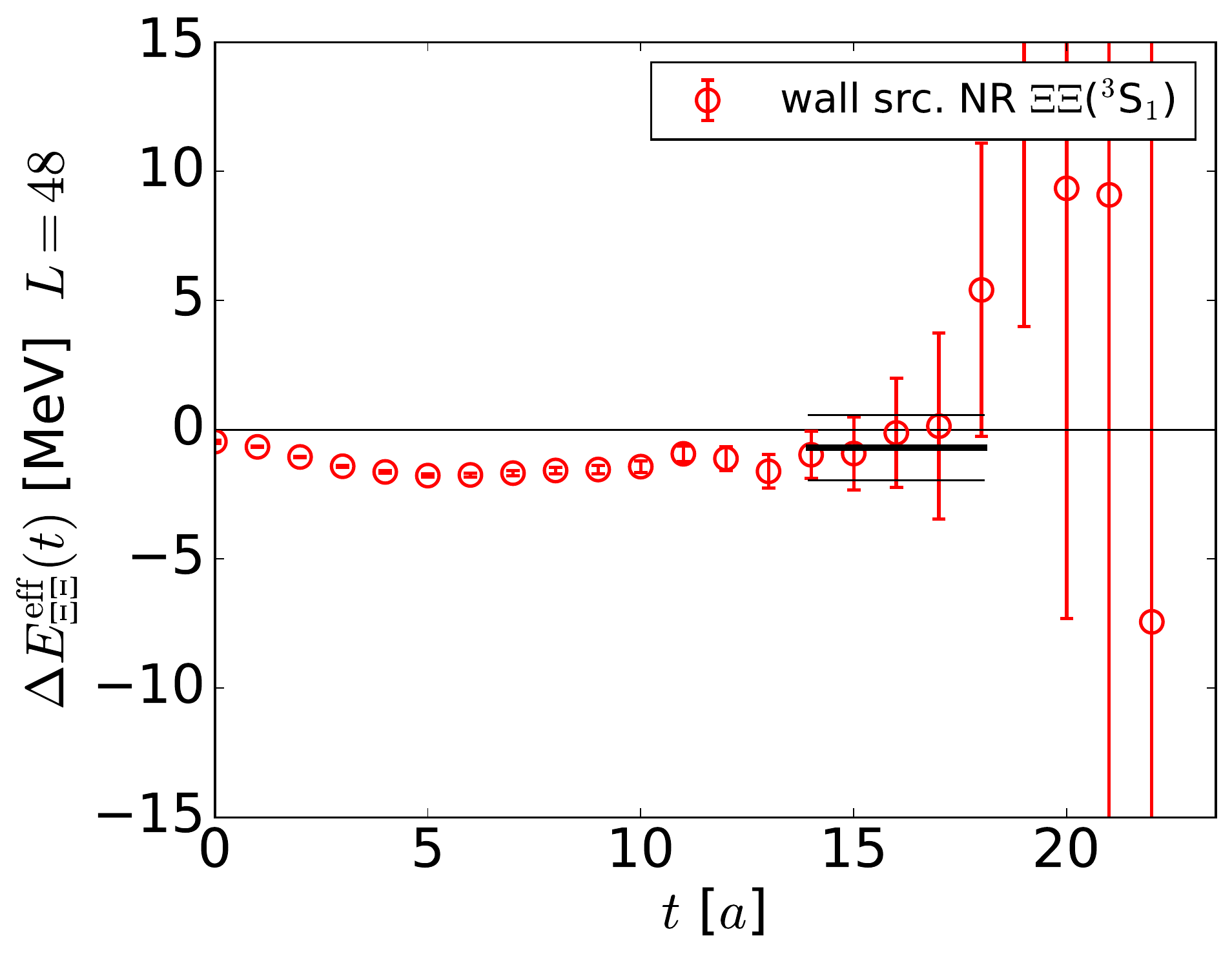}\\
  \includegraphics[width=0.47\textwidth]{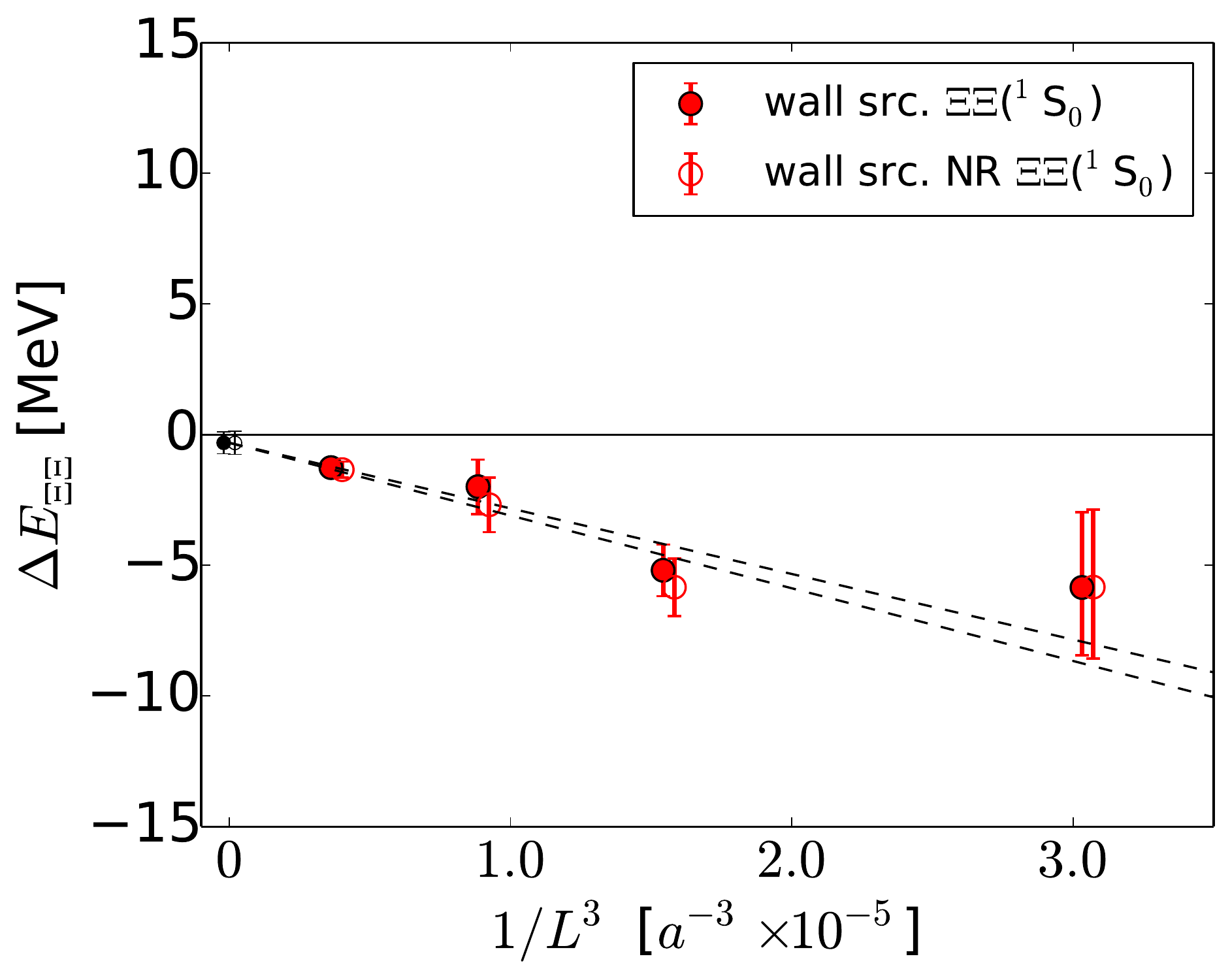}
  \includegraphics[width=0.47\textwidth]{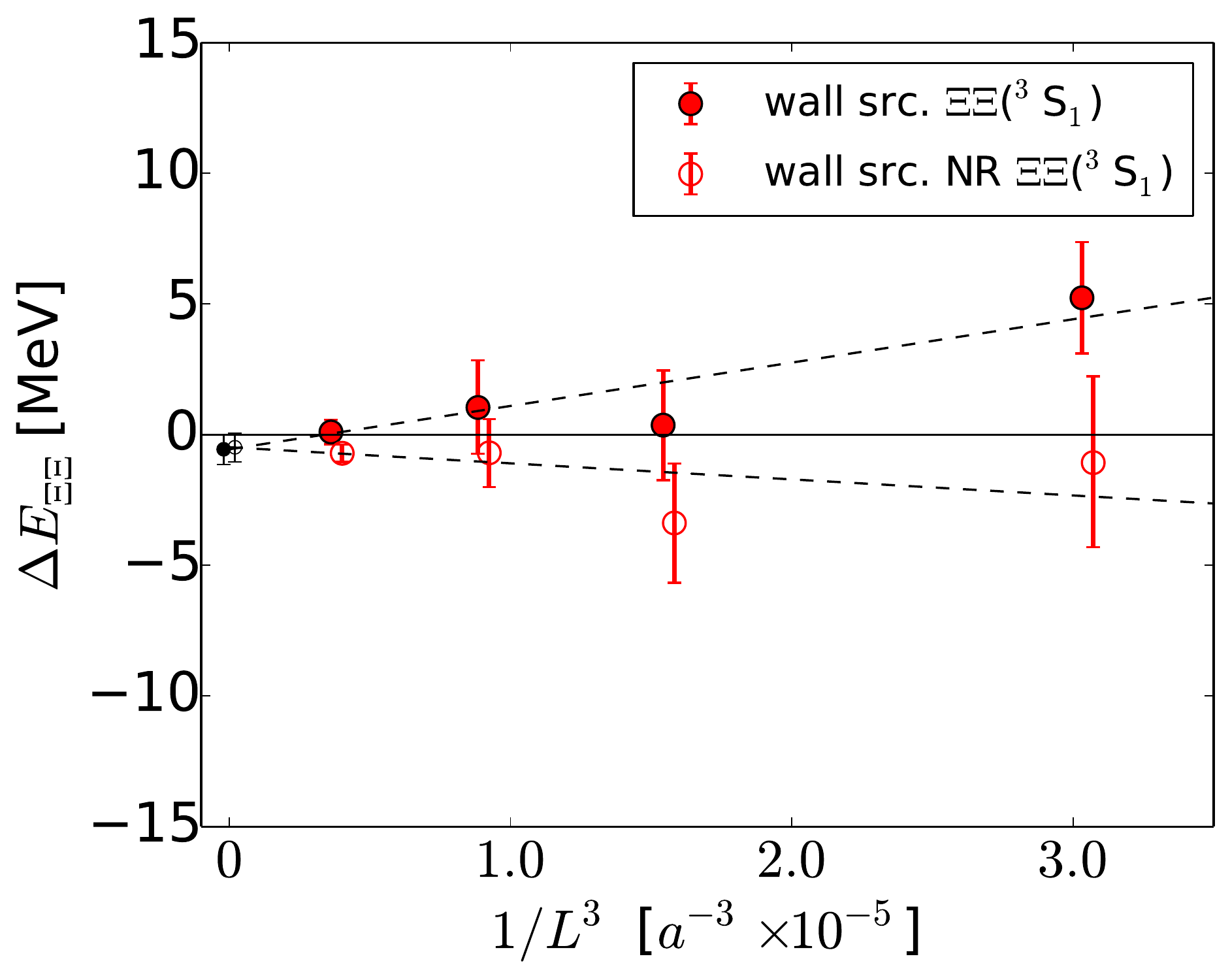}
  \caption{(Upper left) Effective mass $2\meffXi(t)$ (black bar) and effective energy $\EeffXiXi(t)$ (red triangle)
  in the $\Xi\Xi({}^1S_0)$ channel as a function of $t/a$ on the $48^3\times 48$
  lattice  from the wall  source with the non-relativistic operator.
  (Middle left) Effective energy shift $\DelEeffXiXi(t) \equiv \EeffXiXi(t) - 2 \meffXi (t)$,
  together with the fit (statistical only)  in the $\Xi\Xi({}^1S_0)$ channel.
  (Lower left)  The energy shift $\DelEXiXi$ in the $\Xi\Xi({}^1S_0)$ channel
   as a function of $1/L^3$
  from the wall source with the non-relativistic operator (open circle) as well as the relativistic one (solid circle),
  together with their infinite volume extrapolations. 
  The errors are obtained from statistical and systematic errors added in quadrature.
 (Upper right, Middle right, Lower right) Same quantities  in the $\Xi\Xi({}^3S_1)$ channel. 
}
 \label{fig:Eff_Xi_wall}
\end{figure}

To backup the conclusion obtained  with the smeared source, let us now  
 analyze the lattice data with  the wall source.
 Fig.~\ref{fig:Eff_Xi_wall} (Upper left) shows $ 2 \meffXi(t)$ (black bar) and $\EeffXiXi(t)$ (red triangle)  in the $^1S_0$ channel
 for non-relativistic interpolating operators
on the $48^3 \times 48$ lattice, while Fig.~\ref{fig:Eff_Xi_wall} (Middle left) shows the 
errors and fluctuations of the effective energy shift, $\DelEeffXiXi(t)$ in the same channel.
Again we fit the ``plateau" in the range $14 \le t/a \le 18$ just before the explosion of noise over the signal.
Lowering $\tmin$ of the window does not  change the result, although it is not recommended from the 
stability of $\meffXi(t)$.

We perform the similar analysis in other volumes as well as for the relativistic operators.
Shown in Fig.~\ref{fig:Eff_Xi_wall} (Lower left) are $\DelEXiXi (^1S_0)$ as a function of $1/L^3$, 
together with the linear infinite volume extrapolations in $1/L^3$, where statistical and systematic errors are added in quadrature.
The results of $\DelEXiXi (^1S_0)$ are given in Table~\ref{tab:summary_XiXi}
with the infinite volume limit,
$ \DelEXiXi (^1S_0) = -[0.31(0.44)(^{+0.00}_{-0.05})]$ MeV (non-relativistic operator) and
$ -[0.31(0.42)(^{+0.00}_{-0.03})]$  MeV (relativistic operator).

For the $\DelEXiXi (^3S_1)$ channel, one finds again that a ``plateau''-like behavior in $14 \le t/a \le 18$ before the explosion of the noise over the signal in larger $t$ as shown in Fig.~\ref{fig:Eff_Xi_wall} (Middle right) for the non-relativistic interpolating operators.
We perform the same analysis in other volumes and also for the relativistic operators.
In Fig.~\ref{fig:Eff_Xi_wall} (Lower right),  $\DelEXiXi (^3S_1)$ with the wall source
is plotted as a function of $1/L^3$, together with the values at infinite volume obtained by linear extrapolation.
The results are summarized in Table~\ref{tab:summary_XiXi} with the infinite volume limit,
$\DelEXiXi (^3S_1) =  -[0.48(0.54)(^{+0.07}_{-0.10})]$ MeV (non-relativistic operator) and
$ -[0.56(0.53)(^{+0.23}_{-0.24})]$  MeV (relativistic operator).

In Table \ref{tab:summary-box},  we summarize the results $\DelEXiXi $ in all four cases which we have studied 
 in this section.  
 The positive $\Delta E_{\Xi\Xi} (^3S_1)$ for the smeared source is not allowed physically,
 and there are apparent inconsistencies between the results of the smeared source
 and those of the wall source.
 These are convincing enough that
 the previous works on temporal correlations  have been looking at the fake plateaux just before the explosion of the
 noise over the signal as discussed in Sec.~\ref{sec:general}.


\begin{table}[th]
  \centering
  \begin{tabular}{|cc|rc|rc|} 
\hline 
\multicolumn{2}{|c|}{$\Xi\Xi (^1S_0)$} & \multicolumn{2}{c}{smeared source} & \multicolumn{2}{|c|}{wall source} \\
\hline 
volume & operator & $\Delta E$ [MeV] & fit range & $\Delta E$ [MeV] & fit range\\
\hline
$32^3$ & rela.     & $-8.57(1.79)^{+0.38}_{-0.19}$ & 11-16 & $-5.85(2.54)^{+1.37}_{-0.53}$ & 12-16 \\
       & non-rela. & $-10.98(1.96)^{+0.37}_{-0.21}$ & 11-16 & $-5.83(2.69)^{+1.26}_{-0.50}$ & 12-16 \\
$40^3$ & rela.     & $-7.20(1.23)^{+0.43}_{-0.10}$ & 11-17 & $-5.19(0.98)^{+0.15}_{-0.14}$ & 11-16 \\
       & non-rela. & $-9.25(1.24)^{+0.17}_{-0.16}$ & 11-17 & $-5.84(1.09)^{+0.13}_{-0.14}$ & 11-16 \\
$48^3$ & rela.     & $-7.98(0.73)^{+0.37}_{-0.19}$ & 12-19 & $-1.99(1.02)^{+0.18}_{-0.26}$ & 14-18 \\
       & non-rela. & $-9.36(0.54)^{+0.39}_{-0.21}$ & 10-18 & $-2.68(1.04)^{+0.07}_{-0.13}$ & 14-18 \\
       $64^3$ & rela.     & $-4.79(0.81)^{+0.27}_{-0.06}$ & 10-18 & $-1.26(0.28)^{+0.09}_{-0.07}$ & 13-18 \\
         & non-rela. & $-6.93(1.05)^{+0.30}_{-0.19}$ & 11-17 & $-1.34(0.29)^{+0.09}_{-0.08}$ & 13-18 \\
       \hline
       $\infty$ & rela. & $-5.44(0.82)^{+0.28}_{-0.09}$ & & $-0.31(0.42)^{+0.00}_{-0.03}$ & \\
     & non-rela. & $-7.70(0.89)^{+0.37}_{-0.20}$ & & $-0.31(0.44)^{+0.00}_{-0.05}$ & \\
\hline \hline
\multicolumn{2}{|c|}{$\Xi\Xi (^3S_1)$} & \multicolumn{2}{c}{smeared source} & \multicolumn{2}{|c|}{wall source} \\
\hline 
volume & operator & $\Delta E$ [MeV] & fit range & $\Delta E$ [MeV] & fit range\\
\hline
$32^3$ & rela.     & $10.24(1.61)^{+0.72}_{-0.24}$ & 11-15 & $5.24(2.09)^{+0.42}_{-0.39}$ & 11-16 \\
       & non-rela. & $2.84(1.86)^{+0.22}_{-0.38}$ & 11-15 & $-1.07(3.17)^{+0.94}_{-0.65}$ & 12-16 \\
$40^3$ & rela.     & $10.49(2.01)^{+0.46}_{-0.51}$ & 12-17 & $0.37(2.06)^{+0.35}_{-0.44}$ & 12-16 \\
       & non-rela. & $4.76(1.89)^{+0.38}_{-0.19}$ & 11-16 & $-3.38(2.24)^{+0.35}_{-0.48}$ & 12-16 \\
$48^3$ & rela.     & $11.00(0.80)^{+0.24}_{-0.03}$ & 12-19 & $1.04(1.75)^{+0.42}_{-0.26}$ & 15-19 \\
       & non-rela. & $5.59(0.75)^{+0.33}_{-0.30}$ & 11-18 & $-0.69(1.26)^{+0.24}_{-0.33}$ & 14-18 \\
       $64^3$ & rela.     & $12.60(1.05)^{+0.10}_{-0.26}$ & 11-18 & $0.10(0.39)^{+0.26}_{-0.27}$ & 14-18 \\
       & non-rela. & $6.38(1.28)^{+0.64}_{-0.70}$ & 11-16 & $-0.71(0.33)^{+0.03}_{-0.04}$ & 13-18 \\
\hline
$\infty$ & rela. & $+12.20(0.94)^{+0.02}_{-0.12}$ & & $-0.56(0.53)^{+0.23}_{-0.24}$ & \\
& non-rela. & $+6.81(1.04)^{+0.52}_{-0.48}$ & & $-0.48(0.54)^{+0.07}_{-0.10}$ & \\
\hline
  \end{tabular}
  \caption{Summary of $\DelEXiXi$ for both ${}^1S_0$ (upper) and 
    ${}^3S_1$ (lower) channels from smeared and wall sources
    with range of an exponential fit, together with infinite volume
    extrapolations. On each volume, results from both relativistic and non-relativistic baryon operators are given.}
  \label{tab:summary_XiXi}
\end{table}


\begin{table}[tb]
\begin{center}
\begin{tabular}{|c|l|l|}
\hline 
 & smeared source & wall source  \\
\hline  \hline
$\DelEXiXi(^1S_0)$   & $< 0$ (bound state) &  $\simeq 0$ (no bound state) \\
\hline 
$\DelEXiXi(^3S_1)$ & $>0$ (physically not allowed)   & $\simeq 0$ (no bound state) \\
\hline 
\end{tabular}
\caption{Comparison of $\DelEXiXi$ for different channels and different sources
  at infinite volume.
Those are obtained by fitting ``plateau"-like structure of $\DelEeffXiXi(t)$
 in the region of $t$  just before explosion of the signal to noise ratio. }
\label{tab:summary-box}
\end{center}
\end{table}%

\section{$NN$ systems}
\label{sec:NN}

After clarifying the problem of fitting fake plateaux in $\Xi\Xi$  systems, 
let us now turn our discussions to the $NN$ systems in the $NN({}^1S_0)$ and $NN({}^3S_1)$ channels
to show that the same problem takes place.
Interpolating operators for these  channels are given by 
$N_{1}N_{2} - N_{2}N_{1}$ with $N=p, n$ for $NN({}^1S_0)$, which belongs to {\bf 27} in the irreducible
representation of the flavor SU(3), and by
$ p_{\alpha}n_{\alpha} - n_{\alpha} p_{\alpha}$ with $\alpha=1,2$ for $NN({}^3S_1)$,
which belongs to {\bf 10}$^*$ representation.
Note that the $NN({}^1S_0)$ is in the same flavor-SU(3) multiplet  with $\Xi\Xi({}^1S_0)$,
while $NN({}^3S_1)$ belongs to  the different flavor-SU(3) multiplet  with $\Xi\Xi({}^3S_1)$,
so that we do not expect qualitative resemblance between  $NN({}^3S_1)$ and $\Xi\Xi({}^3S_1)$.


\begin{figure}[tb]
  \centering
  \includegraphics[width=0.45\textwidth]{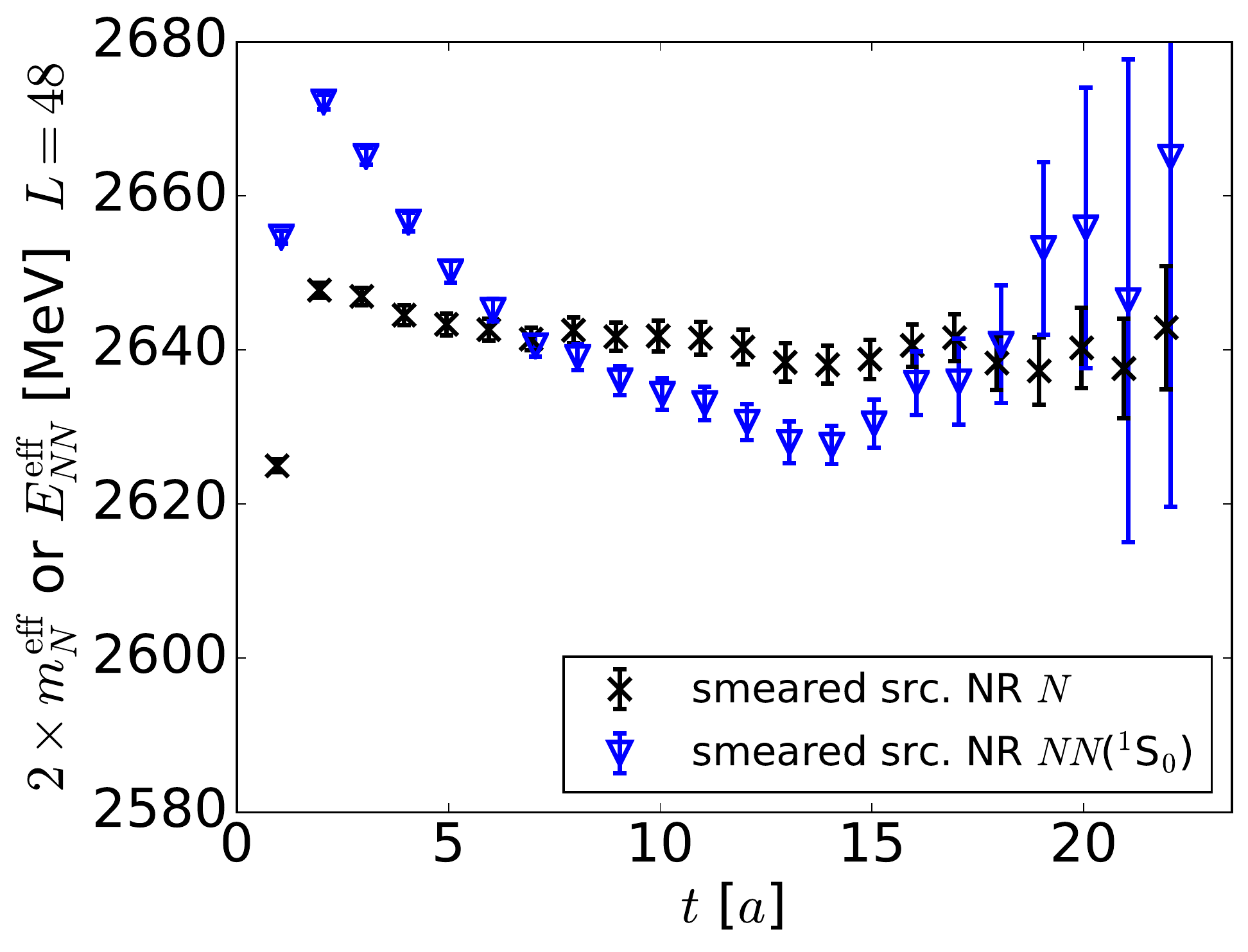}
  \includegraphics[width=0.45\textwidth]{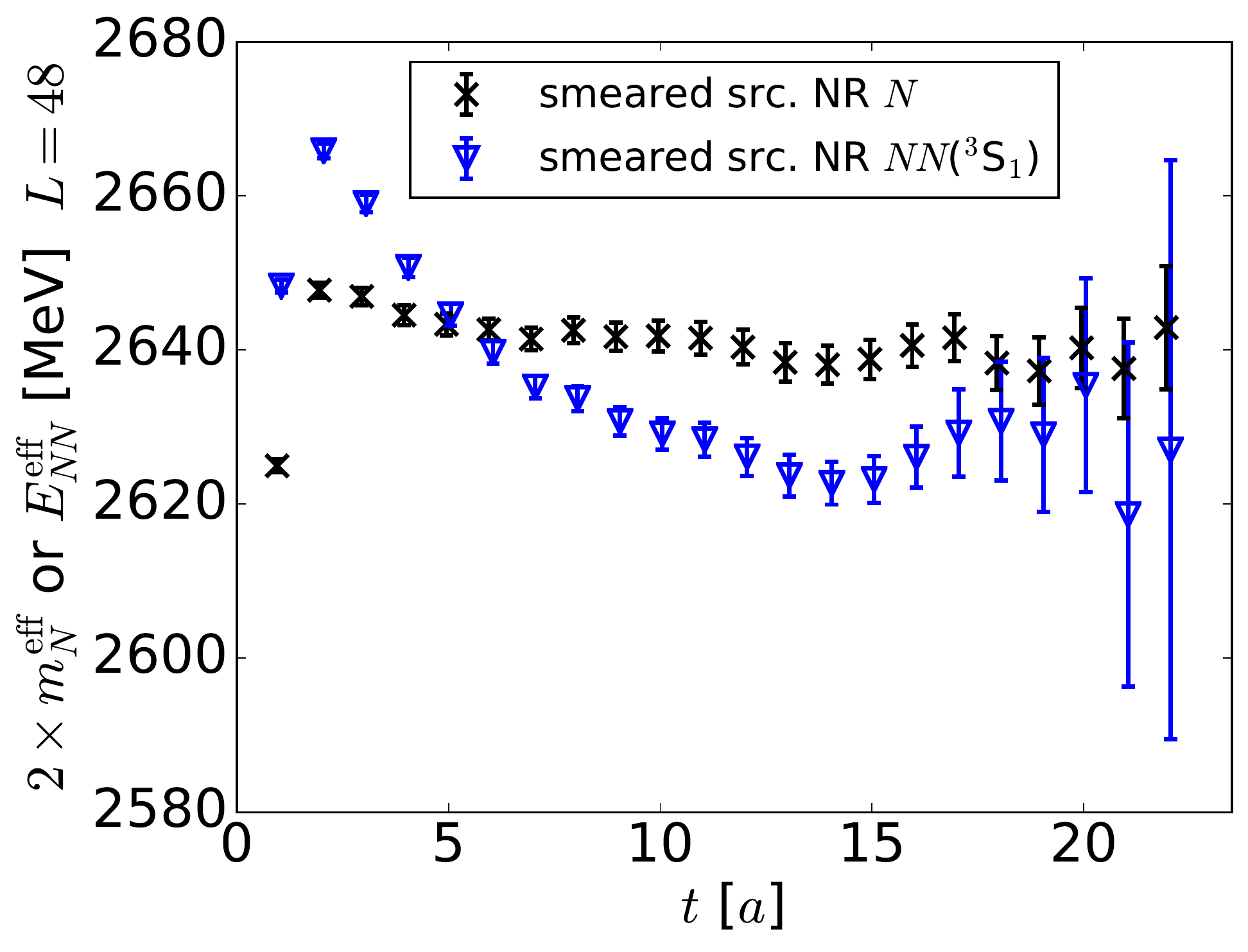}\\
  \includegraphics[width=0.45\textwidth]{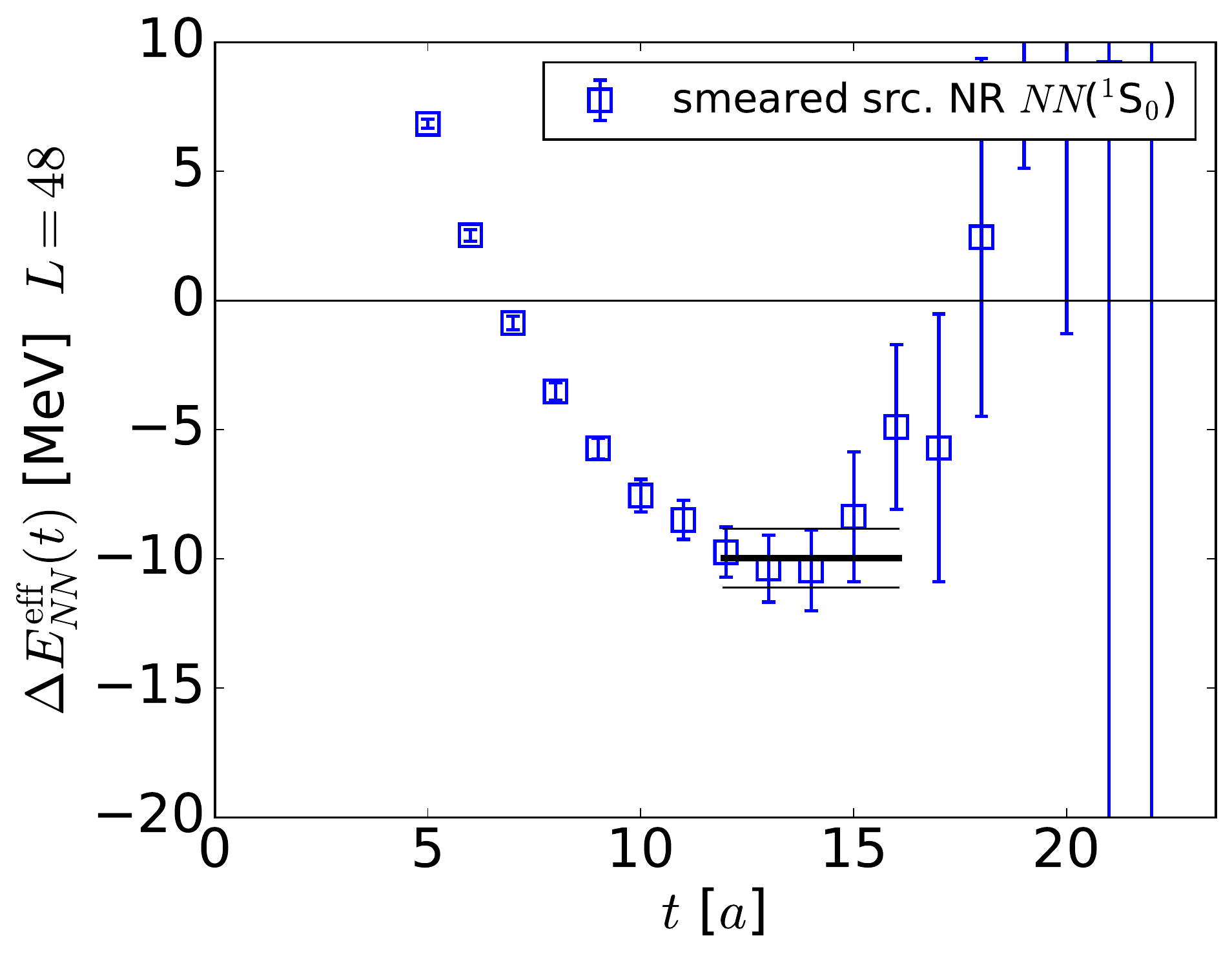}
  \includegraphics[width=0.45\textwidth]{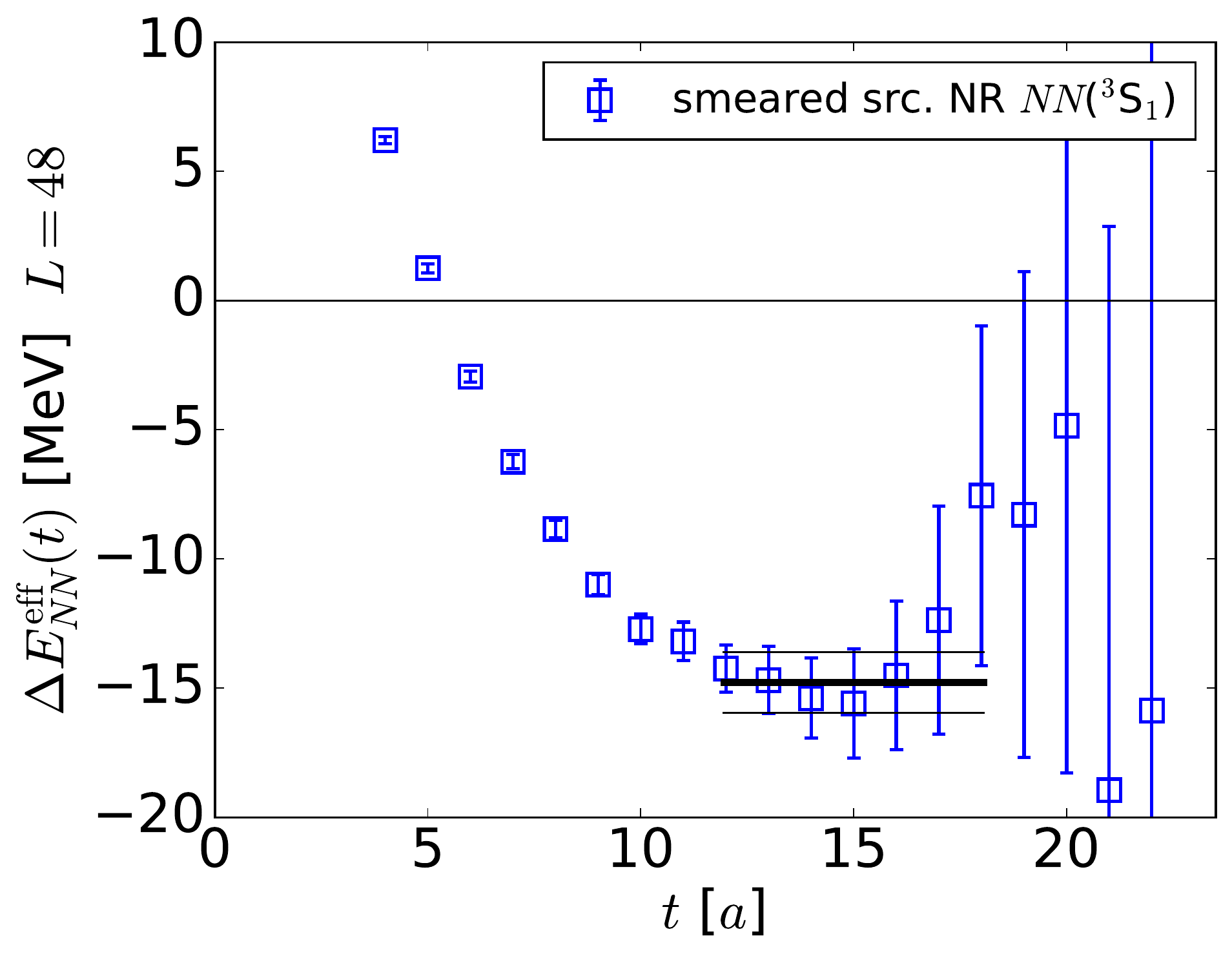}\\
  \includegraphics[width=0.45\textwidth]{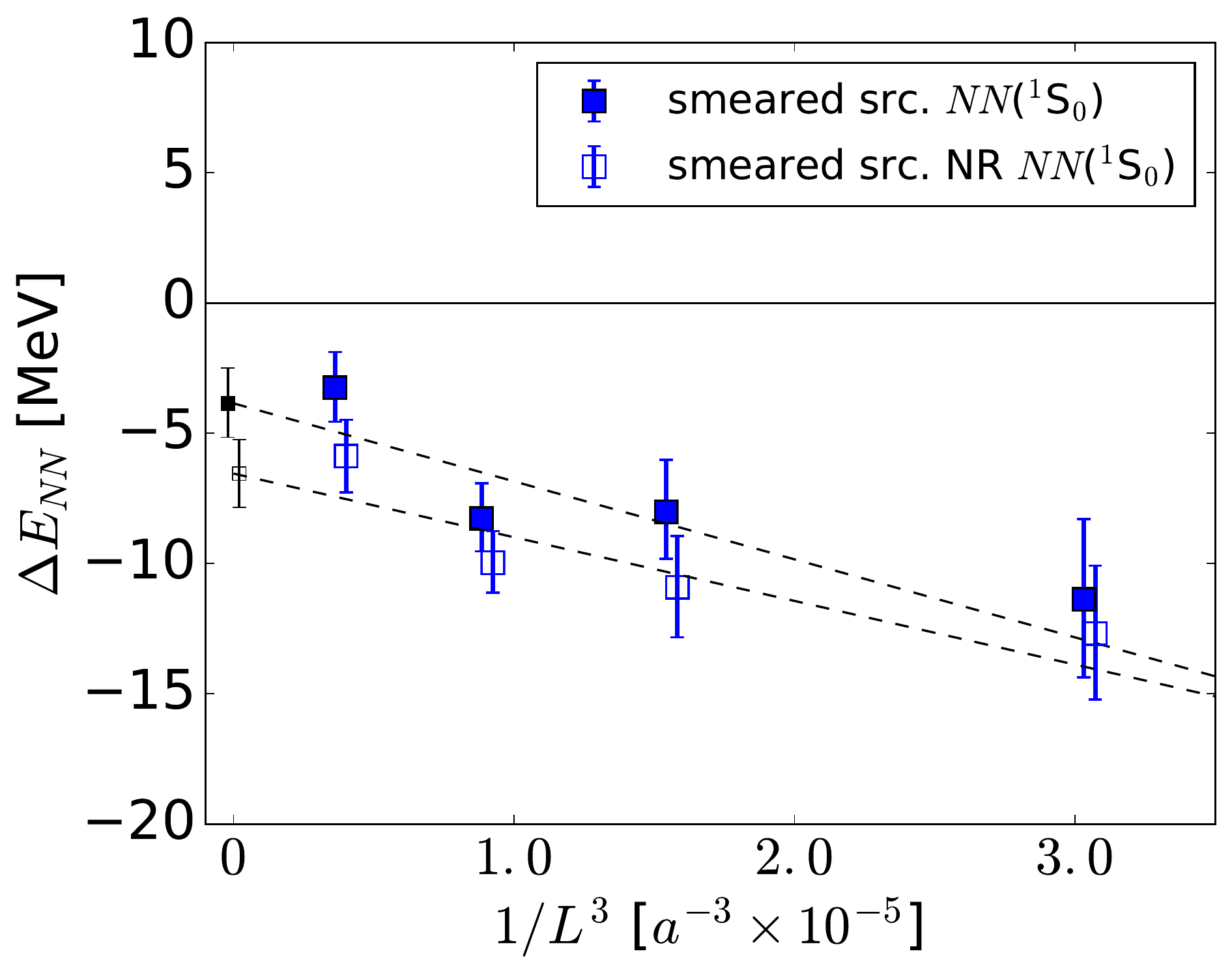}
  \includegraphics[width=0.45\textwidth]{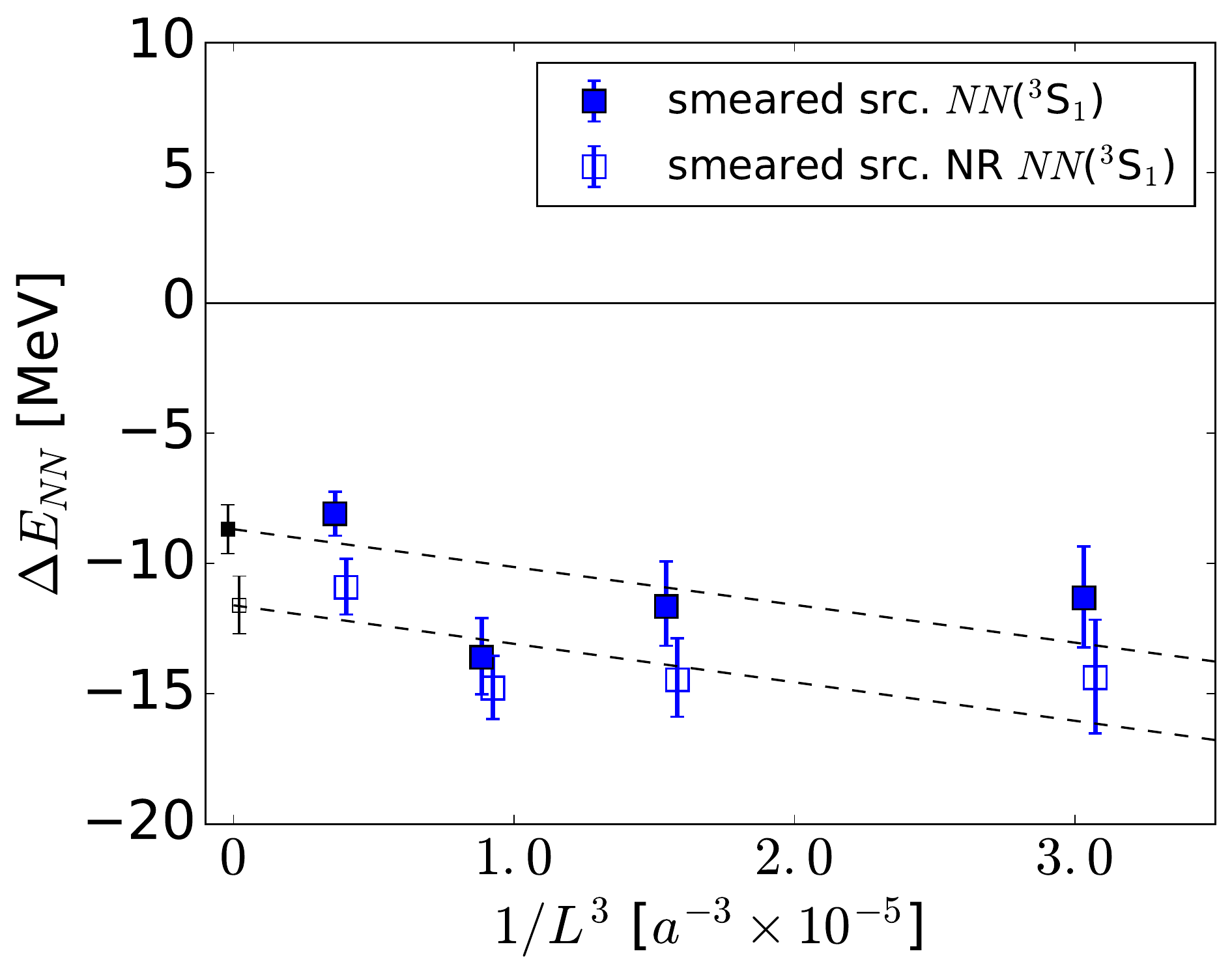}
  \caption{
  (Upper left) $2\meffN(t)$ (black cross) and the effective energy $\EeffNN(t)$ (blue triangle)
  in the ${}^{1}S_0$ channel as a function of $t/a$ on the $48^3\times 48$ lattice from the smeared source with the non-relativistic operator. 
  (Upper right)  Same in the ${}^{3}S_1$ channel.
  (Middle left) Effective energy shift $\DelEeffNN (t)$, together with the fit (statistical only)  in the ${}^{1}S_0$ channel with the same lattice setup.
  (Middle right) Same     in the ${}^{3}S_1$ channel.
  (Lower left) Energy shift $\DelENN$ in the ${}^{1}S_0$ channel as a function of $1/L^3$  
    from the smeared source
    with both non-relativistic (open square) and relativistic operators (solid square).
    Shown together are the linear extrapolation in $1/L^3$ to the infinite volume. 
    The errors are obtained from statistical and systematic errors added in quadrature.
  (Lower right) Same   in the ${}^{3}S_1$ channel.
    }
  \label{fig:E_NN}
\end{figure}

The upper two panels of Fig.~\ref{fig:E_NN}  shows $2\meffN(t)$ and 
the effective energy $\EeffNN(t)$ 
for the  smeared source  with the non-relativistic nucleon operator
on the $48^3\times 48$ lattice  in
the $NN({}^1S_0)$ channel (Left) and in the $NN({}^3S_1)$ channel (Right).

The effective energy shifts from the smeared source on the $48^3\times 48$ lattice
are shown in the middle two panels in Fig.~\ref{fig:E_NN}: Left (Right) panel for 
the ${}^1S_0$ (${}^3S_1$) channel with the non-relativistic nucleon operator.
The explosion in the noise to signal ratio takes place for smaller $t/a$ than that for $\DelEeffXiXi(t)$
 due to larger statistical errors in the $NN$ case.
 We try to fit the plateau-like structure just before the explosion
 typically in the range $12 \le t/a \le 16$.  Obviously,  we already knew from the discussions in the 
 previous sections that such a plateau-like structure is fake. Our aim here (as in the case of the $\Xi\Xi$) 
 is to show that the results of such fitting procedure adopted in previous literature do not make much sense.

  In Table~\ref{tab:summary_NN}, results of  $\DelENN$ on four volumes for the smeared 
  source and for non-relativistic and relativistic operators are summarized 
  in the middle column.
    The fitting range for $NN$ is relatively earlier
 than that for $\Xi\Xi$ due to larger statistical errors. 
 Systematic errors are estimated by changing the upper and lower limit of the fitting window 
 by one unit of $t/a$ as we have done  in the case of $\Xi\Xi$.

 The lower panels of Fig.~\ref{fig:E_NN} shows $\DelENN$ 
 in the ${}^1S_0$ channel (Left) and in the ${}^3S_1$ channel (Right)
  as a function of $1/L^3$, together with the linear extrapolation in $1/L^3$ to the infinite volume.
In each figure,  results of the non-relativistic and relativistic operators are plotted with 
the numerical data  given in Table~\ref{tab:summary_NN}.
The result of the smeared source for non-relativistic operator turns out to be
 $\DelENN (^1S_0) = - [ 6.54(1.29)(^{+0.11}_{-0.00})]$ MeV 
 and 
  $\DelENN (^3S_1) = - [ 11.60(1.06)(^{+0.36}_{-0.24})]$ MeV, 
which agrees with 
$\DelENN (^1S_0) = - [ 7.4(1.3)(0.6)] $ MeV 
and 
$\DelENN (^3S_1) = - [ 11.5(1.1)(0.6)] $ MeV  in the previous work~\cite{Yamazaki:2012hi}. 
This agreement simply implies that the our present analysis and the previous analysis are consistent with each other
 and does not necessarily  imply that there is indeed a  bound state in these channels.  
This can be seen explicitly by the results of the wall source as shown below.

\begin{figure}[htb]
  \centering
  \includegraphics[width=0.45\textwidth]{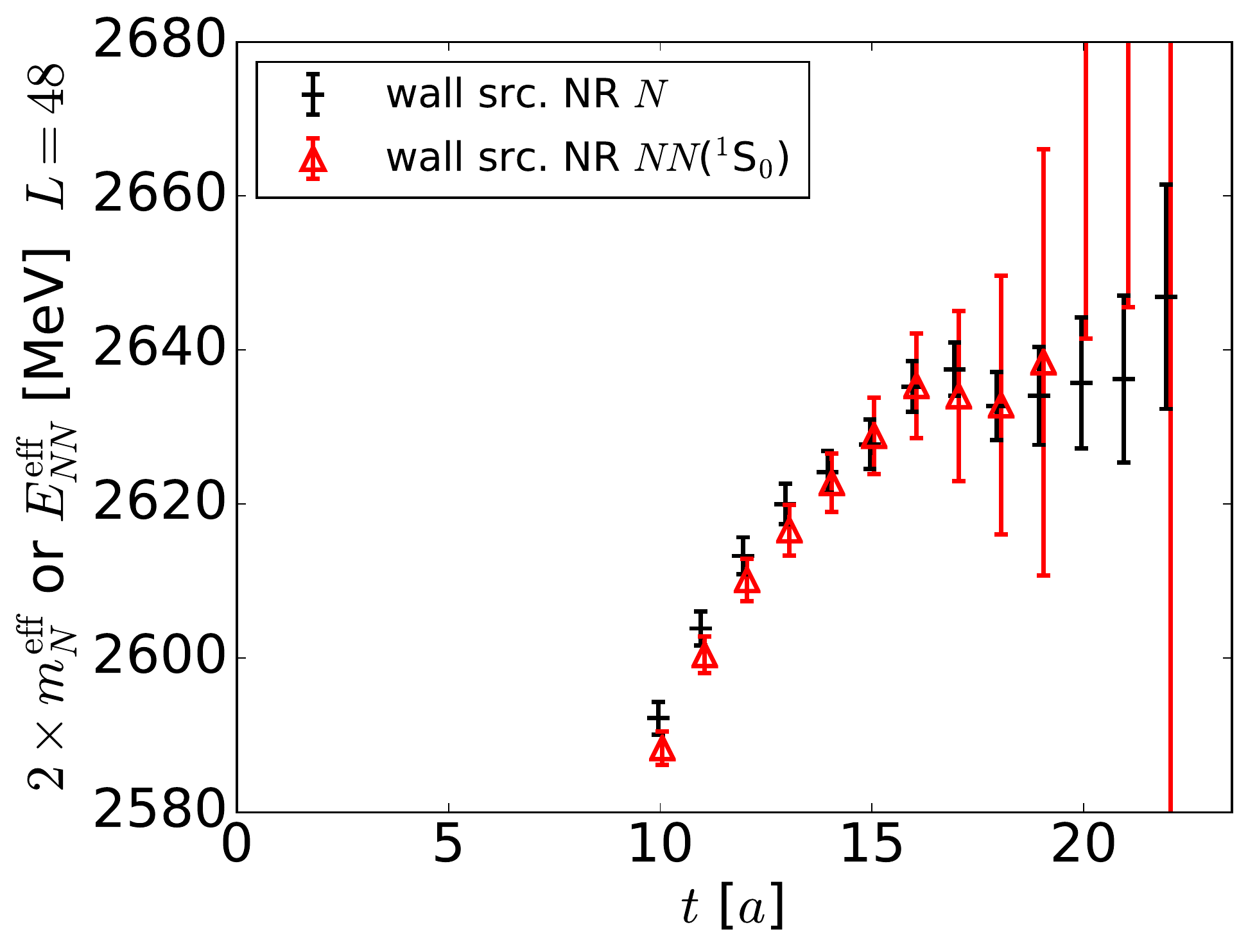}
  \includegraphics[width=0.45\textwidth]{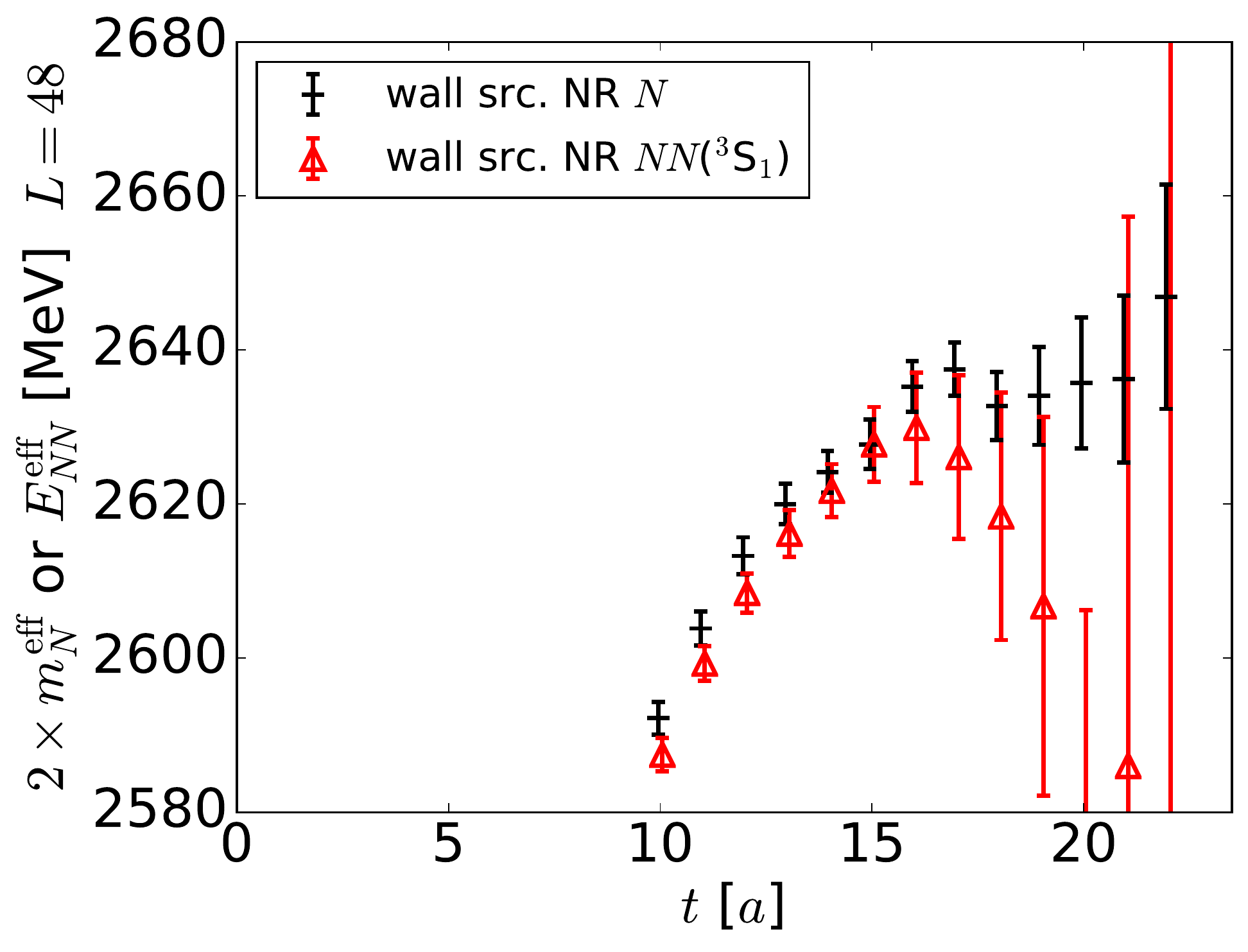}\\
  \includegraphics[width=0.45\textwidth]{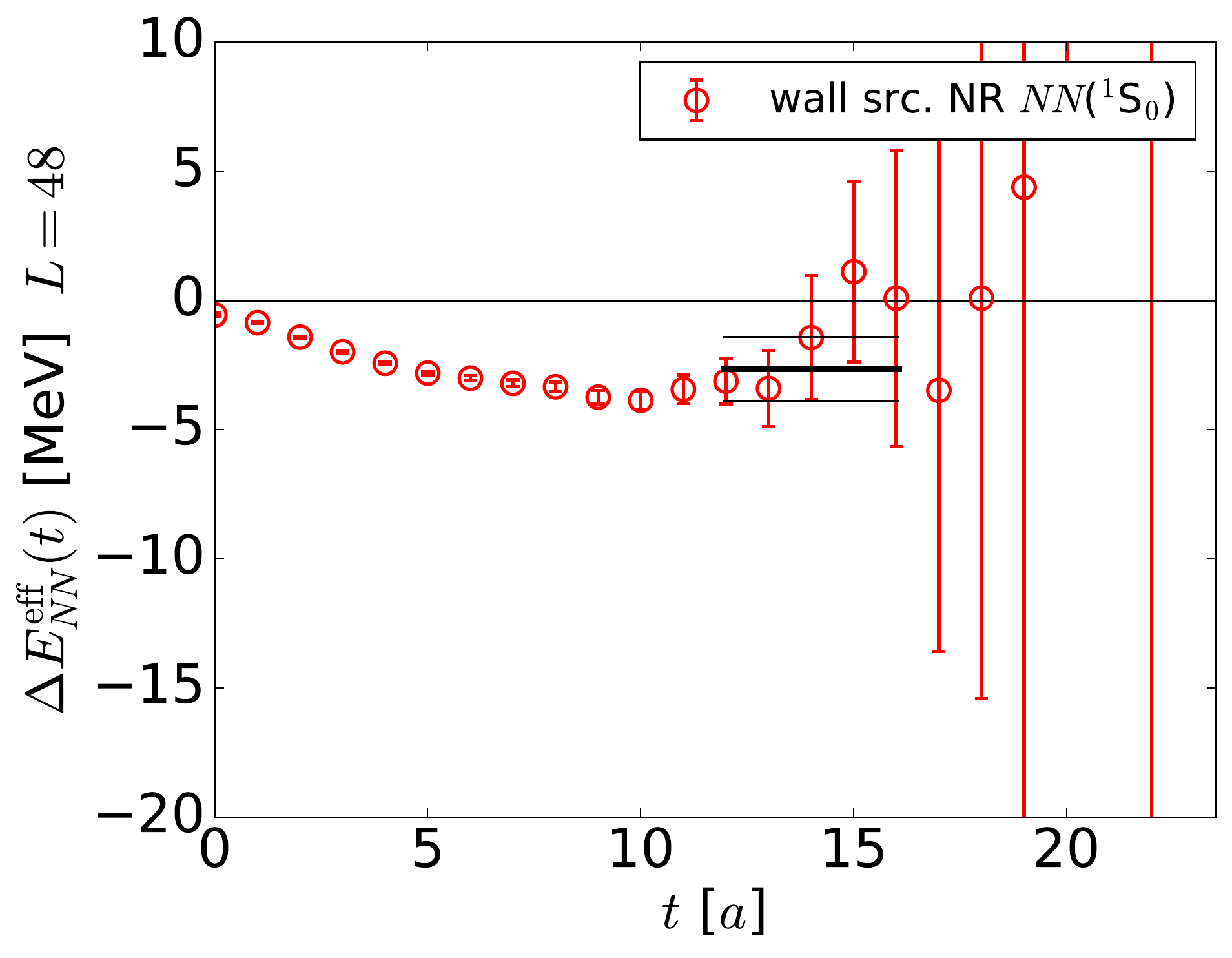}
  \includegraphics[width=0.45\textwidth]{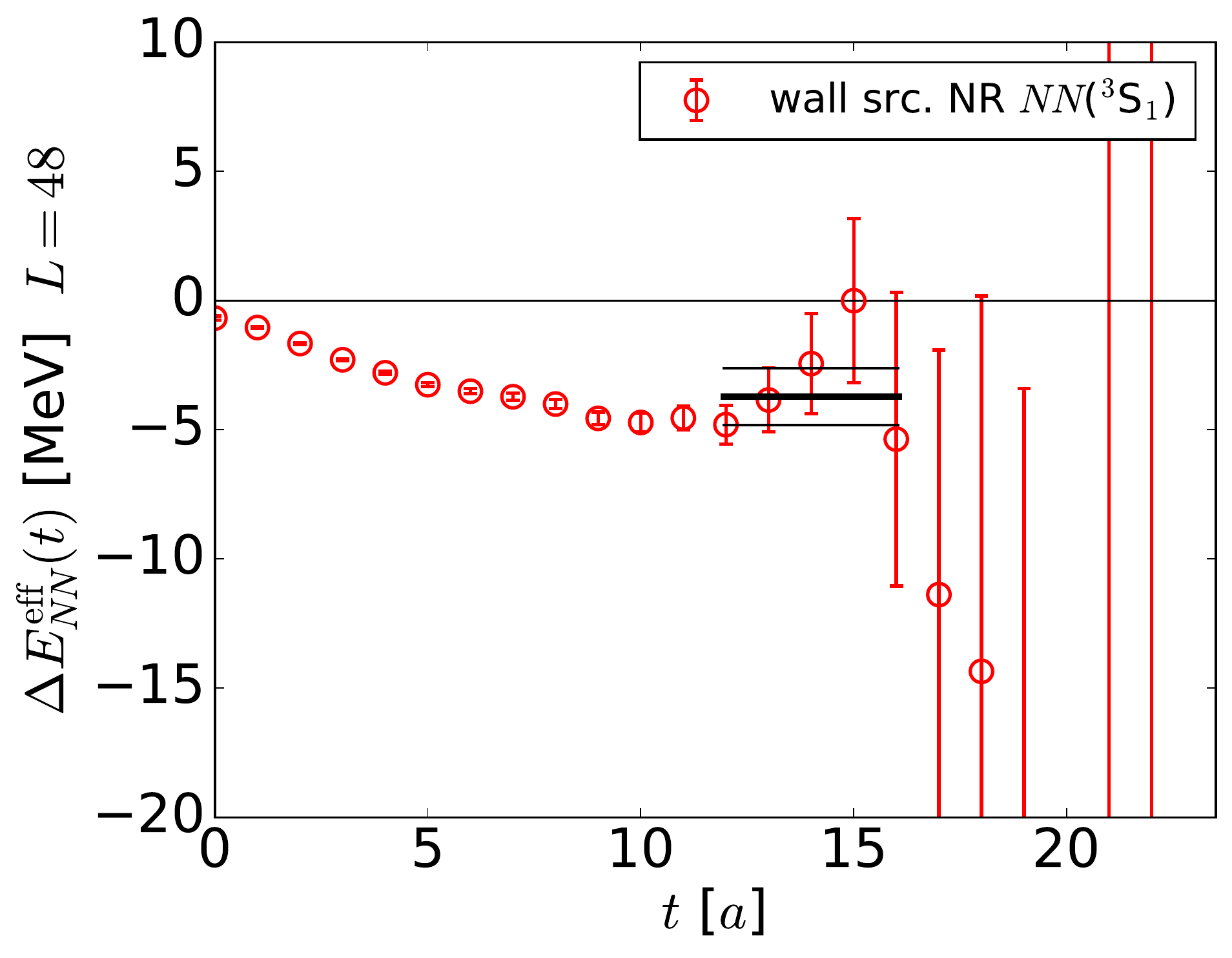}\\
  \includegraphics[width=0.45\textwidth]{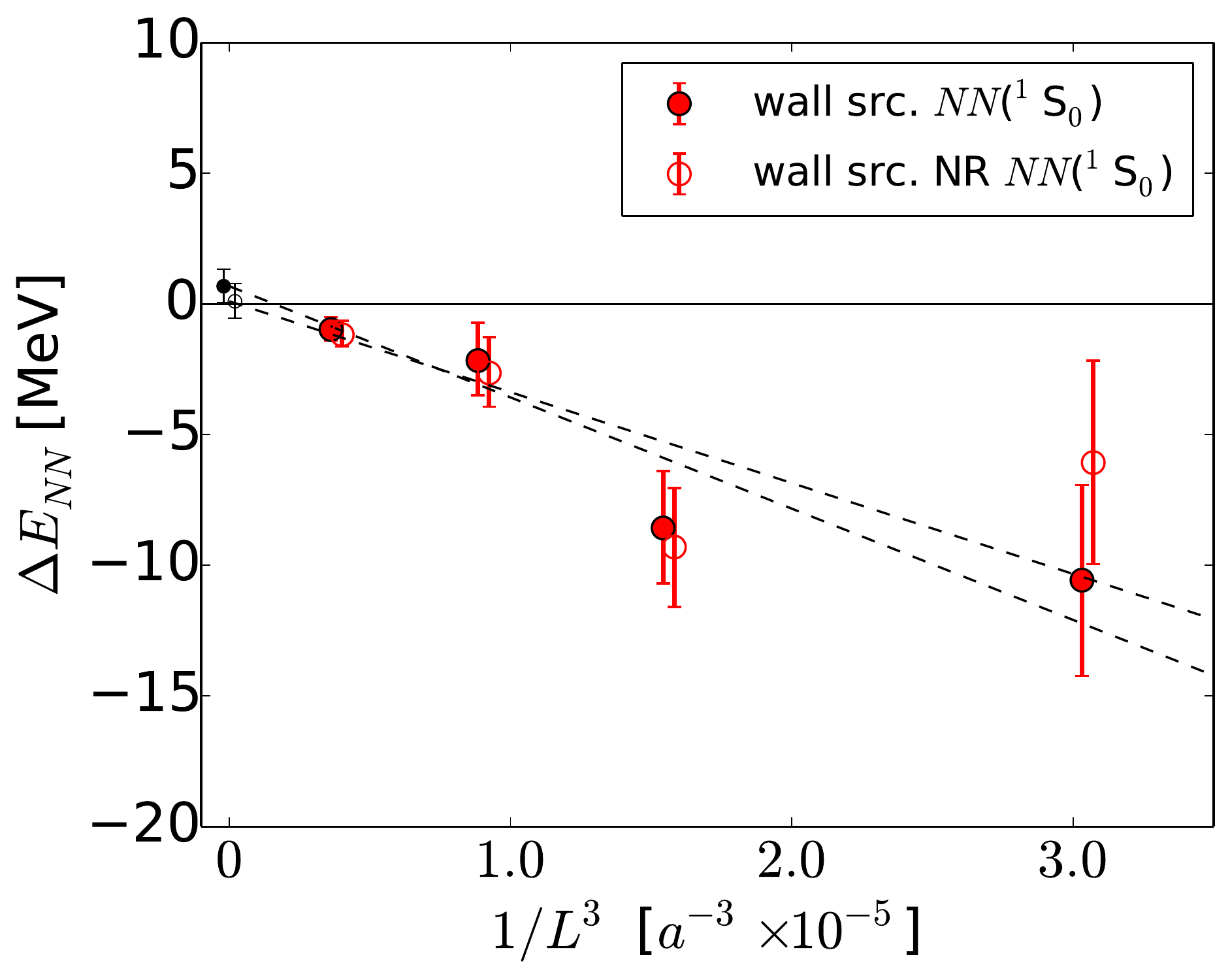}
  \includegraphics[width=0.45\textwidth]{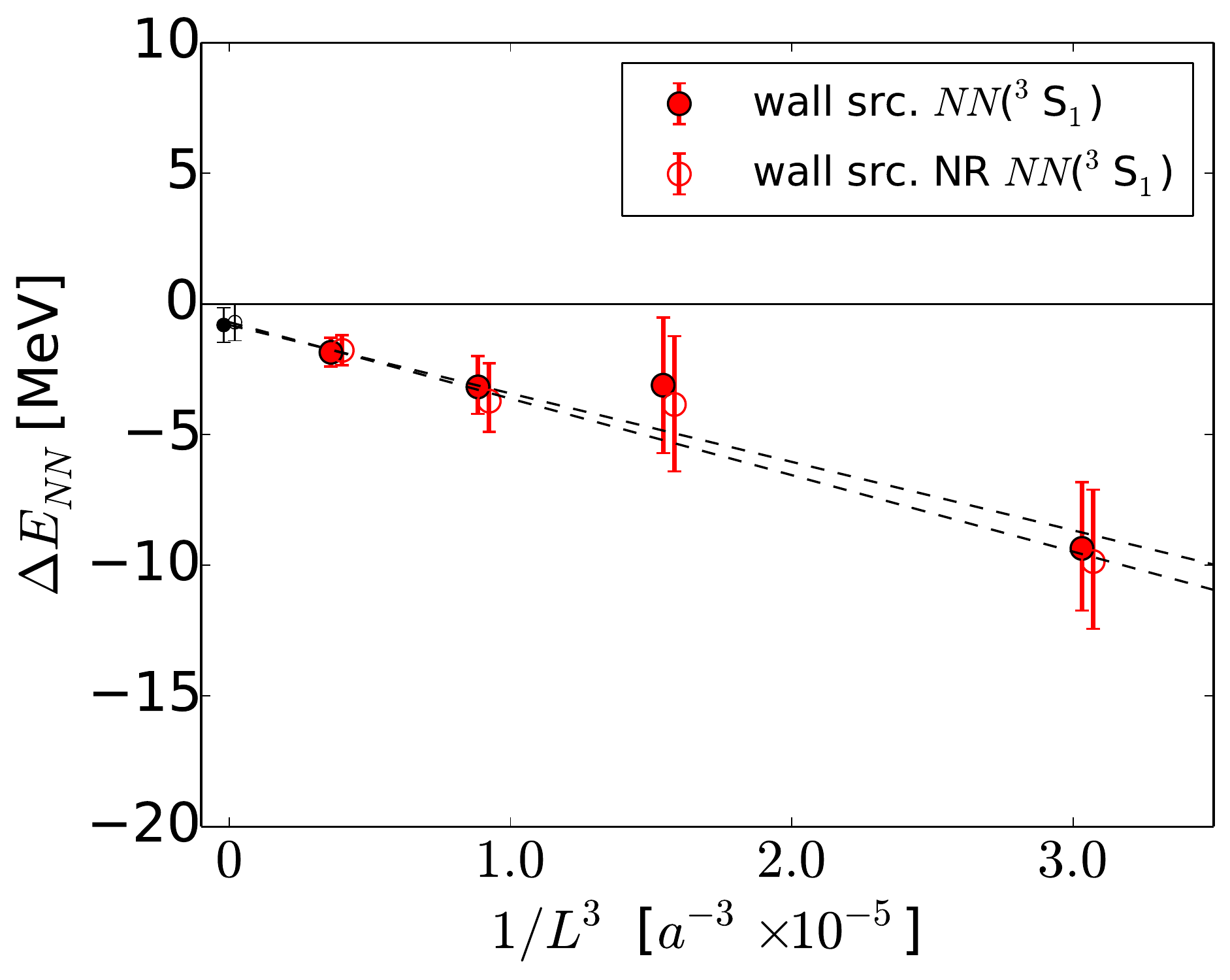}
  \caption{
  (Upper left) $2\meffN(t)$ (black bar) and the effective energy $\EeffNN(t)$ (red triangle)
  in the ${}^{1}S_0$ channel as a function of $t/a$ on the $48^3\times 48$ lattice from the wall source with the non-relativistic operator. 
  (Upper right)  Same in the ${}^{3}S_1$ channel.
  (Middle left) Effective energy shift $\DelEeffNN (t)$, together with the fit (statistical only)  in the ${}^{1}S_0$ channel with the same lattice setup.
  (Middle right) Same     in the ${}^{3}S_1$ channel.
  (Lower left) Energy shift $\DelENN$ in the ${}^{1}S_0$ channel as a function of $1/L^3$  
    from the wall source
    with both non-relativistic (open circle) and relativistic operators (solid circle).
    Shown together are the linear extrapolation in $1/L^3$ to the infinite volume. 
    The errors are obtained from statistical and systematic errors added in quadrature.
  (Lower right) Same   in the ${}^{3}S_1$ channel.
    }
  \label{fig:E_NN_wall}
\end{figure}

We now repeat the same analyses
by changing the smeared source to the wall source.
 The results are summarized in Fig.~\ref{fig:E_NN_wall}  with the 
 data in the right column in  Table~\ref{tab:summary_NN}.
 The lower  panels of Fig.~\ref{fig:E_NN_wall} indicate that (i) the numbers are
 significantly different from those  with the smeared source and  (ii)
  there is no strong evidence of the bound states in both ${}^1S_0$ and ${}^3S_1$
  channels.  In fact, we obtain, for the wall source with non-relativistic operator,
 $\DelENN (^1S_0) = + [ 0.10(0.65)(^{+0.19}_{-0.01})]$ MeV 
 and 
 $\DelENN (^3S_1) = - [ 0.69(0.71)(^{+0.07}_{-0.00})]$ MeV.

\begin{table}
  \centering
  \begin{tabular}{|cc|rc|rc|}
\hline
\multicolumn{2}{|c|}{$NN (^1S_0)$} & \multicolumn{2}{c}{smeared source} & \multicolumn{2}{|c|}{wall source} \\
\hline 
volume & operator & $\Delta E$ [MeV] & fit range & $\Delta E$ [MeV] & fit range\\
\hline
$32^3$ & rela.     & $-11.37(2.79)^{+1.31}_{-1.10}$ & 11-16 & $-10.57(3.61)^{+0.39}_{-0.57}$ & 11-15 \\
       & non-rela. & $-12.68(2.30)^{+1.23}_{-1.05}$ & 10-16 & $-6.07(3.68)^{+1.31}_{-1.25}$ & 11-15 \\
$40^3$ & rela.     & $-8.02(1.72)^{+1.02}_{-0.53}$ & 11-15 & $-8.57(2.08)^{+0.67}_{-0.45}$ & 11-15 \\
       & non-rela. & $-10.91(1.89)^{+0.55}_{-0.35}$ & 11-17 & $-9.30(2.15)^{+0.71}_{-0.79}$ & 11-15 \\
$48^3$ & rela.     & $-8.27(1.09)^{+0.81}_{-0.63}$ & 12-16 & $-2.16(1.21)^{+0.80}_{-0.54}$ & 12-16 \\
       & non-rela. & $-9.96(1.14)^{+0.40}_{-0.16}$ & 12-16 & $-2.64(1.24)^{+0.59}_{-0.37}$ & 12-16 \\
       $64^3$ & rela.     & $-3.25(1.28)^{+0.48}_{-0.24}$ & 10-16 & $-0.97(0.39)^{+0.27}_{-0.14}$ & 12-16 \\
       & non-rela. & $-5.87(1.39)^{+0.14}_{-0.10}$ & 10-16 & $-1.18(0.42)^{+0.33}_{-0.17}$ & 12-17 \\
\hline
$\infty$ & rela. & $-3.85(1.28)^{+0.45}_{-0.24}$ & & $+0.68(0.62)^{+0.20}_{-0.05}$ & \\
         & non-rela. & $-6.54(1.29)^{+0.11}_{+0.00}$ & & $+0.10(0.65)^{+0.19}_{-0.01}$ & \\
& Ref.\cite{Yamazaki:2012hi}  (non-rela.) & $- 7.4(1.3)(0.6)$ & & $-$ & \\
\hline \hline
\multicolumn{2}{|c|}{$NN (^3S_1)$} & \multicolumn{2}{c}{smeared source} & \multicolumn{2}{|c|}{wall source} \\
\hline 
volume & operator & $\Delta E$ [MeV] & fit range & $\Delta E$ [MeV] & fit range\\
\hline
$32^3$ & rela.     & $-11.31(1.85)^{+0.68}_{-0.45}$ & 10-14 & $-9.35(2.09)^{+1.43}_{-1.12}$ & 10-14 \\
       & non-rela. & $-14.38(2.12)^{+0.65}_{-0.19}$ & 10-15 & $-9.86(2.27)^{+1.56}_{-1.23}$ & 10-14 \\
$40^3$ & rela.     & $-11.64(1.41)^{+1.01}_{-0.54}$ & 11-15 & $-3.11(2.49)^{+0.71}_{-0.74}$ & 11-15 \\
       & non-rela. & $-14.46(1.40)^{+0.78}_{-0.27}$ & 11-15 & $-3.84(2.44)^{+0.95}_{-0.76}$ & 11-15 \\
$48^3$ & rela.     & $-13.60(1.39)^{+0.58}_{-0.30}$ & 13-18 & $-3.17(0.99)^{+0.63}_{-0.27}$ & 12-16 \\
       & non-rela. & $-14.78(1.18)^{+0.38}_{-0.16}$ & 12-18 & $-3.72(1.10)^{+0.95}_{-0.42}$ & 12-16 \\
       $64^3$ & rela.     & $-8.08(0.82)^{+0.18}_{-0.18}$ & 10-16 & $-1.85(0.53)^{+0.15}_{-0.11}$ & 13-18 \\
     & non-rela. & $-10.91(1.01)^{+0.42}_{-0.26}$ & 10-16 & $-1.77(0.56)^{+0.16}_{-0.09}$ & 13-18 \\
\hline
$\infty$ & rela. & $-8.68(0.92)^{+0.19}_{-0.16}$ & & $-0.80(0.66)^{+0.05}_{-0.01}$ & \\
         & non-rela. & $-11.60(1.06)^{+0.36}_{-0.24}$ & & $-0.69(0.71)^{+0.07}_{-0.00}$ & \\
& Ref.\cite{Yamazaki:2012hi}  (non-rela.) & $- 11.5(1.1)(0.6)$ & & $-$ & \\
\hline
  \end{tabular}
  \caption{A summary of $\DelENN$ for smeared and wall sources 
    with both relativistic and non-relativistic operators on four volumes
    and corresponding exponential fit ranges, 
  together with infinite volume extrapolations. The result of the previous work with the same lattice setup is shown  
in the column  \cite{Yamazaki:2012hi}  (non-rela.). }
  \label{tab:summary_NN}
\end{table}

\section{$^3$He and $^4$He systems}
\label{sec:He}
We now  consider $^3$He (2 protons and 1 neutron) and  $^4$He (2 protons and 2 neutrons).
Since $m_u=m_d$ in 2+1 flavor QCD, $^3$He  is identical to triton, $^3$H (1 proton and 2 neutrons),
as far as its mass is concerned.

\begin{figure}[tb]
  \centering
  \includegraphics[width=0.45\textwidth]{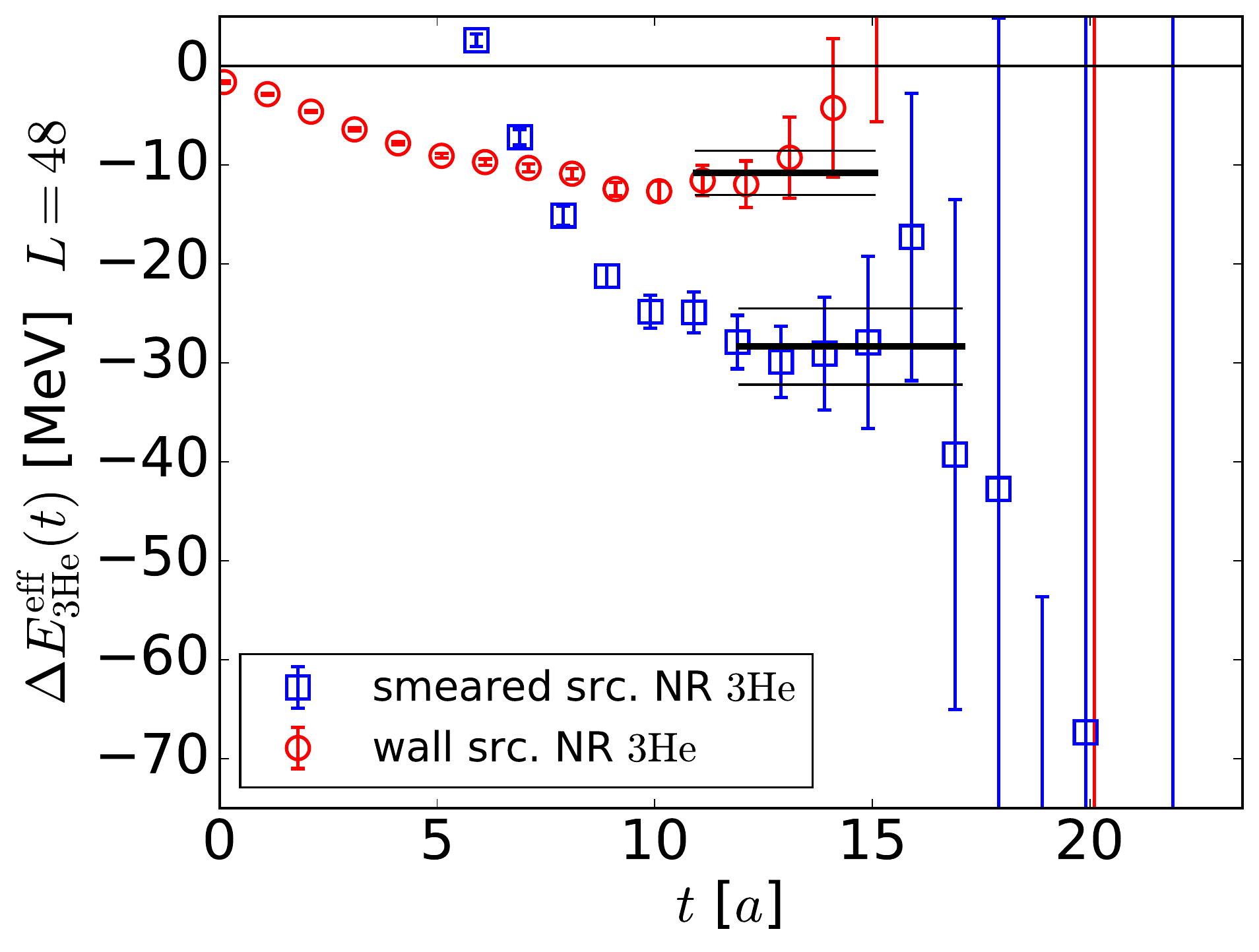}
  \includegraphics[width=0.45\textwidth]{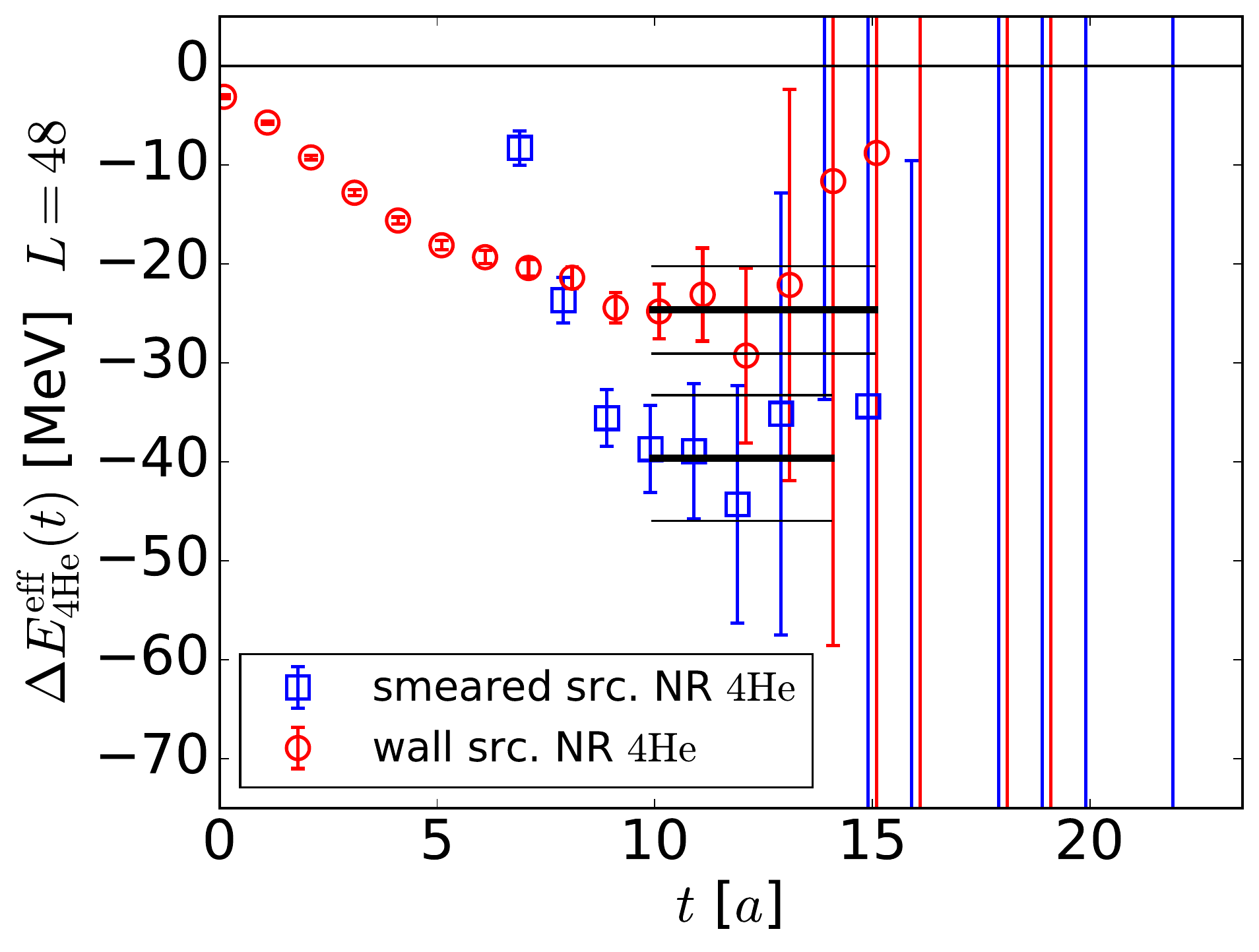}\\
  \includegraphics[width=0.45\textwidth]{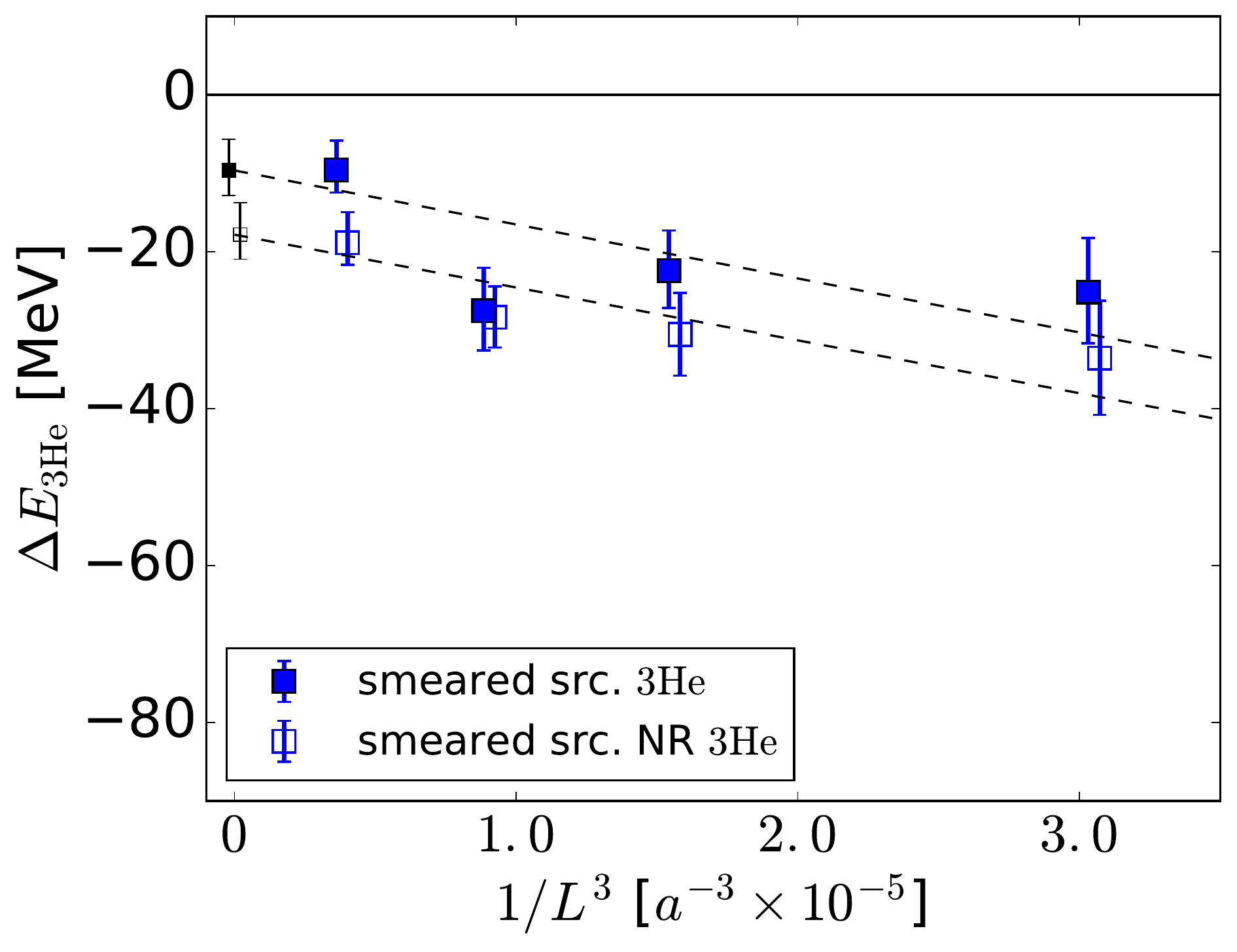}
  \includegraphics[width=0.45\textwidth]{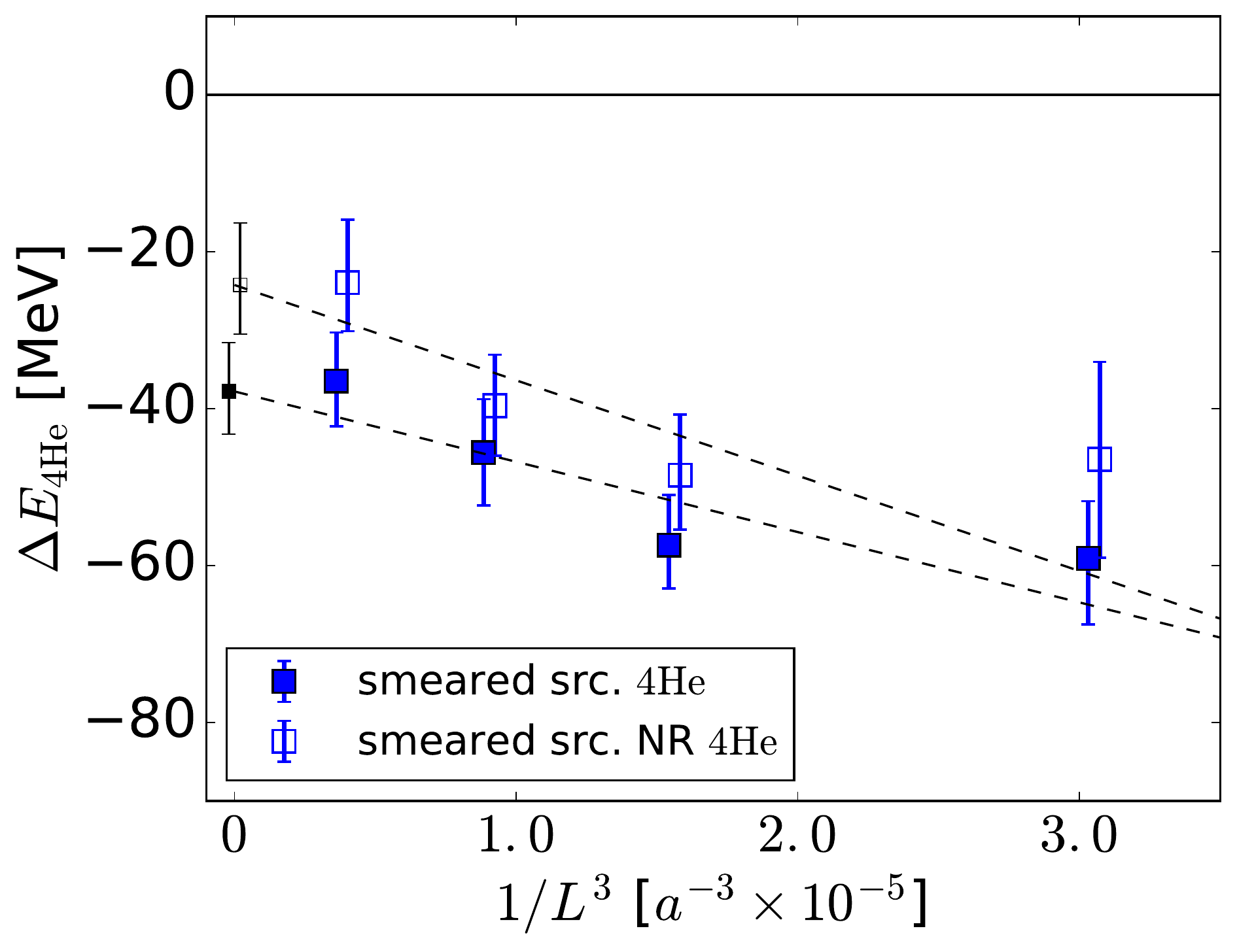} \\
  \includegraphics[width=0.45\textwidth]{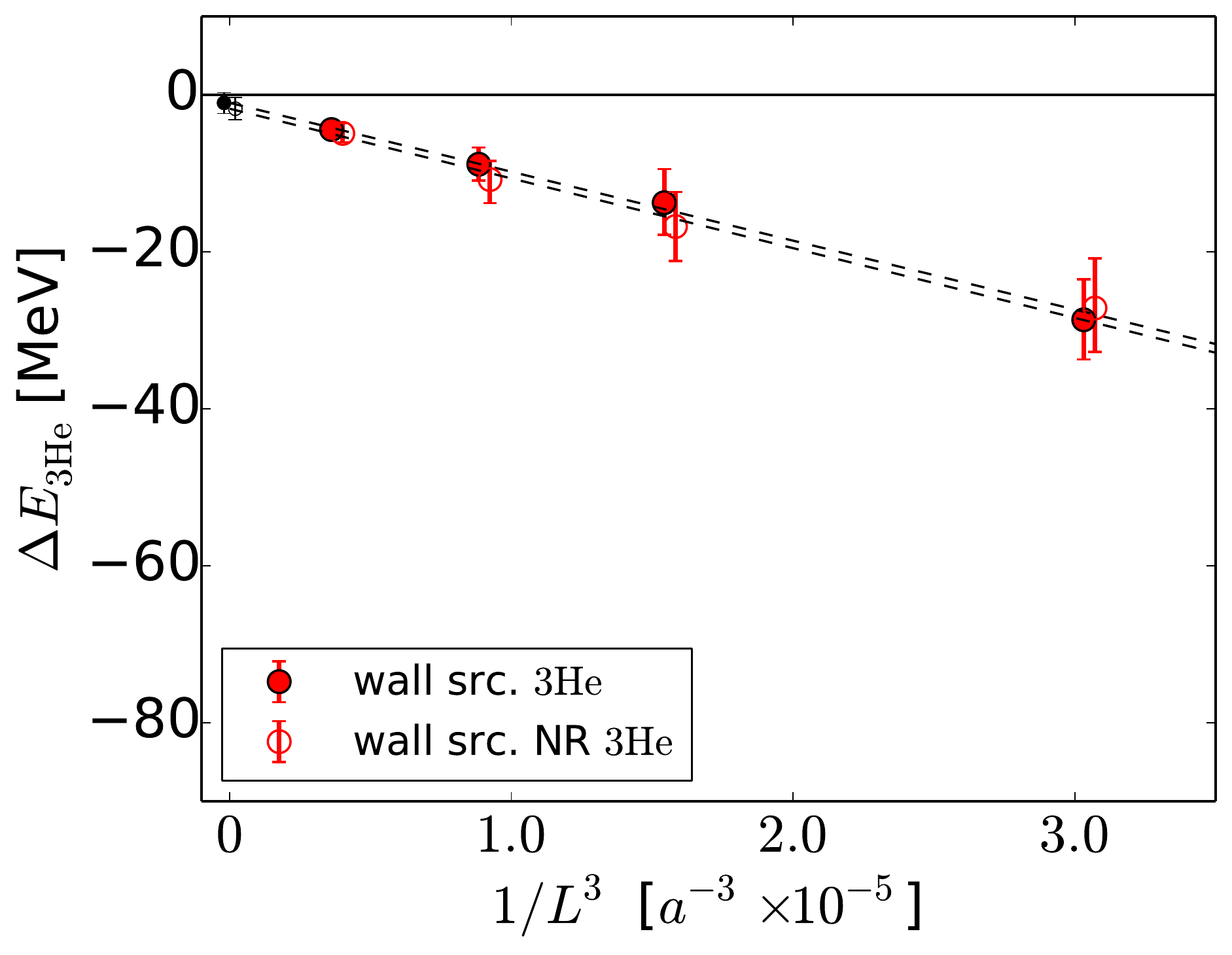}
  \includegraphics[width=0.45\textwidth]{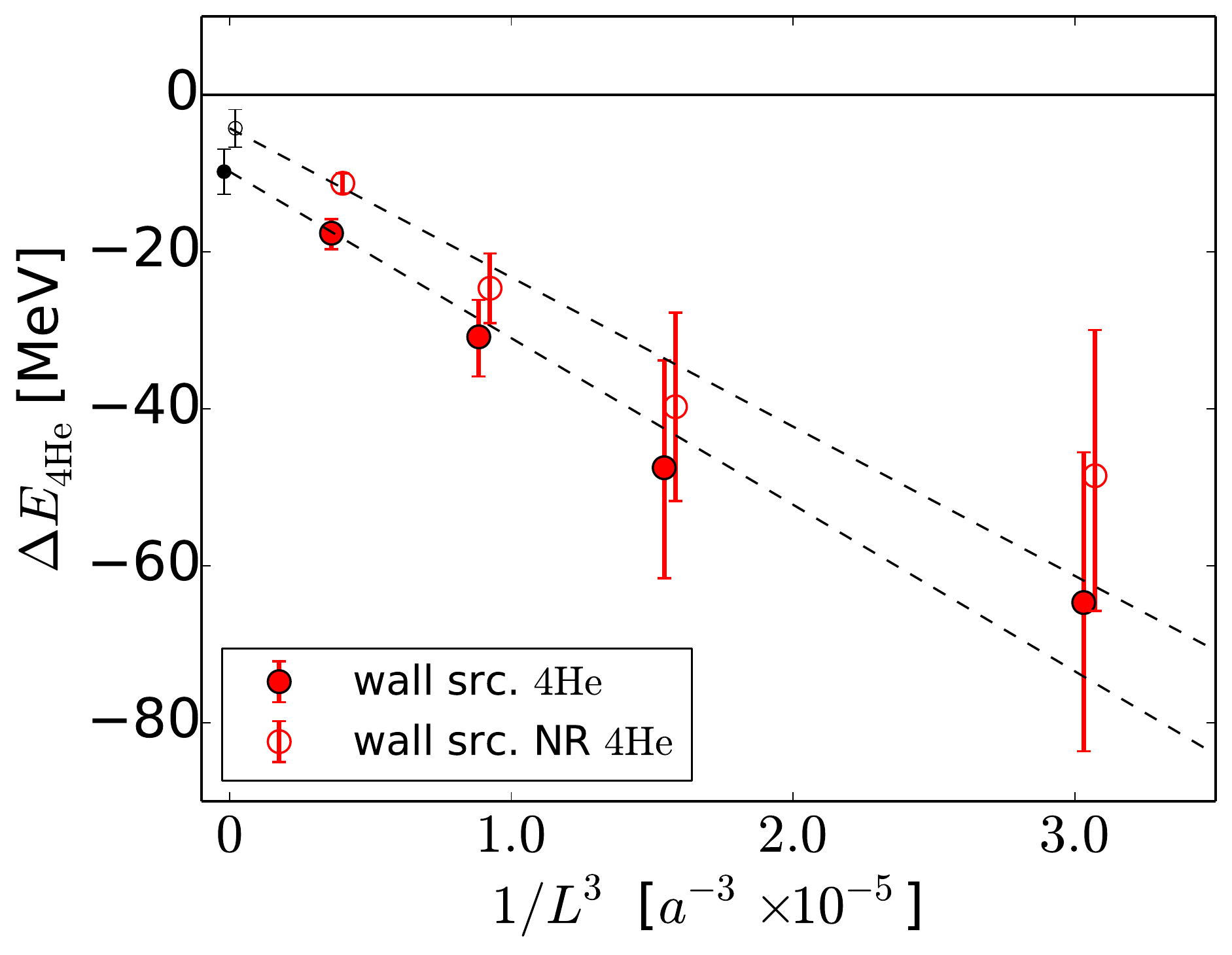}
  \caption{ (Upper left) The effective energy shift $\DelEeffTri(t)$ on the $48^3 \times 48$ lattice
 for both smeared (blue squares) and wall (red circles) sources with non-relativistic operators,
 together with the fit (statistical only).
 (Middle left)  Energy shift $\DelETri$ as a function of $1/L^3$  
    from the smeared source
    with both non-relativistic (open square) and relativistic operators (solid square).
    Shown together are the linear extrapolation in $1/L^3$ to the infinite volume. 
    The errors are obtained from statistical and systematic errors added in quadrature.
(Lower left) Energy shift $\DelETri$ as a function of $1/L^3$  from the wall source
 together with the linear extrapolation in $1/L^3$ to the infinite volume. 
 (Upper right, Middle right and Lower right)  Same as the left figures for ${}^4{\rm He}$.
}
  \label{fig:34He}
\end{figure}

Upper left (right)  panel of Fig.~\ref{fig:34He} shows the effective energy
 shift $\DelEeffTri(t)=\EeffTri(t) - 3 \meffN(t)$  ($\DelEeffHe(t)=\EeffHe(t) - 4 \meffN(t)$)
 on the $48^3\times 48$ lattice for both smeared and wall sources with the non-relativistic operator.
 The explosion in the noise to signal ratio
 from even smaller $t/a$ than that of the two-nucleon case. 
  We try to fit the plateau-like structure just before the explosion typically in the range $10 \le t/a \le 14$.  
  In Table~\ref{tab:summary_34He}, results of $\DelETri$  and $\DelEHe$   on four volumes for smeared as well as wall
    sources and for non-relativistic as well as relativistic operators are summarized.
    Systematic errors are estimated by changing the upper and lower limit of the fitting window  by one unit of $t/a$.

Middle left (right)  panel of Fig.~\ref{fig:34He}  shows $\DelETri$ ($\DelEHe$) from the smeared source 
as a function of $1/L^3$,  together with the linear extrapolation in $1/L^3$ to the infinite volume,
for both non-relativistic and relativistic operators.
Lower left (right)  panel of Fig.~\ref{fig:34He}  shows $\DelETri$ ($\DelEHe$)  from the wall source
as a function of $1/L^3$,  together with the linear extrapolation in $1/L^3$ to the infinite volume,
for both non-relativistic and relativistic operators.
As in the case of $NN$, the result of the smeared source and that of the wall source
do not agree:  The former indicates the bound states for both ${}^3{\rm He}$ and ${}^4{\rm He}$
as suggested in ~\cite{Yamazaki:2012hi},
while the latter shows  no strong evidence of such bound states.

\begin{table}
  \centering
  \begin{tabular}{|cc|rc|rc|}
\hline  
\multicolumn{2}{|c|}{$^3$He(=$^3$H)} & \multicolumn{2}{c}{smeared source} & \multicolumn{2}{|c|}{wall source} \\
\hline 
volume & operator & $\Delta E$ [MeV] & fit range & $\Delta E$ [MeV] & fit range\\
\hline
$32^3$ & rela.     & $-25.14(6.40)^{+2.49}_{-1.36}$ & 10-14 & $-28.66(5.05)^{+0.96}_{-0.30}$ & 9-13 \\
       & non-rela. & $-33.57(7.13)^{+1.63}_{-1.00}$ & 10-14 & $-27.16(4.93)^{+3.96}_{-2.67}$ & 9-13 \\
$40^3$ & rela.     & $-22.41(4.75)^{+1.92}_{-0.37}$ & 11-16 & $-13.75(3.97)^{+1.60}_{-0.94}$ & 9-13 \\
       & non-rela. & $-30.55(5.20)^{+0.93}_{-0.61}$ & 11-16 & $-16.78(4.03)^{+1.81}_{-1.69}$ & 9-13 \\
$48^3$ & rela.     & $-27.52(4.90)^{+2.43}_{-1.23}$ & 13-17 & $-8.84(2.02)^{+0.72}_{-0.46}$ & 11-15 \\
       & non-rela. & $-28.35(3.85)^{+0.88}_{-0.50}$ & 12-17 & $-10.80(2.22)^{+0.90}_{-1.99}$ & 11-15 \\
$64^3$ & rela.     & $-9.59(2.47)^{+2.81}_{-1.45}$ & 9-13 & $-4.42(0.99)^{+0.21}_{-0.36}$ & 12-15 \\
  & non-rela. & $-18.85(2.33)^{+3.13}_{-1.61}$ & 8-13 & $-4.91(1.11)^{+0.65}_{-0.29}$ & 12-16 \\
\hline
$\infty$ & rela. & $-9.64(2.85)^{+2.81}_{-1.43}$ & & $-1.02(1.32)^{+0.06}_{-0.34}$ & \\
         & non-rela. & $-17.83(2.73)^{+3.02}_{-1.53}$ & & $-1.77(1.42)^{+0.16}_{-0.04}$ & \\
& Ref.\cite{Yamazaki:2012hi} (non-rela.)   & $- 20.3(4.0)(2.0)$ & & $-$ &  \\
\hline \hline
\multicolumn{2}{|c|}{$^4$He} & \multicolumn{2}{c}{smeared source} & \multicolumn{2}{|c|}{wall source} \\
\hline 
volume & operator & $\Delta E$ [MeV] & fit range & $\Delta E$ [MeV] & fit range\\
\hline
$32^3$ & rela.     & $-59.09(7.25)^{+0.80}_{-4.26}$ & 8-11 & $-64.68(17.95)^{+6.64}_{-6.10}$ & 9-12 \\
       & non-rela. & $-46.47(12.37)^{+1.17}_{-1.93}$ & 9-13 & $-48.52(16.61)^{+8.32}_{-4.65}$ & 9-12 \\
$40^3$ & rela.     & $-57.39(4.59)^{+4.45}_{-3.11}$ & 8-12 & $-47.51(12.98)^{+4.23}_{-5.46}$ & 9-13 \\
       & non-rela. & $-48.48(5.54)^{+5.42}_{-4.19}$ & 9-12 & $-39.74(11.99)^{+0.41}_{-0.75}$ & 9-13 \\
$48^3$ & rela.     & $-45.60(6.66)^{+1.40}_{-0.91}$ & 10-14 & $-30.83(4.43)^{+1.57}_{-2.38}$ & 10-14 \\
       & non-rela. & $-39.62(6.35)^{+1.42}_{-0.75}$ & 10-14 & $-24.64(4.42)^{+0.08}_{-0.14}$ & 10-15 \\
       $64^3$ & rela.     & $-36.47(5.79)^{+2.19}_{-0.01}$ & 8-12 & $-17.63(1.66)^{+0.66}_{-1.19}$ & 11-13 \\
     & non-rela. & $-23.94(5.19)^{+6.11}_{-3.35}$ & 8-11 & $-11.27(1.24)^{+0.15}_{-0.28}$ & 10-13 \\
\hline
$\infty$ & rela. & $-37.81(5.45)^{+2.96}_{+0.00}$ & & $-9.79(2.85)^{+0.00}_{-0.33}$ & \\
         & non-rela. & $-24.24(5.63)^{+5.59}_{-2.78}$ & & $-4.25(2.37)^{+0.09}_{-0.55}$ & \\
& Ref.\cite{Yamazaki:2012hi} (non-rela.) & $- 43(12)(8)$ &  & $-$ &  \\
\hline
  \end{tabular}
  \caption{A summary of $\DelETri$ and $\DelEHe$ for smeared and wall sources with both relativistic and non-relativistic operators on four volumes
    and corresponding exponential fit ranges,
    together with infinite volume extrapolations. 
  }
  \label{tab:summary_34He}
\end{table}

\section{Conclusions}
\label{sec:conclusion}

In this paper, we have addressed the issue of the single state saturation
of the temporal correlation function for the multi-baryons by  employing
(2+1)-flavor lattice QCD at $m_{\pi}=0.51$ GeV on four lattice volumes with
$L=$ 2.9, 3.6, 4.3 and 5.8 fm.  A major difference between the single baryon
and multi-baryons  on the lattice is that there appears energy levels
corresponding to the elastic scattering for the latter.  The level spacings
become smaller as $L$ becomes larger, since they correspond to the continuum
states for $L\rightarrow \infty$.  Therefore, it is required to take large
temporal distance $t$  between the source and sink operators to isolate the
ground state of  multi-baryons.  This is, however, very difficult due to the
exponential growth of the noise over the signal which has been known to be a
characteristic feature of the multi-baryon correlations.  In such a situation,
one may be misled by a  fake plateau of the effective energy shift $\Delta
E^{\rm eff}(t)$ at intermediate values of $t$  before the explosion of the
noise takes place.
  
  We have demonstrated,   by using the mock data, 
  that the above situation can easily happen with a slight contamination of
 the elastic scattering state. Then we analyzed
the lattice data in  $\Xi\Xi (^1S_0)$ and $\Xi\Xi (^3S_1)$  channels to show explicitly that
 the same situation indeed  takes place for the real data. 
  By adopting the smeared source operator used in~\cite{Yamazaki:2012hi}
  and the wall source operator,  we fit  the plateau-like structure 
 around $t/a\sim 15$ and find that the results of $\Delta E_{\Xi\Xi}$
 at each $L$ as well as those extrapolated
  to $L \rightarrow \infty$ turn out to disagree with each other between two sources. This implies that
   the ground state saturation is not achieved in such intermediate values of $t$.  
   Moreover, we found that
     $\Delta E_{\rm \Xi\Xi} (^3S_1) > 0$  for the smeared source at $L\rightarrow \infty$, which is not 
     physically acceptable.

     One may suspect that the above disagreement originates from slower temporal convergence of single
     baryon for the wall source than the smeared source.
     However, this is not necessarily the case,
     since the plateau of the effective energy shift shows much stronger dependence on
     the change of the two-baryon sink operators for the smeared source as shown in Appendix~\ref{sec:app:sink-dep}.
     In fact, one can explicitly show, by using the HAL QCD method, 
     that the smeared source has significantly larger
     contamination from the two-baryon excited states.
     The details will be reported in a forthcoming paper~\cite{iritani1}.

We have applied the same analysis also to $NN(^1S_0)$, $NN(^3S_1)$, $^3$He and $^4$He,
although  the statistical errors become lager for non-strange baryons than those for $\Xi\Xi$.
 Again, the results of the two sources do not agree with each other: The smeared source indicates
 that there are bound states in all these channels, while no definite signatures 
 of the bound states  are found for the wall source.
 
 By combining the general theoretical considerations and the numerical evidences, we conclude that
 the plateau-like structure seen at the moderate values of $t$ in the temporal correlation for multi-baryons should be considered as
  a  ``mirage" in the sense that the true signals  are located in much larger  $t$ with different values of $\Delta E^{\rm eff}(t)$.
  This also casts strong doubt on the recent works on the basis 
  of the plateau fitting of the temporal correlations
  \cite{Yamazaki:2012hi,Yamazaki:2015asa,Yamazaki:2011nd,Yamazaki:2009ua,Beane:2009py,Beane:2010hg,Beane:2011iw,Beane:2012ey,Beane:2012vq,Beane:2013br,Beane:2014ora,Beane:2014oea,Beane:2015yha,Chang:2015qxa,Detmold:2015daa,Orginos:2015aya,Berkowitz:2015eaa}, almost all of which
  claim the existence of bound multi-baryons (such as dineutron, deuteron, $^{3}$He, $^{4}$He, and other strange multi-baryons).
  At least, one should use more sophisticated approaches 
  than the plateau fitting, such as the Bayesian fitting, Black box, or variational methods
  to extract the ground state energy from the temporal correlators 
  (see e.g. Ref.~\cite{Lin:2011ti} for the review of these methods
  and the applications to single-hadron spectroscopy.)
  A trustable way to examine the validity of these results   
  is to study the $L$-dependence of $\Delta E$  \`{a} la L\"{u}scher's finite volume formula.
  Detailed analysis along this line will be reported in another forthcoming paper \cite{iritani2}. 
 
  It should also be noted that the use of the full space-time correlations (HAL QCD method)
  instead of only the temporal
 correlations has been shown to solve the single-state saturation problem discussed in this paper~\cite{HALQCD:2012aa}.
 Detailed examination between the results from the temporal correlation alone and those from the space-time correlation
  by using the same lattice data as those in the present  paper  will be also
 reported in a forthcoming paper~\cite{iritani1}.


\acknowledgments

We thank the authors of Ref.~\cite{Yamazaki:2012hi} for providing the gauge configurations
and the detailed account on the smearing source used in~\cite{Yamazaki:2012hi}.
T.D. thanks Dr. T.~Izubuchi for valuable discussions on the AMA technique.
We also thank ILDG/JLDG~\cite{conf:ildg/jldg}, 
which serves as an essential infrastructure in this study.
Several lattice QCD codes are used in this study,
CPS~\cite{CPS}, Bridge++~\cite{bridge++} and the modified version thereof by Dr. H.~Matsufuru, 
cuLGT~\cite{Schrock:2012fj} and domain-decomposed quark solver~\cite{Boku:2012zi,Teraki:2013}.
We are grateful for authors and maintainers of these computational codes.
The numerical calculations have been performed on BlueGene/Q and SR16000 at KEK, HA-PACS at University of Tsukuba and 
FX10 at the University of Tokyo.
Part of the data are also obtained as the by-product of HPCI project (hp150085, hp160093)
using the K computer at RIKEN, AICS.
This work is supported in part by the Japanese Grant-in-Aid for Scientific Research (No. 24740146, 25287046, 26400281, 15K17667,16H03978 ), 
by MEXT Strategic Program for Innovative Research (SPIRE) Field 5,  
by a priority issue (Elucidation of the fundamental laws and evolution of the universe) to be tackled by using Post ``K" Computer, 
and by Joint Institute for Computational Fundamental Science (JICFuS).  T.H. were partially supported 
by RIKEN iTHES Project.



\bibliographystyle{JHEP}

\begin{thebibliography}{99}

\bibitem{Luscher:1990ux} 
  M.~Luscher,
  Nucl.\ Phys.\ B {\bf 354}, 531 (1991).
  
\bibitem{Yamazaki:2012hi}
  T.~Yamazaki, K.~i.~Ishikawa, Y.~Kuramashi and A.~Ukawa,
  Phys.\ Rev.\ D {\bf 86} (2012) 074514
  [arXiv:1207.4277 [hep-lat]].

\bibitem{Yamazaki:2015asa}
  T.~Yamazaki, K.~i.~Ishikawa, Y.~Kuramashi and A.~Ukawa,
  Phys.\ Rev.\ D {\bf 92} (2015) 1,  014501
  [arXiv:1502.04182 [hep-lat]].
  
\bibitem{Ishii:2006ec}
  N.~Ishii, S.~Aoki and T.~Hatsuda,
Phys.\ Rev.\ Lett. {\bf 99} (2007) 022001 
  [arXiv:nucl-th/0611096].
  
  \bibitem{Aoki:2008hh}
  S.~Aoki, T.~Hatsuda and N.~Ishii,
Comput.\ Sci.\ Dis.   {\bf 1} (2008) 015009
  [arXiv:0805.2462 [hep-ph]].

\bibitem{Aoki:2009ji}
  S.~Aoki, T.~Hatsuda and N.~Ishii,
Prog.\ Theor.\ Phys.  {\bf 123} (2010) 89
  [arXiv:0909.5585 [hep-lat]].
  
\bibitem{Aoki:2011ep} 
  S.~Aoki for HAL QCD Collaboration,
  Prog.\ Part.\ Nucl.\ Phys. {\bf 66} (2011) 687
  [arXiv:1107.1284 [hep-lat]].

\bibitem{Aoki:2012tk} 
  S.~Aoki {\it et al.}  [HAL QCD Collaboration],
 Prog.\ Theor.\ Exp.\ Phys. {\bf 2012} (2012) 01A105
  [arXiv:1206.5088 [hep-lat]].

\bibitem{HALQCD:2012aa}
  N.~Ishii {\it et al.}  [HAL QCD Collaboration],
  Phys.\ Lett.\  \textbf{B712}  (2012) 437. 
  
 \bibitem{Inoue:2010es}
  T.~Inoue {\it et al.}  [HAL QCD Collaboration],
  Phys.\ Rev.\ Lett.\  {\bf 106} (2011) 162002
  [arXiv:1012.5928 [hep-lat]].
    
\bibitem{Inoue:2011ai}
  T.~Inoue {\it et al.}  [HAL QCD Collaboration],
  Nucl.\ Phys.\ A {\bf 881} (2012) 28
  [arXiv:1112.5926 [hep-lat]].

\bibitem{Parisi:1983ae}
  G.~Parisi,
  Phys.\ Rept.\  {\bf 103} (1984) 203.

\bibitem{Lepage:1989hd}
  G.~P.~Lepage,
  in {\it From Actions to Answers: Proceedings of the TASI 1989},
  edited by T.~Degrand and D.~Toussaint
  (World Scientific, Singapore, 1990).

\bibitem{Aoki:2005et}
  S.~Aoki {\it et al.} [CP-PACS and JLQCD Collaborations],
  Phys.\ Rev.\ D {\bf 73} (2006) 034501
  [hep-lat/0508031].
  

\bibitem{Doi:2012xd}
  T.~Doi and M.~G.~Endres,
  Comput.\ Phys.\ Commun.\  {\bf 184} (2013) 117
  [arXiv:1205.0585 [hep-lat]].

\bibitem{Yamazaki:2009ua}
  T.~Yamazaki {\it et al.} [PACS-CS Collaboration],
  Phys.\ Rev.\ D {\bf 81} (2010) 111504
  [arXiv:0912.1383 [hep-lat]].

\bibitem{Detmold:2012eu}
  W.~Detmold and K.~Orginos,
  Phys.\ Rev.\ D {\bf 87} (2013) no.11,  114512
  [arXiv:1207.1452 [hep-lat]].

\bibitem{Gunther:2013xj}
  J.~G\"{u}nter, B.~C.~Toth and L.~Varnhorst,
  Phys.\ Rev.\ D {\bf 87} (2013) no.9,  094513
  [arXiv:1301.4895 [hep-lat]].

\bibitem{Nemura:2015yha}
  H.~Nemura,
  arXiv:1510.00903 [hep-lat].

\bibitem{Blum:2012uh}
  T.~Blum, T.~Izubuchi and E.~Shintani,
  Phys.\ Rev.\ D {\bf 88} (2013) no.9,  094503
  [arXiv:1208.4349 [hep-lat]].

\bibitem{Shintani:2014vja}
  E.~Shintani, R.~Arthur, T.~Blum, T.~Izubuchi, C.~Jung and C.~Lehner,
  Phys.\ Rev.\ D {\bf 91} (2015) no.11,  114511
  [arXiv:1402.0244 [hep-lat]].

\bibitem{iritani1}
  T.~Iritani {\it et al.}, {\it in preparation.}

\bibitem{Yamazaki:2011nd}
  T.~Yamazaki {\it et al.} [PACS-CS Collaboration],
  Phys.\ Rev.\ D {\bf 84} (2011) 054506
  [arXiv:1105.1418 [hep-lat]].


\bibitem{Beane:2009py}
  S.~R.~Beane {\it et al.} [NPLQCD Collaboration],
  Phys.\ Rev.\ D {\bf 81} (2010) 054505
  [arXiv:0912.4243 [hep-lat]].
  
\bibitem{Beane:2010hg}
  S.~R.~Beane {\it et al.} [NPLQCD Collaboration],
  Phys.\ Rev.\ Lett.\  {\bf 106} (2011) 162001
  [arXiv:1012.3812 [hep-lat]].
  
\bibitem{Beane:2011iw}
  S.~R.~Beane {\it et al.} [NPLQCD Collaboration],
  Phys.\ Rev.\ D {\bf 85} (2012) 054511
  [arXiv:1109.2889 [hep-lat]].

\bibitem{Beane:2012ey}
  S.~R.~Beane {\it et al.},
  Phys.\ Rev.\ Lett.\  {\bf 109} (2012) 172001
  [arXiv:1204.3606 [hep-lat]].
  
\bibitem{Beane:2012vq}
  S.~R.~Beane {\it et al.} [NPLQCD Collaboration],
  Phys.\ Rev.\ D {\bf 87} (2013) 3,  034506
  [arXiv:1206.5219 [hep-lat]].
  
\bibitem{Beane:2013br}
  S.~R.~Beane {\it et al.} [NPLQCD Collaboration],
  Phys.\ Rev.\ C {\bf 88} (2013) 2,  024003
  [arXiv:1301.5790 [hep-lat]].
  
\bibitem{Beane:2014ora}
  S.~R.~Beane {\it et al.},
  Phys.\ Rev.\ Lett.\  {\bf 113} (2014) 25,  252001
  [arXiv:1409.3556 [hep-lat]].
  
\bibitem{Beane:2014oea}
  S.~R.~Beane, W.~Detmold, K.~Orginos and M.~J.~Savage,
  J.\ Phys.\ G {\bf 42} (2015) 3,  034022
  [arXiv:1410.2937 [nucl-th]].
  
\bibitem{Beane:2015yha}
  S.~R.~Beane {\it et al.} [NPLQCD Collaboration],
  Phys.\ Rev.\ Lett.\  {\bf 115} (2015) 13,  132001
  [arXiv:1505.02422 [hep-lat]].
  
\bibitem{Chang:2015qxa}
  E.~Chang, W.~Detmold, K.~Orginos, A.~Parreno, M.~J.~Savage, B.~C.~Tiburzi and S.~R.~Beane,
  arXiv:1506.05518 [hep-lat].
    
\bibitem{Detmold:2015daa}
  W.~Detmold, K.~Orginos, A.~Parreno, M.~J.~Savage, B.~C.~Tiburzi, S.~R.~Beane and E.~Chang,
  arXiv:1508.05884 [hep-lat].

\bibitem{Orginos:2015aya}
  K.~Orginos, A.~Parreno, M.~J.~Savage, S.~R.~Beane, E.~Chang and W.~Detmold,
  arXiv:1508.07583 [hep-lat].
 
\bibitem{Berkowitz:2015eaa}
  E.~Berkowitz, T.~Kurth, A.~Nicholson, B.~Joo, E.~Rinaldi, M.~Strother, P.~M.~Vranas and A.~Walker-Loud,
  arXiv:1508.00886 [hep-lat].

\bibitem{Lin:2011ti}
  H.~W.~Lin,
  Chin.\ J.\ Phys.\  {\bf 49} (2011) 827
  [arXiv:1106.1608 [hep-lat]].
 

\bibitem{iritani2}
  T.~Iritani {\it et al.}, {\it in preparation.}

\bibitem{conf:ildg/jldg}
  \url{http://www.lqcd.org/ildg},
  \url{http://www.jldg.org}

\bibitem{CPS}
  Columbia Physics System (CPS),
  \url{http://usqcd-software.github.io/CPS.html}

\bibitem{bridge++}
  Bridge++, 
  \url{http://bridge.kek.jp/Lattice-code/}

\bibitem{Schrock:2012fj}
  M.~Schr\"{o}ck and H.~Vogt,
  Comput.\ Phys.\ Commun.\  {\bf 184} (2013) 1907
  [arXiv:1212.5221 [hep-lat]].

\bibitem{Boku:2012zi}
  T.~Boku {\it et al.},
  PoS LATTICE {\bf 2012}, 188 (2012)
  [arXiv:1210.7398 [hep-lat]]. 
 
 \bibitem{Teraki:2013} 
Masaaki Terai, Ken-ichi Ishikawa,
Yoshinori Sugisaki, Kazuo Minami,
Fumiyoshi Shoji, Yoshifumi Nakamura,
Yoshinobu Kuramashi, Mitsuo Yokokawa,
"Performance Tuning of a Lattice QCD code on a node of the K computer,"
IPSJ Transactions on Advanced Computing Systems, Vol.6 No.3 43-57 (Sep. 2013) (in Japanese).

\bibitem{Charron:2013paa}
  B.~Charron [HAL QCD Collaboration],
  PoS LATTICE {\bf 2013} (2014) 223
  [arXiv:1312.1032 [hep-lat]].


\bibitem{bruno}
  B.~Charron {\it et al.}, {\it in preparation.}

 
\end{thebibliography}

\newpage
\appendix

\section{Sink operator dependence}
\label{sec:app:sink-dep}

\begin{figure}[b]
  \centering
  \includegraphics[width=0.47\textwidth,clip]{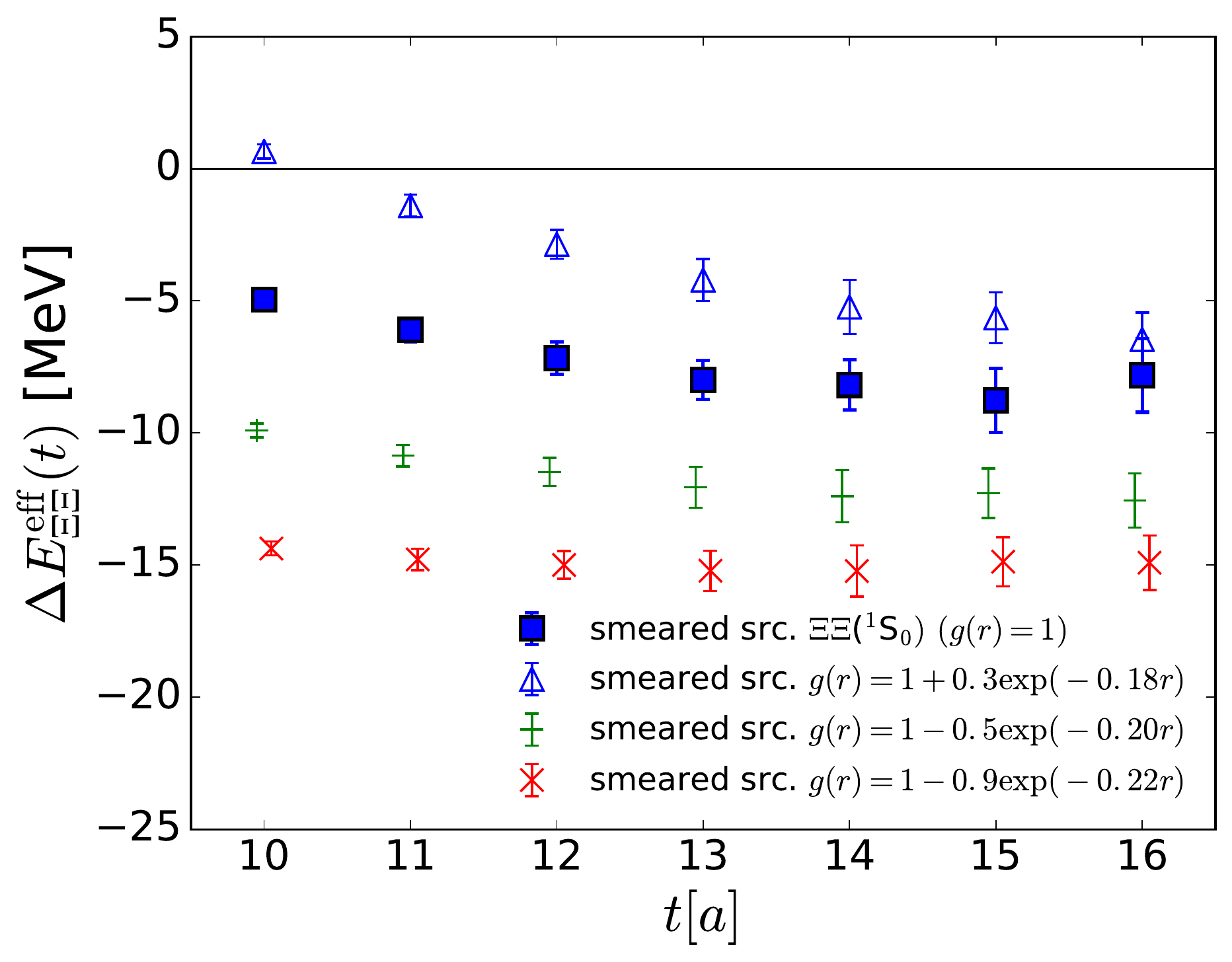}
  \includegraphics[width=0.47\textwidth,clip]{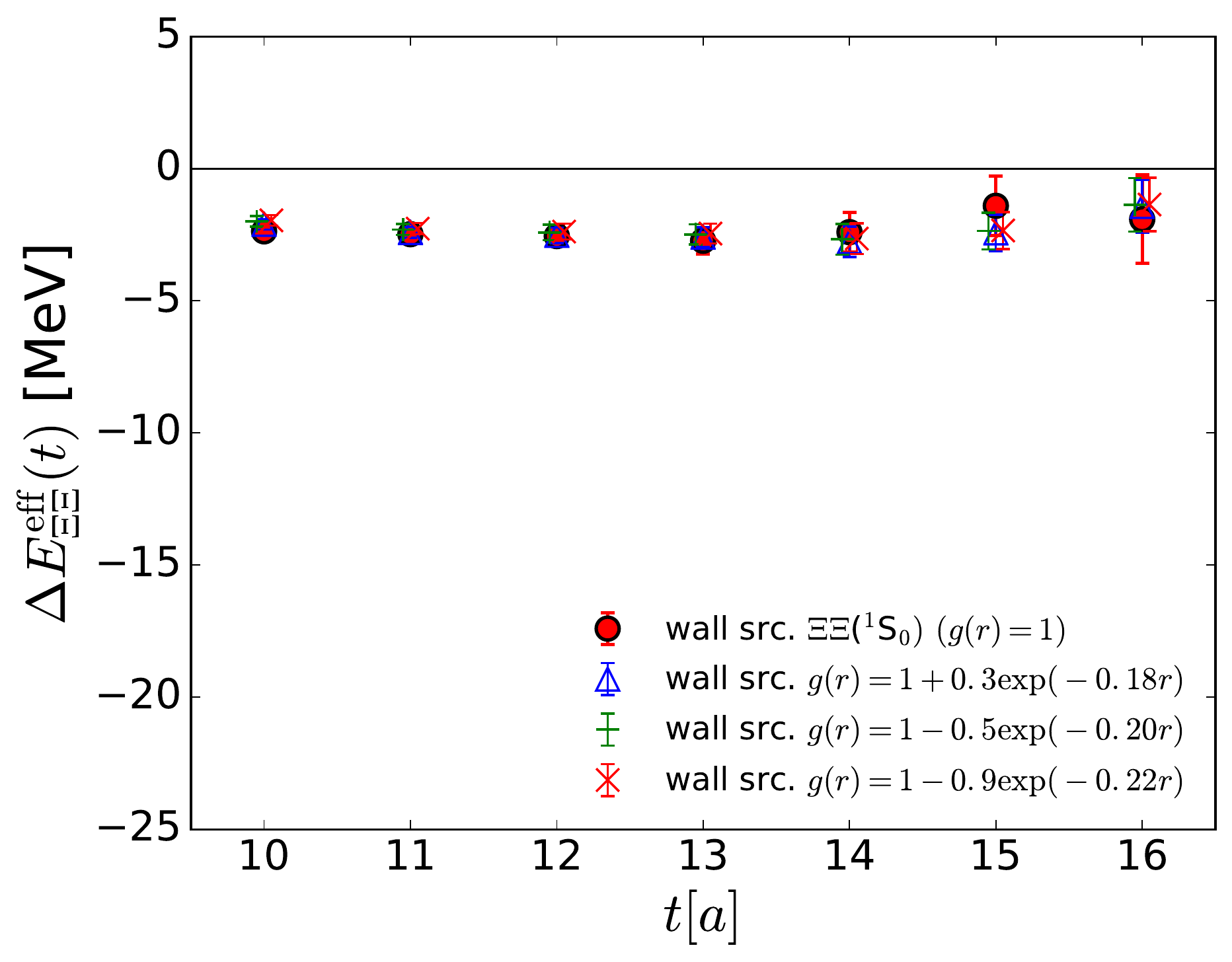}
  \caption{
    \label{fig:sink_proj}
    The effective energy shift
    $\Delta E^\mathrm{eff}_{\Xi\Xi}(t)$
    of $\Xi\Xi$($^1$S$_0$) for $L^3 = 48^3$ 
    with the smeared source (Left) and the wall source (Right).
  }
\end{figure}

In the main text, we investigated the reliability of the plateau-like behavior using
different source operators.  In this Appendix, we make similar analysis 
by using  different sink operators.  We consider
the  $\Xi\Xi$ system in the $^1S_0$ channel as a representative case and 
start with the following temporal correlation function,
\begin{eqnarray}
  C^{(g)}_{\Xi\Xi}(t) = \sum_{\vec{r}} g(|\vec{r}|) 
  \sum_{\vec{R}} \langle \Xi(\vec{R}+\vec{r},t) 
  \Xi(\vec{R},t) \overline{{\cal J}_{\Xi\Xi}}(t=0) \rangle .
  \label{eq:corr_snk}
\end{eqnarray}
The interpolating operator for $\Xi(\vec{x},t)$ is given by Eq.~(\ref{eq:operator}) and 
 we consider only the relativistic operator in this Appendix.
The source operator, ${\cal J}_{\Xi\Xi}$, is taken 
to be the  same as those used in Sec.~\ref{sec:params},
with either of the smeared source or of the wall source.
 The sink operator is a combination of the two local $\Xi$ operators
 with a smearing function $g(r)$~\cite{Charron:2013paa,bruno}.
  The temporal correlation  $C_{\Xi\Xi}(t)$ in Sec.~\ref{sec:params}
  corresponds to the case $g(r) = 1$.
The effective energy, $\EeffXiXi(t)$, and the effective energy shift, 
$\DelEeffXiXi(t) = \EeffXiXi(t) - 2 \meffXi(t)$, are obtained from  $C^{(g)}_{\Xi\Xi}(t)$.

In the following analysis, we adopt $g(r)$ with the following form,
\begin{eqnarray}
  g(r) = 1 + A \exp(-Br) ,
\end{eqnarray}
where four different parameter sets, $(A,B) = (0.3,0.18), (-0.5,0.20), (-0.9,0.22)$ and $(0,0)$,
are considered.

In Fig.~\ref{fig:sink_proj} (Left),
we plot $\DelEeffXiXi(t)$ for four different sink operators in the case of the smeared source.
Although we find a plateau-like behavior for each sink operator, the values of 
 $\DelEeffXiXi(t)$  do not agree among different  sink operators.
Such a large sink-operator dependence  indicates that the contamination from the 
elastic scattering states in the finite volume causes fake plateaux
 as demonstrated in Sec.~\ref{sec:general}.
The true plateau may be identified at  much larger values of $t$, but the explosion of the 
noise prohibits to extract sensible signal at large $t$ as  we discussed in the text. 
Shown in Fig.~\ref{fig:sink_proj} (Right) are $\DelEeffXiXi(t)$ for four different sink operators in the case of the wall source. 
For each sink operator, we find a plateau-like behavior: 
 In this case, it happens that the values of  $\DelEeffXiXi(t)$  agree among different sink operators
  within statistical errors.


\section{Effective energy shifts on various volumes}
\label{sec:app:Eeff}

In this appendix, we give effective energy shifts for various channels on various volume.

\begin{figure}[tbh]
  \centering
  \includegraphics[width=0.45\textwidth]{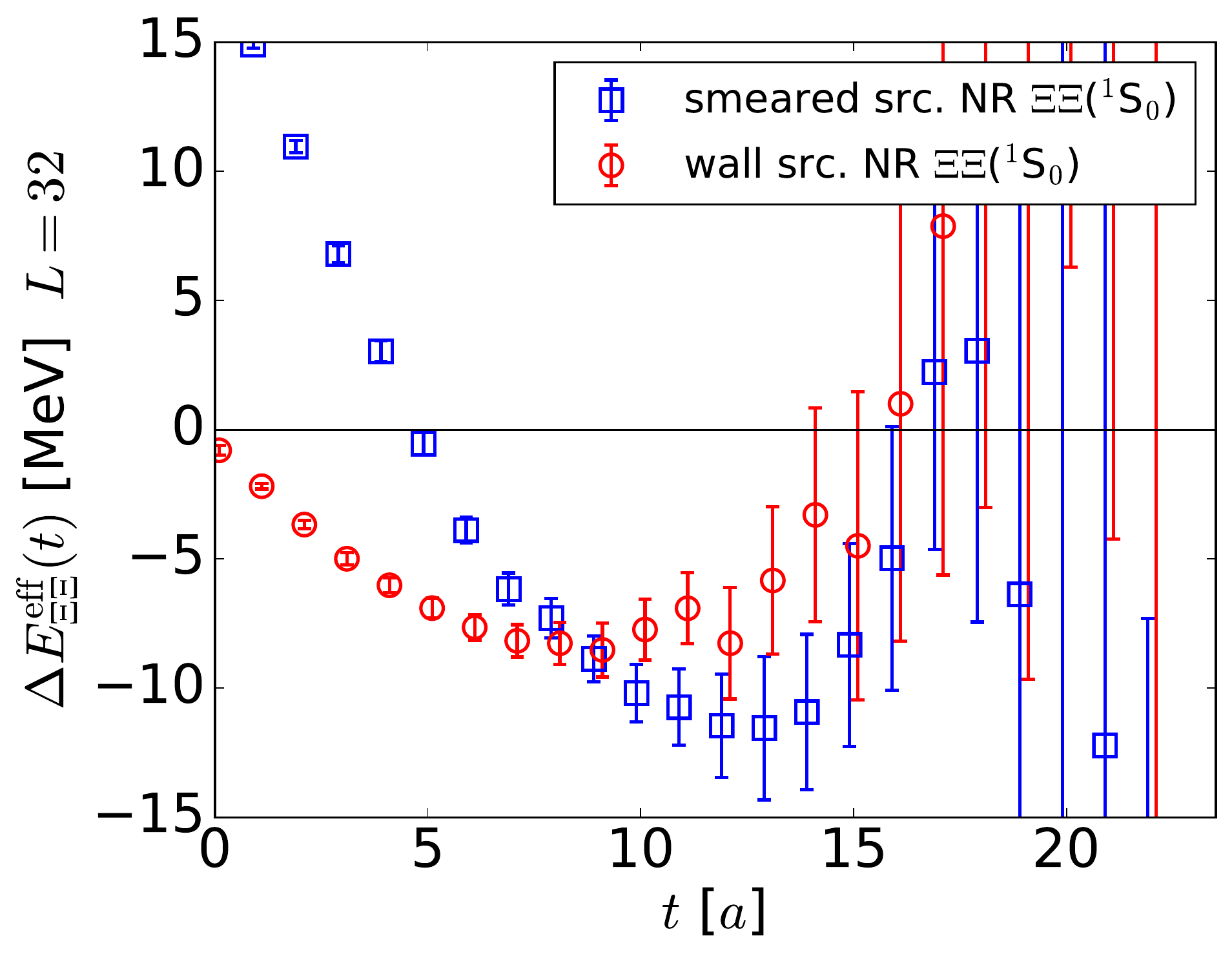}
  \includegraphics[width=0.45\textwidth]{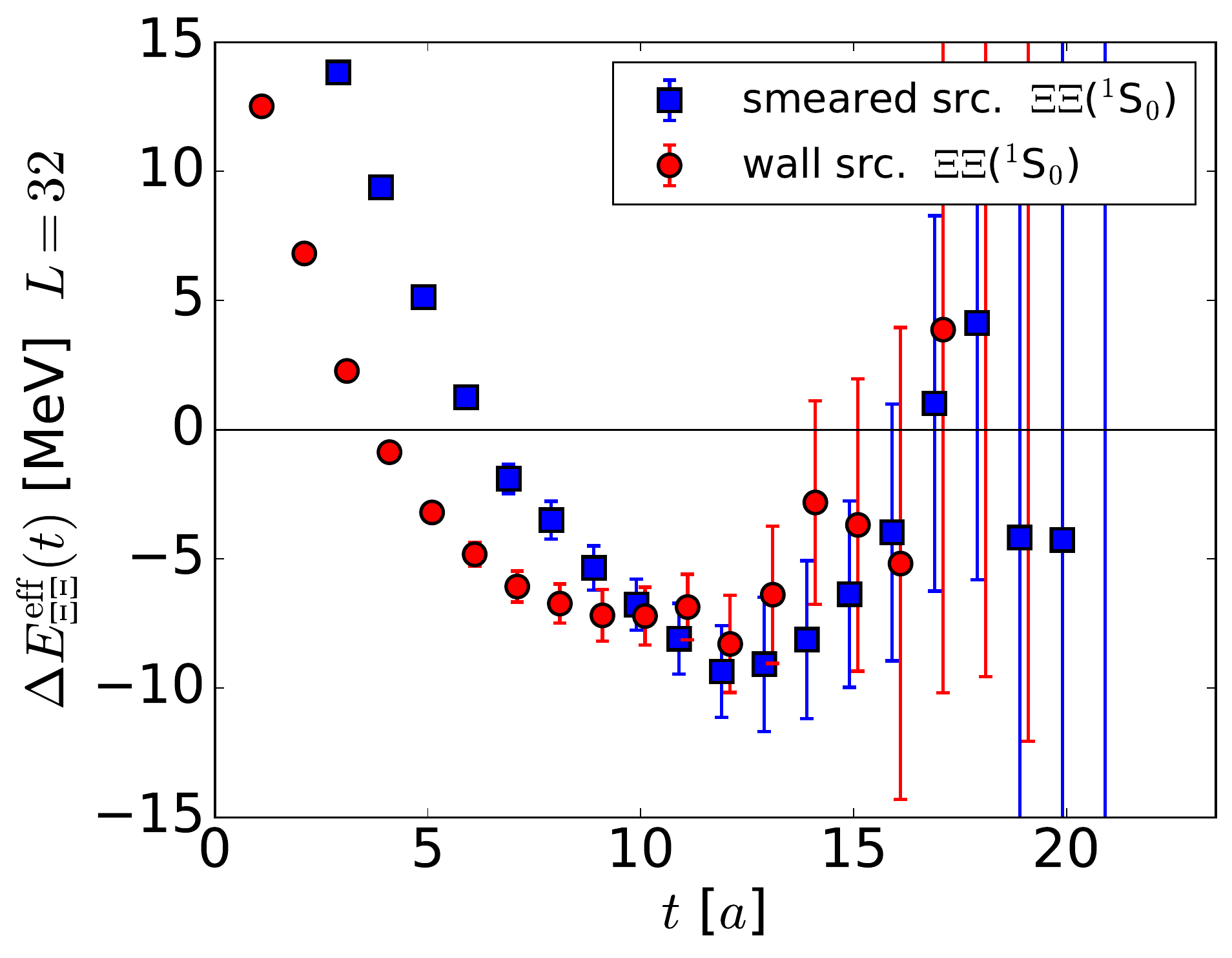}

  \includegraphics[width=0.45\textwidth]{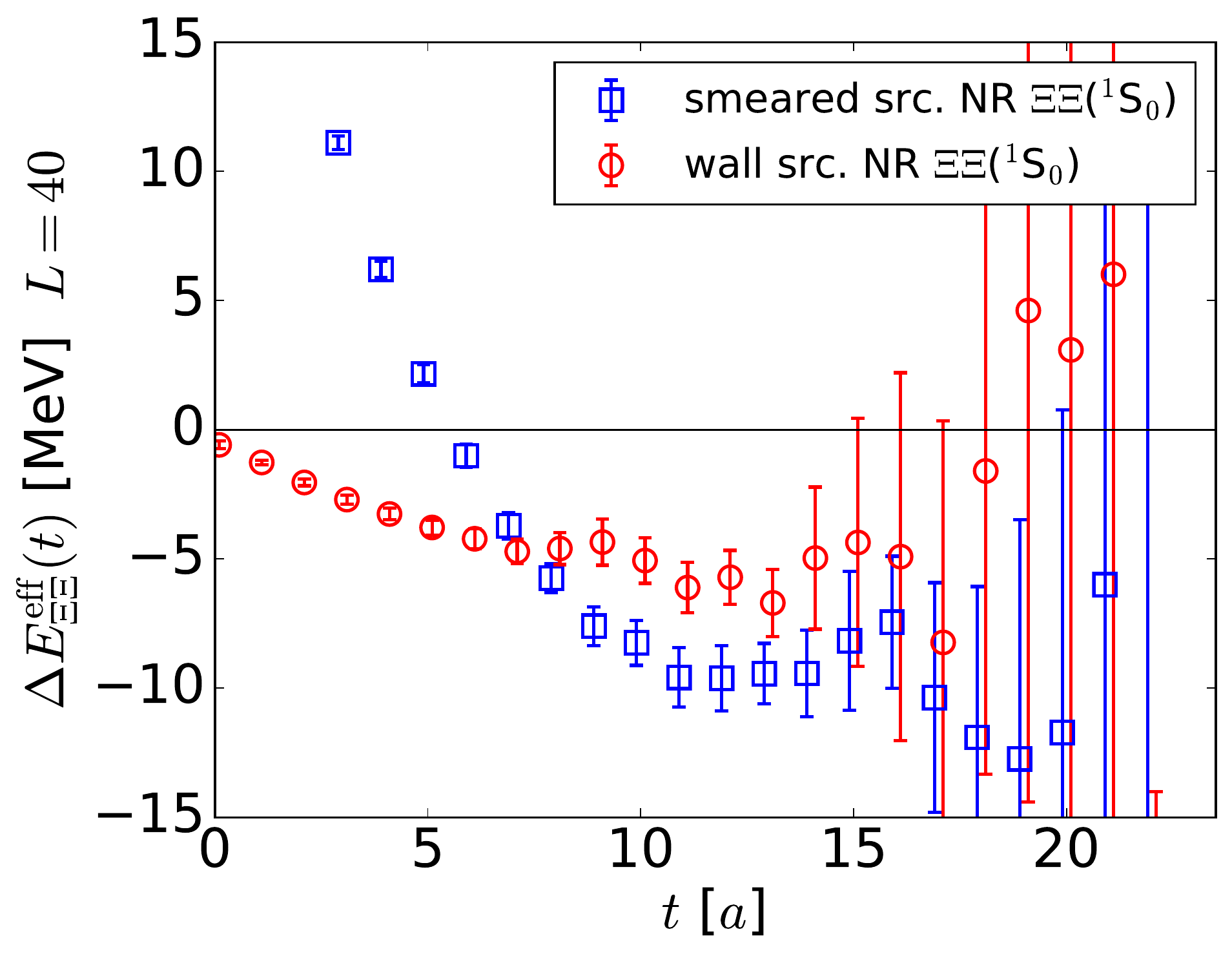}
  \includegraphics[width=0.45\textwidth]{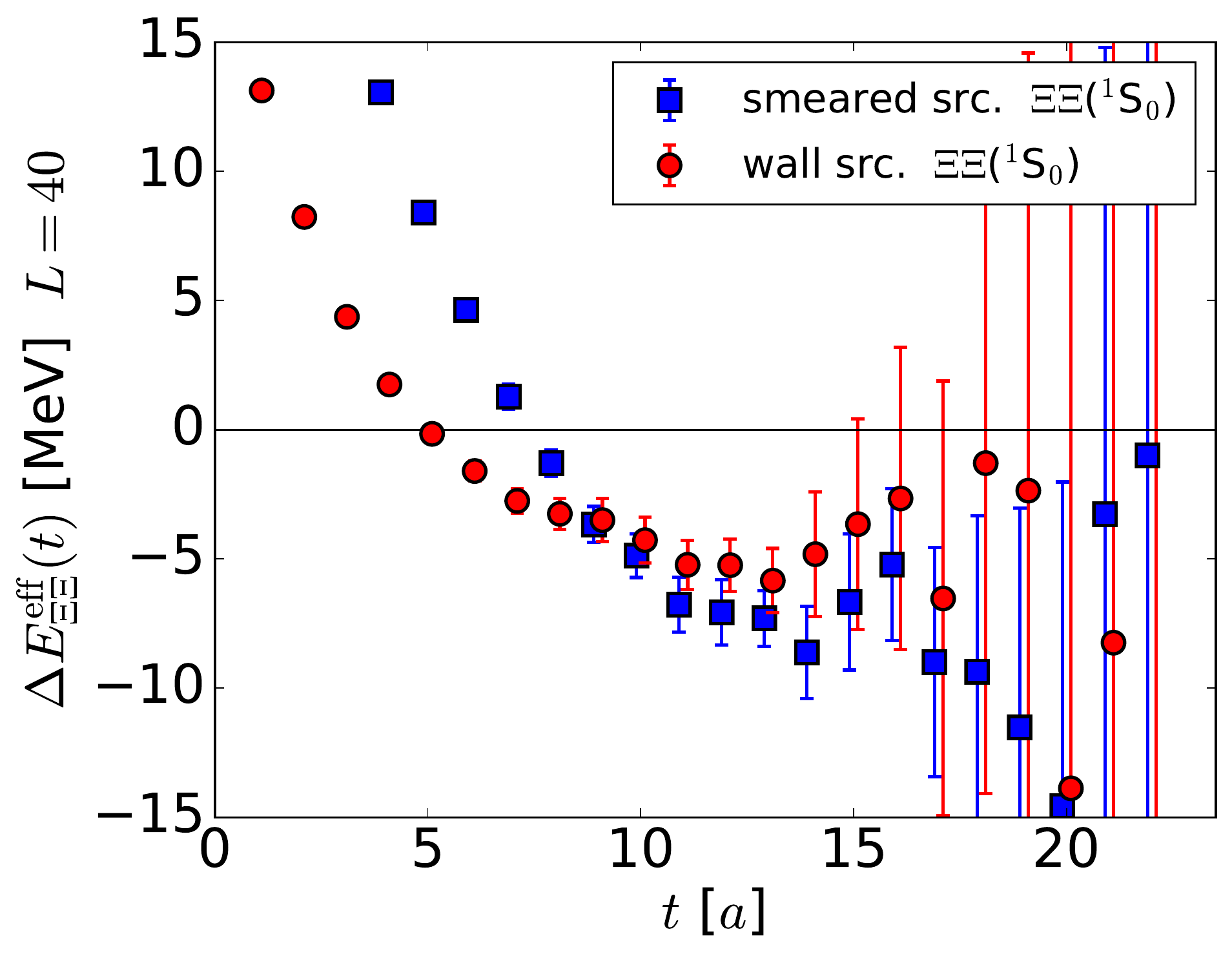}

  \includegraphics[width=0.45\textwidth]{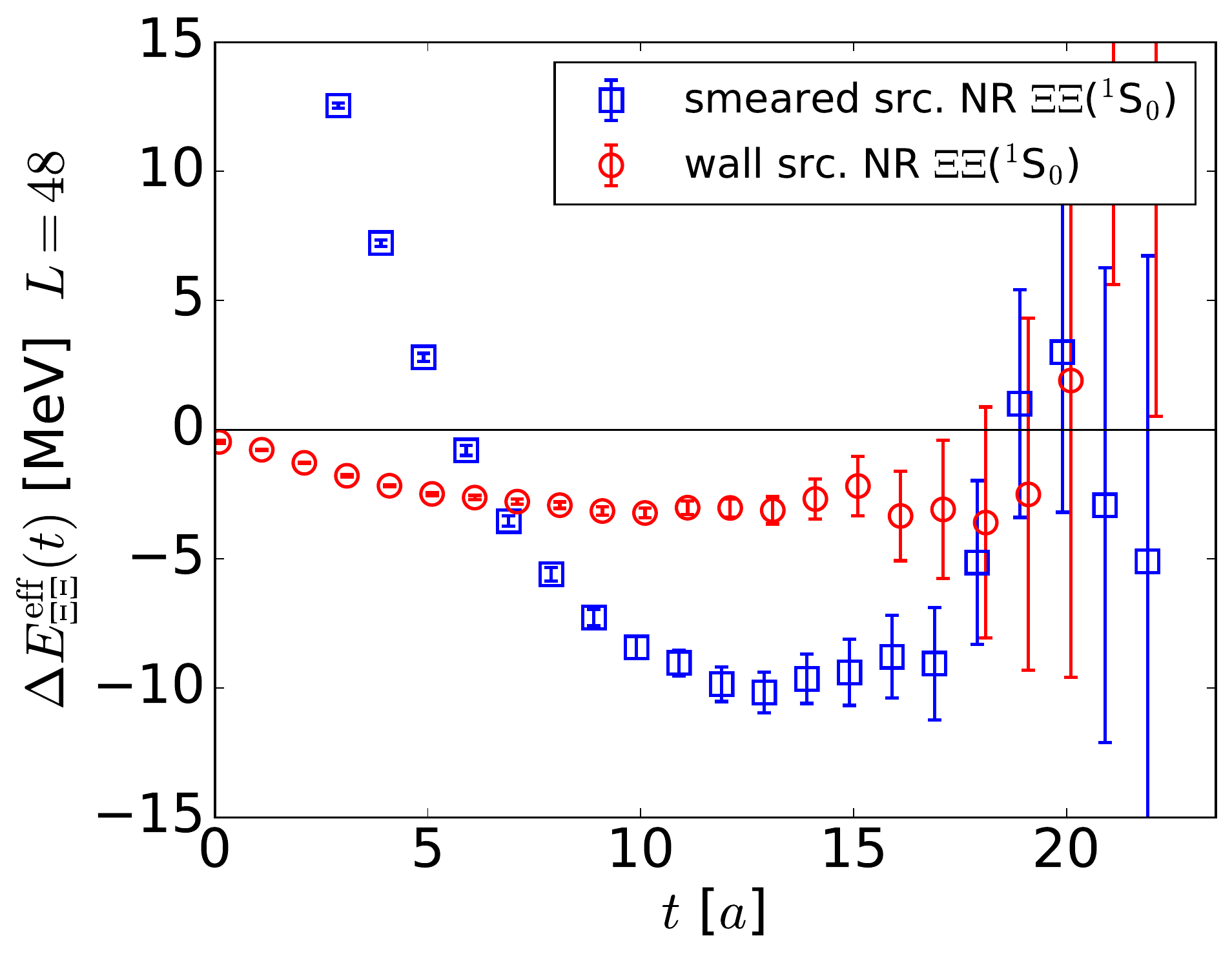}
  \includegraphics[width=0.45\textwidth]{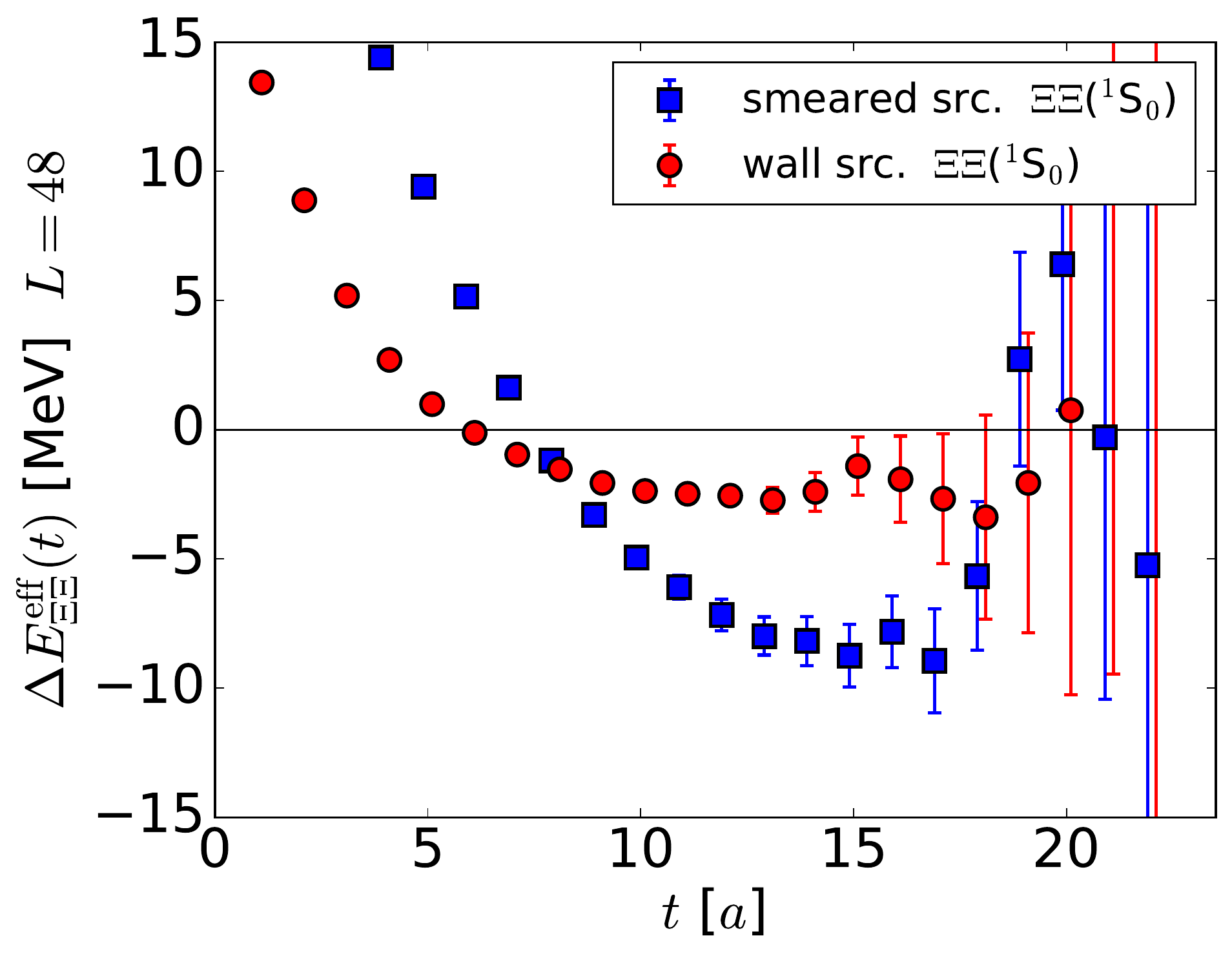}

  \includegraphics[width=0.45\textwidth]{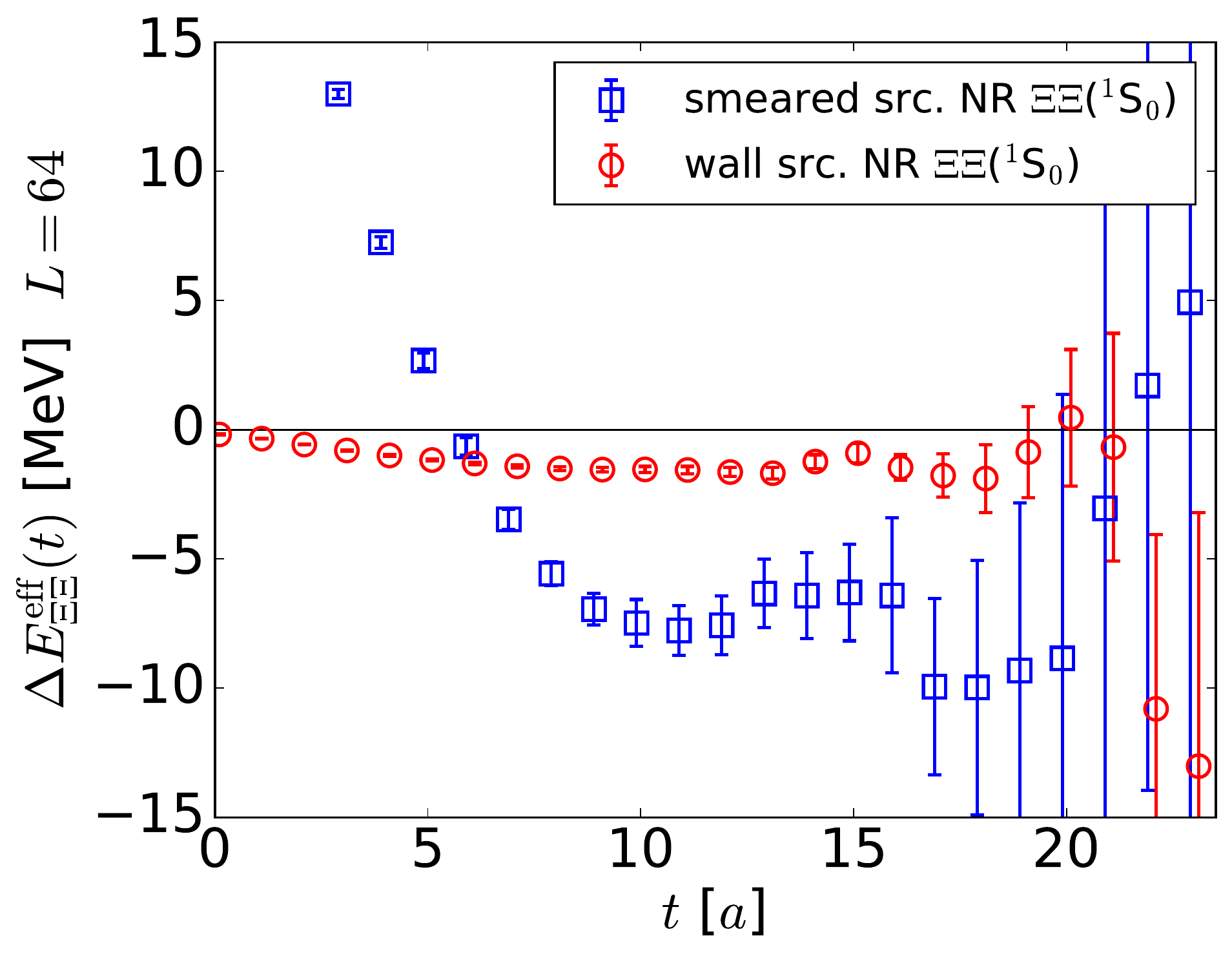}
  \includegraphics[width=0.45\textwidth]{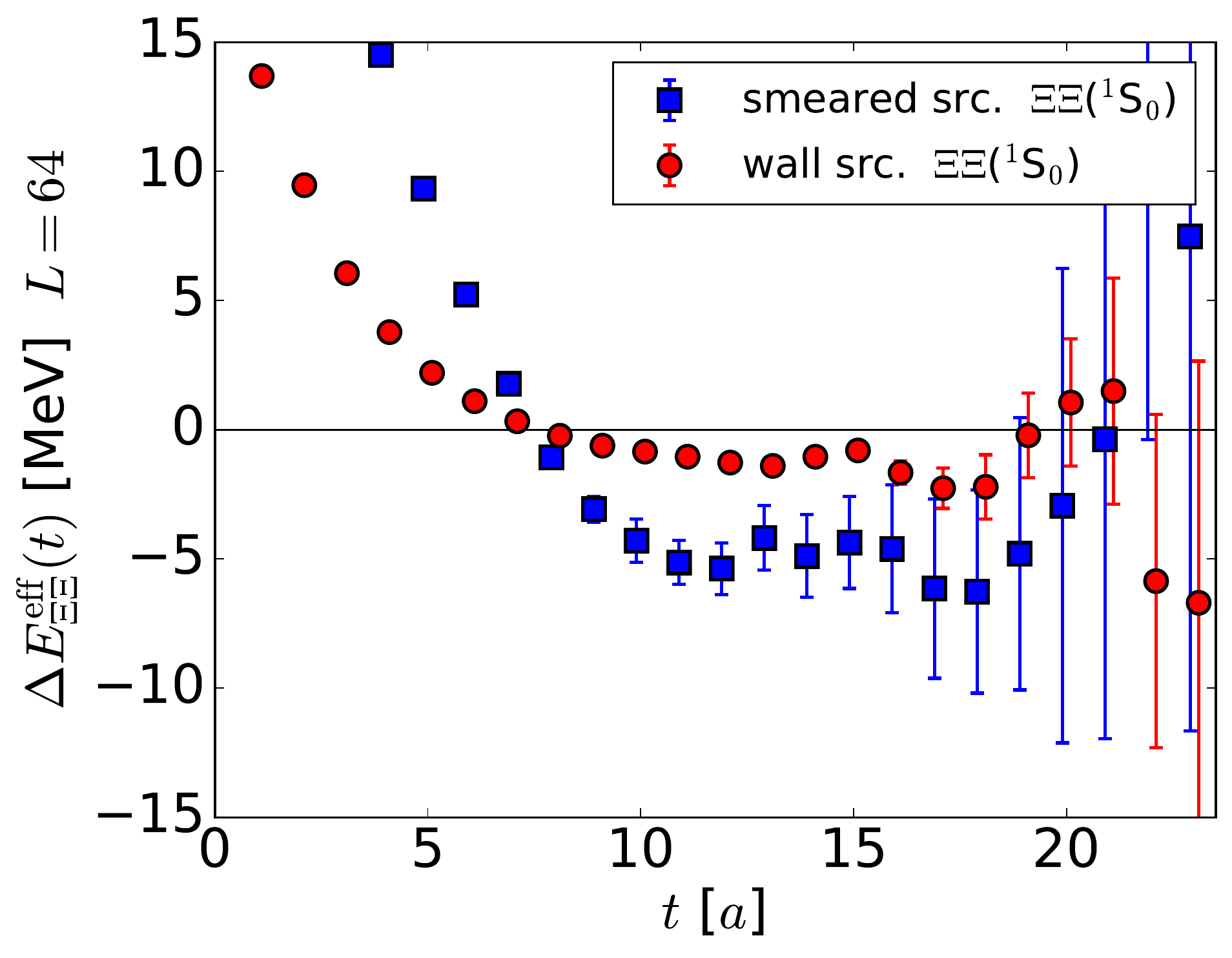}
  \caption{
  The effective energy shift $\DelEeffXiXi(t)$ in the $^1$S$_0$ channel for both smeared and wall sources.
  From the top to bottom, $L^3 = 32^3, 40^3, 48^3, 64^3$.
  (Left)  The results from non-relativistic operators.
  (Right) Those from relativistic operators.
  }
  \label{fig:XiXi1S0}
\end{figure}

\begin{figure}[tbh]
  \centering
  \includegraphics[width=0.45\textwidth]{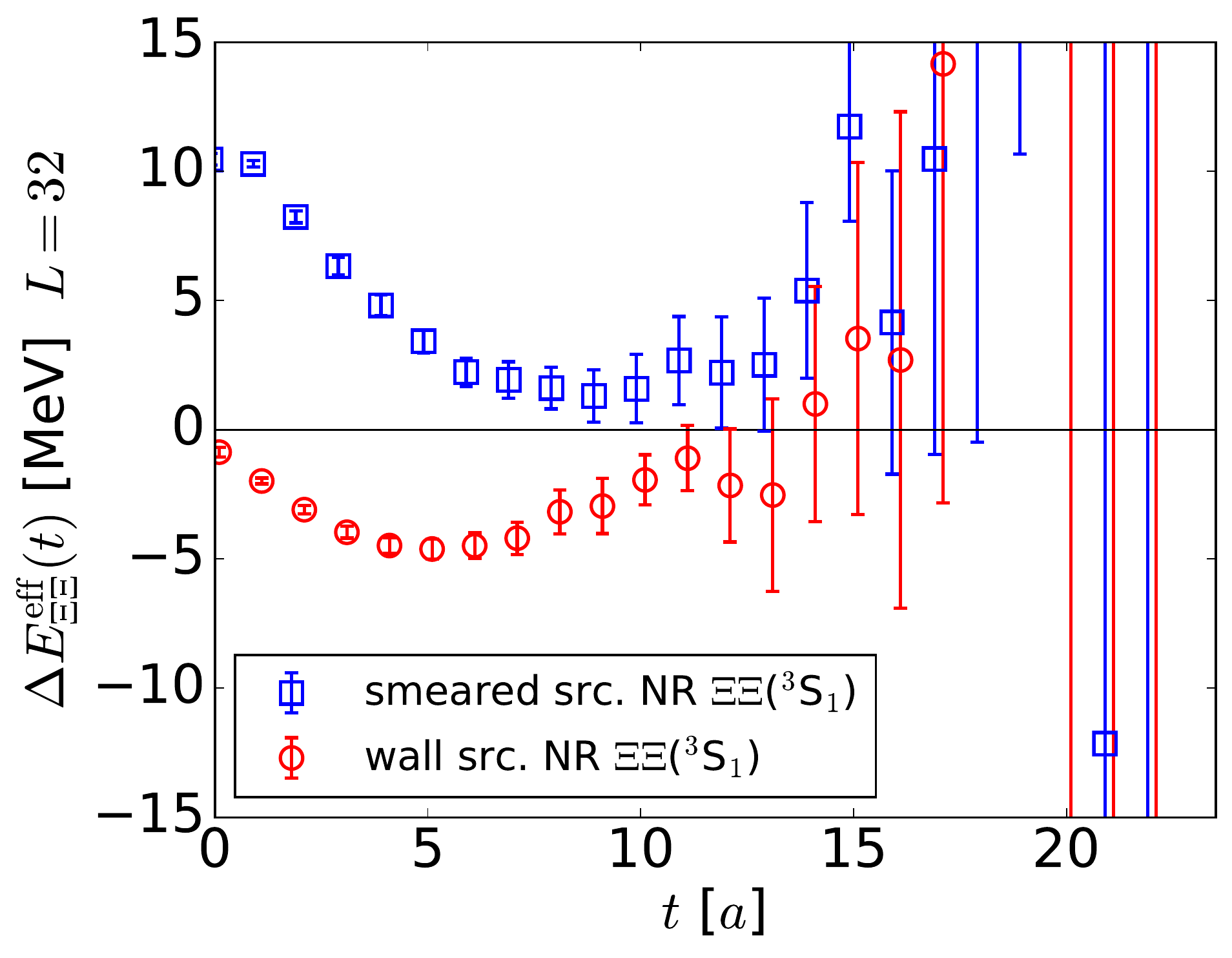}
  \includegraphics[width=0.45\textwidth]{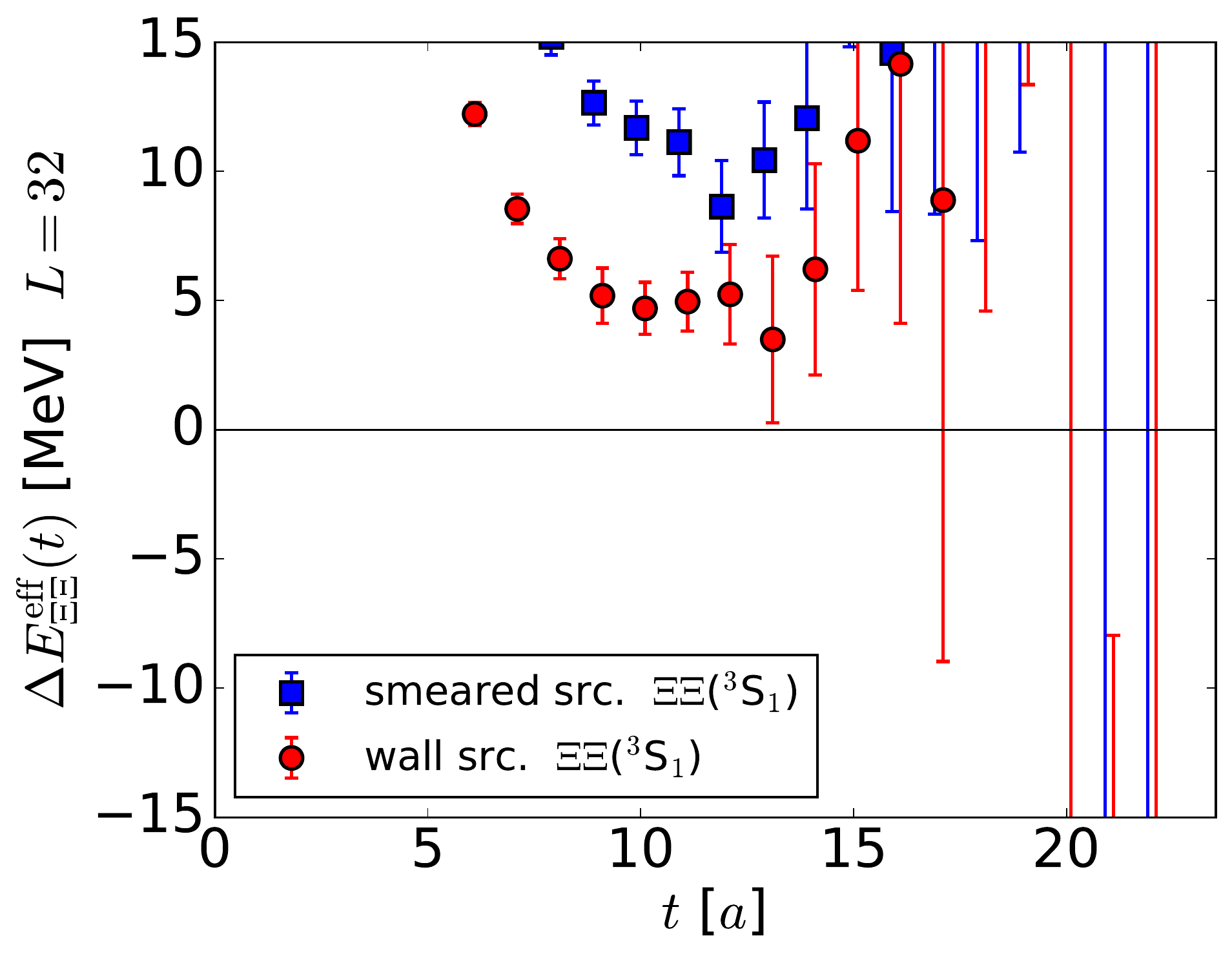}

  \includegraphics[width=0.45\textwidth]{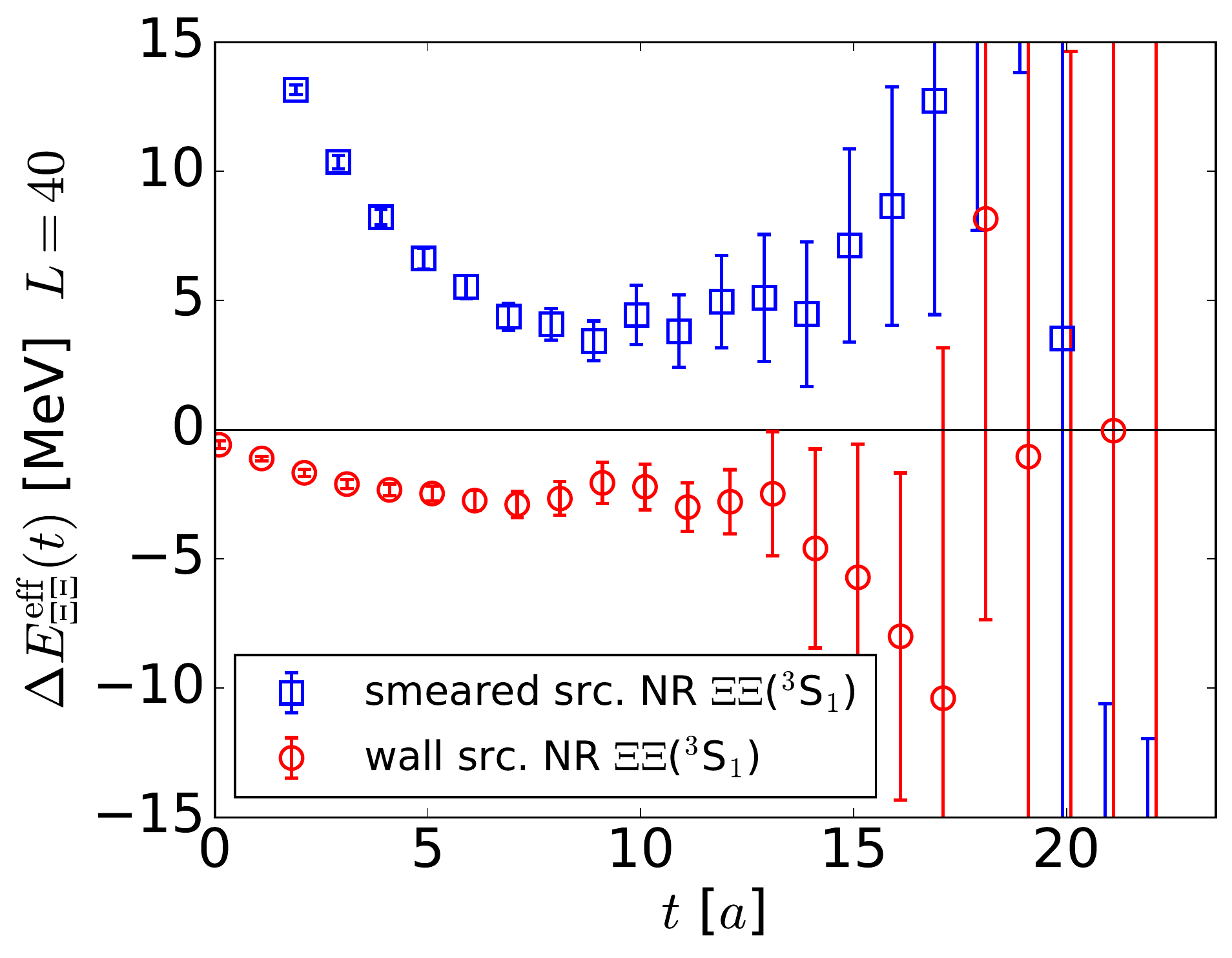}
  \includegraphics[width=0.45\textwidth]{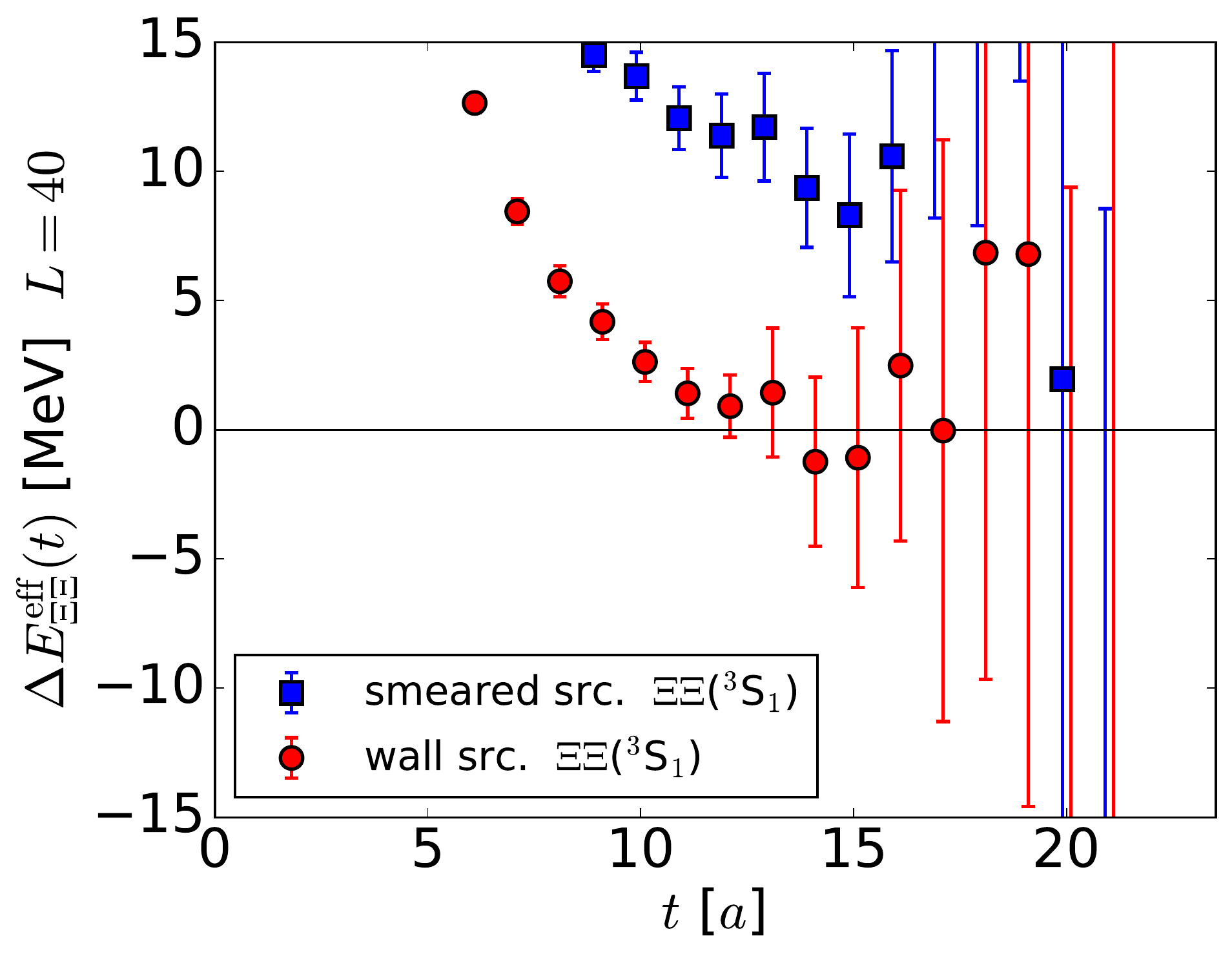}

  \includegraphics[width=0.45\textwidth]{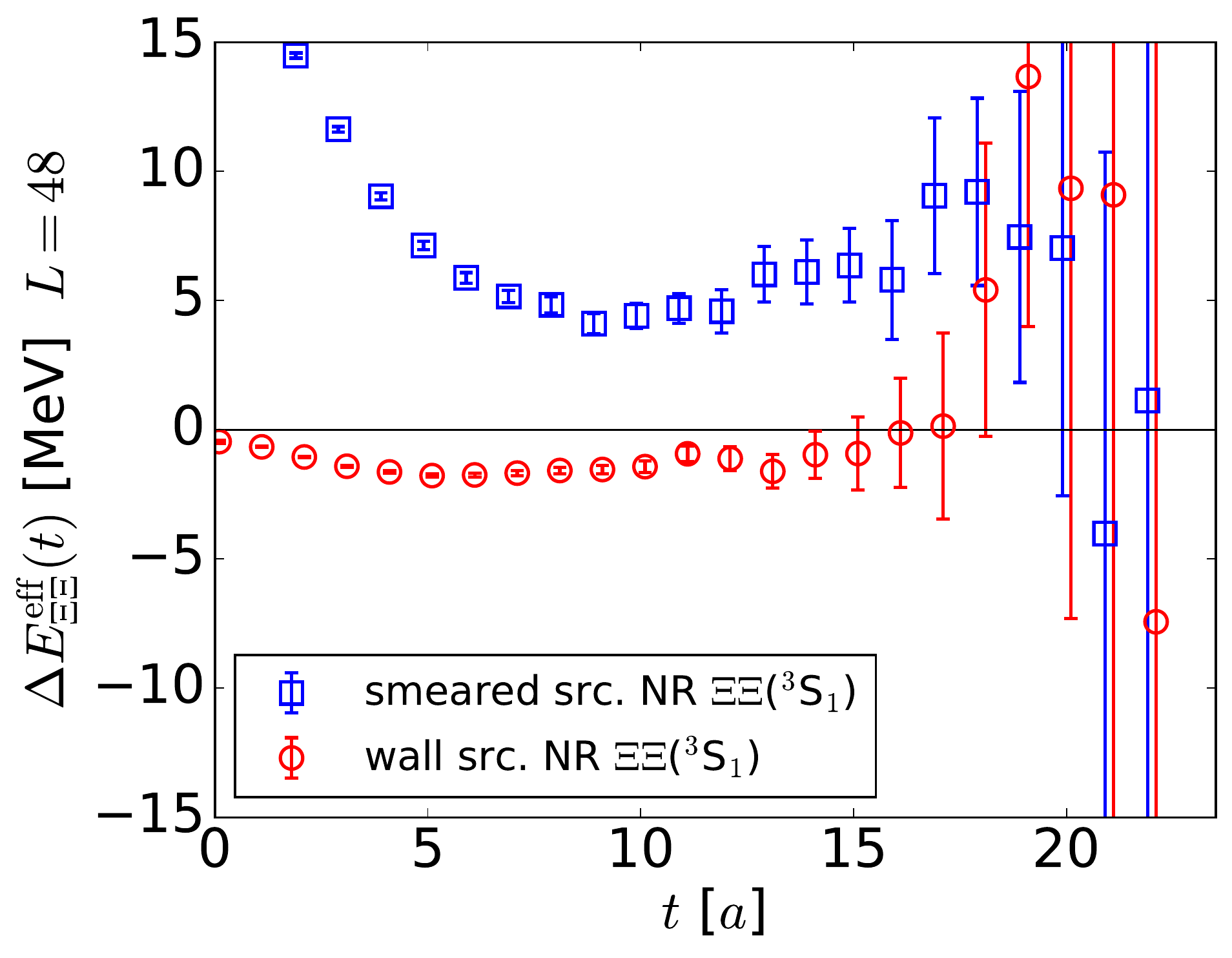}
  \includegraphics[width=0.45\textwidth]{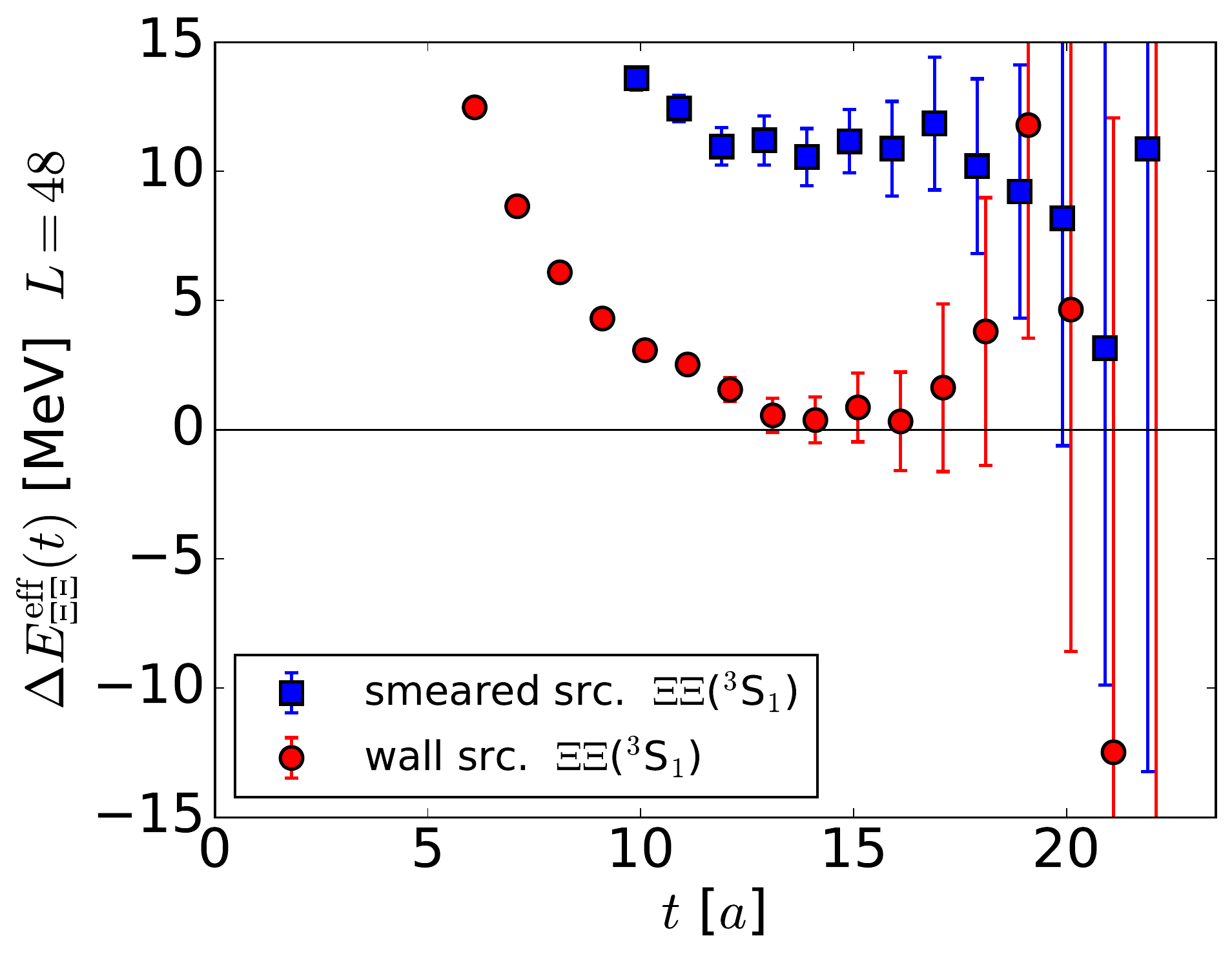}

  \includegraphics[width=0.45\textwidth]{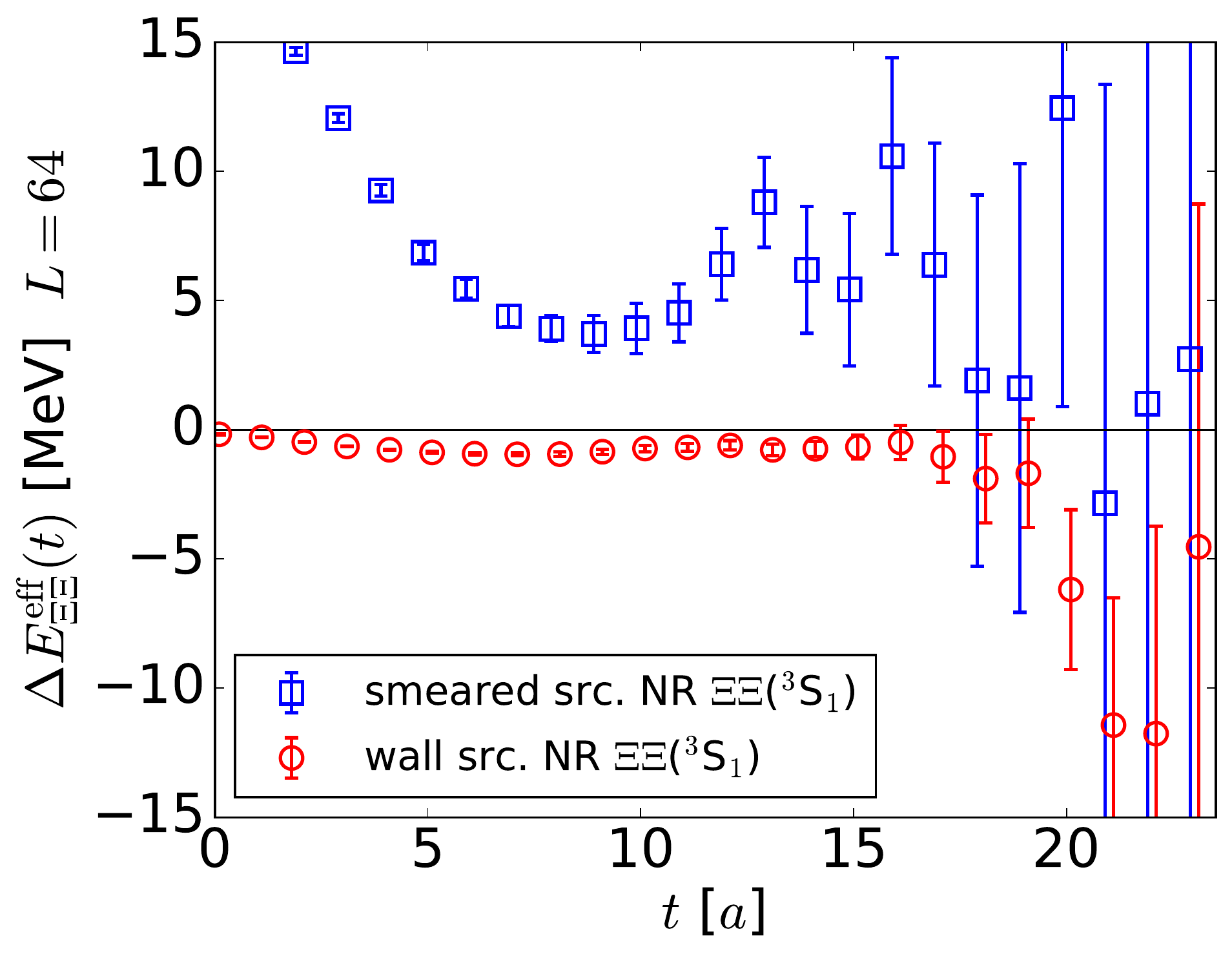}
  \includegraphics[width=0.45\textwidth]{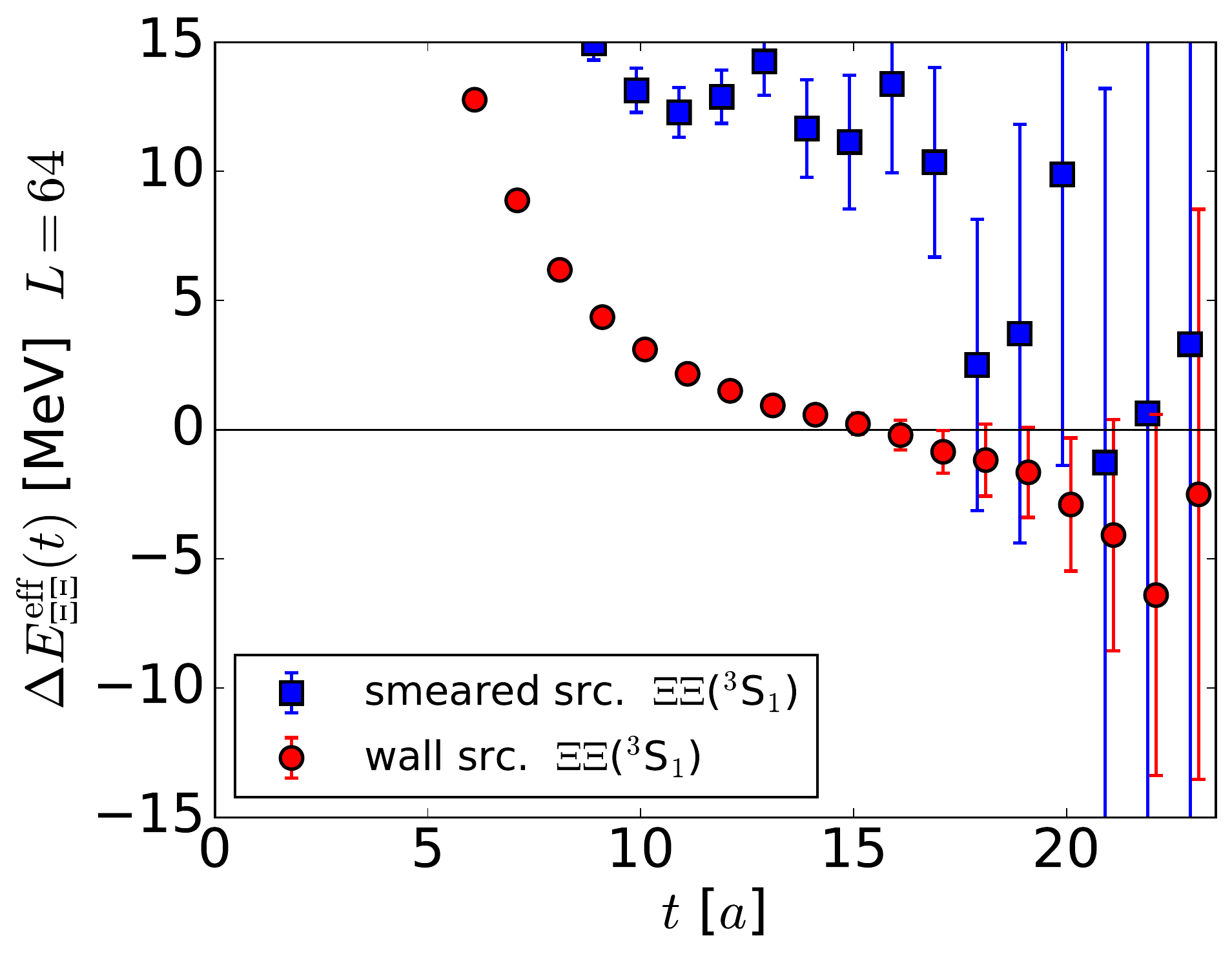}
  \caption{
  The effective energy shift $\DelEeffXiXi(t)$ in the $^3$S$_1$ channel for both smeared and wall sources.
  From the top to bottom, $L^3 = 32^3, 40^3, 48^3, 64^3$.
  (Left)  The results from non-relativistic operators.
  (Right) Those from relativistic operators.
   }
    \label{fig:XiXi3S1}
\end{figure}

\begin{figure}[tbh]
  \centering
  \includegraphics[width=0.45\textwidth]{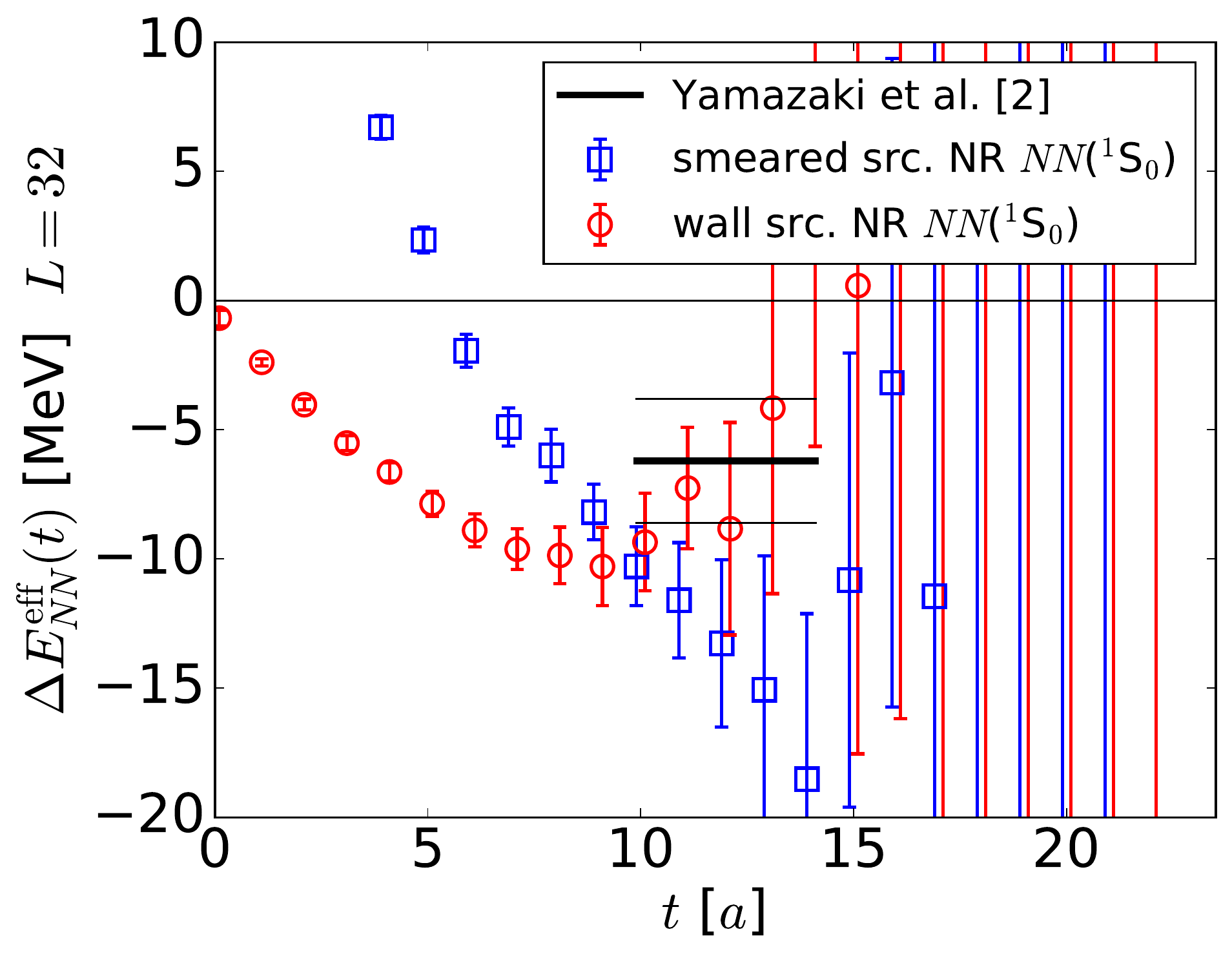}
  \includegraphics[width=0.45\textwidth]{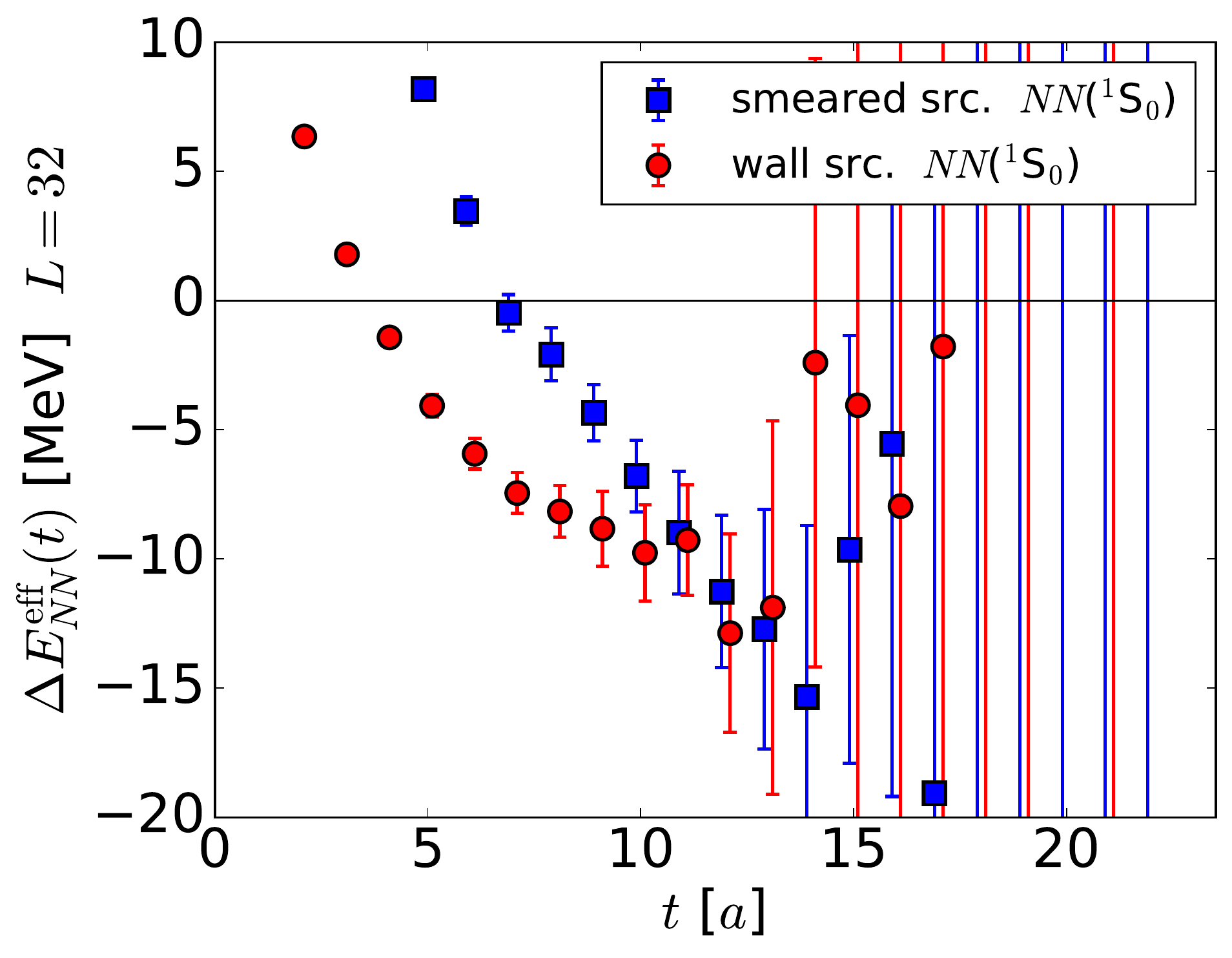}

  \includegraphics[width=0.45\textwidth]{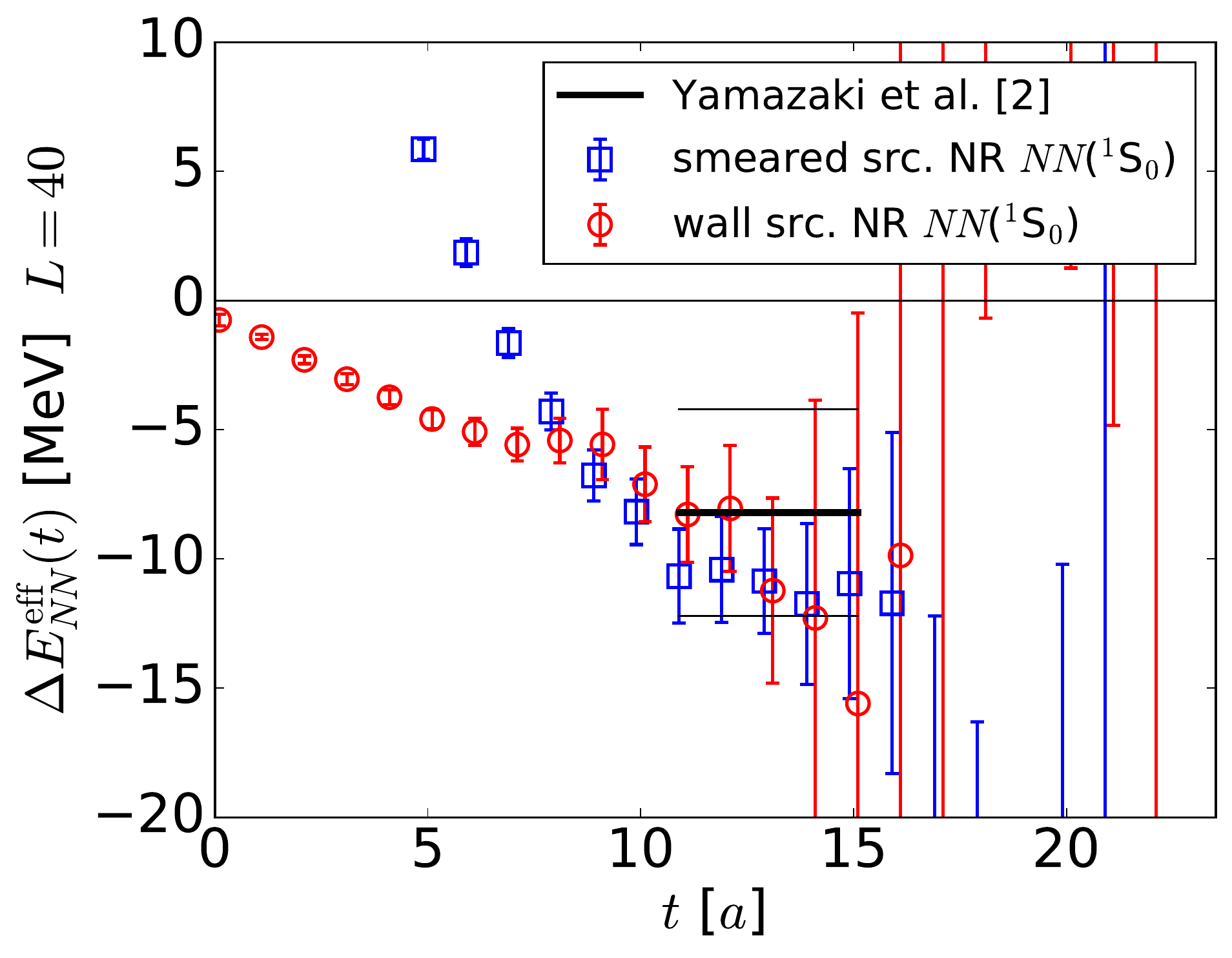}
  \includegraphics[width=0.45\textwidth]{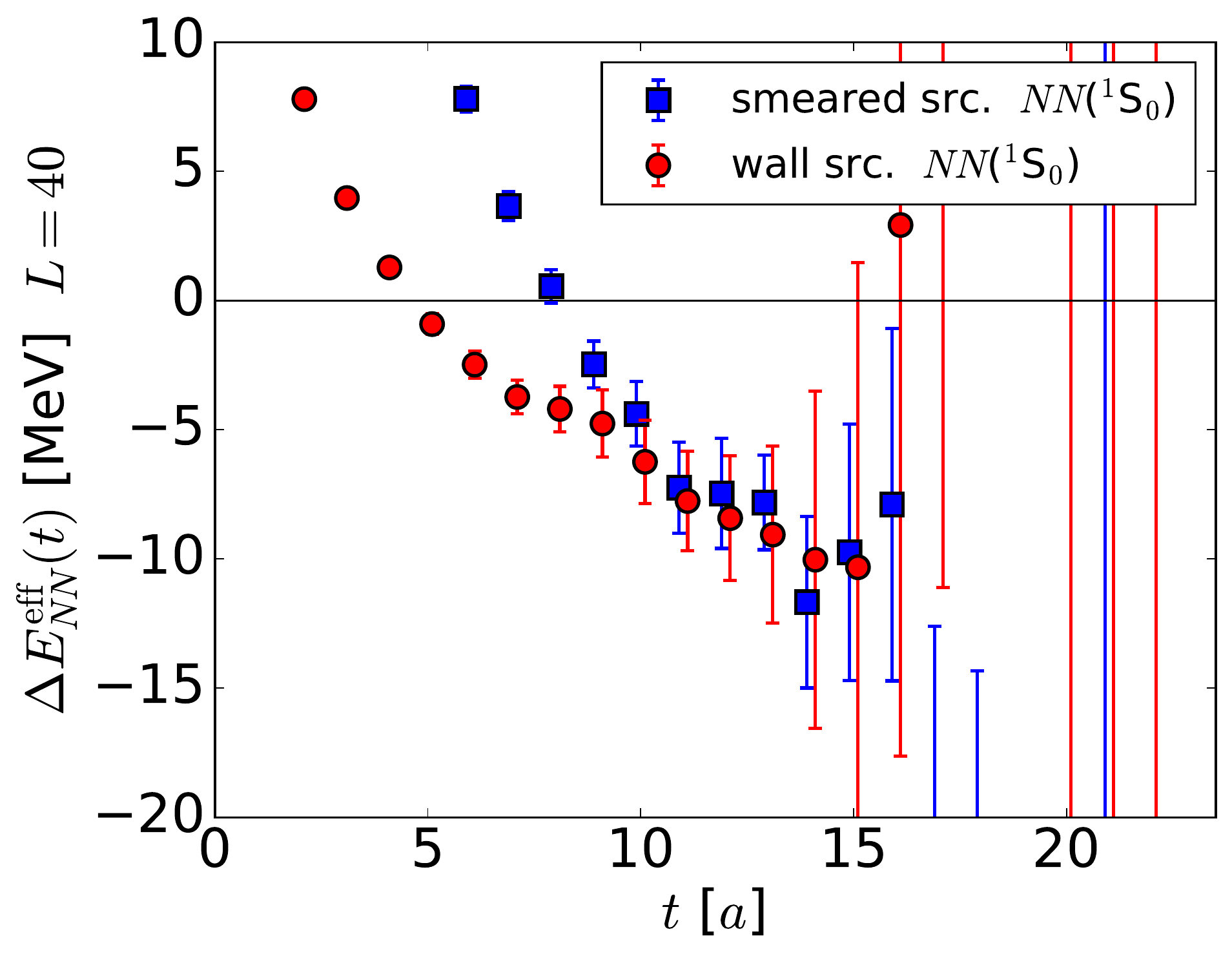}

  \includegraphics[width=0.45\textwidth]{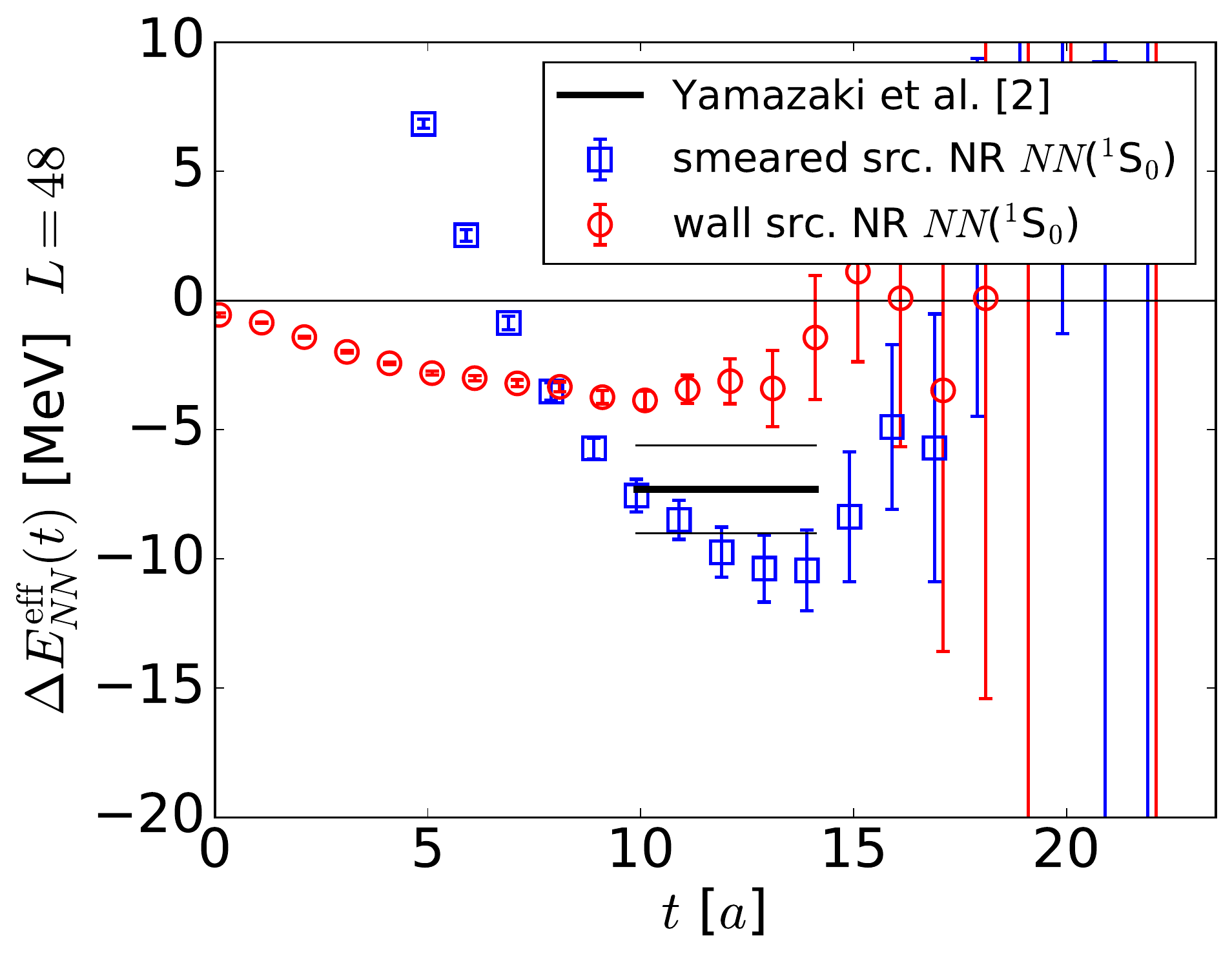}
  \includegraphics[width=0.45\textwidth]{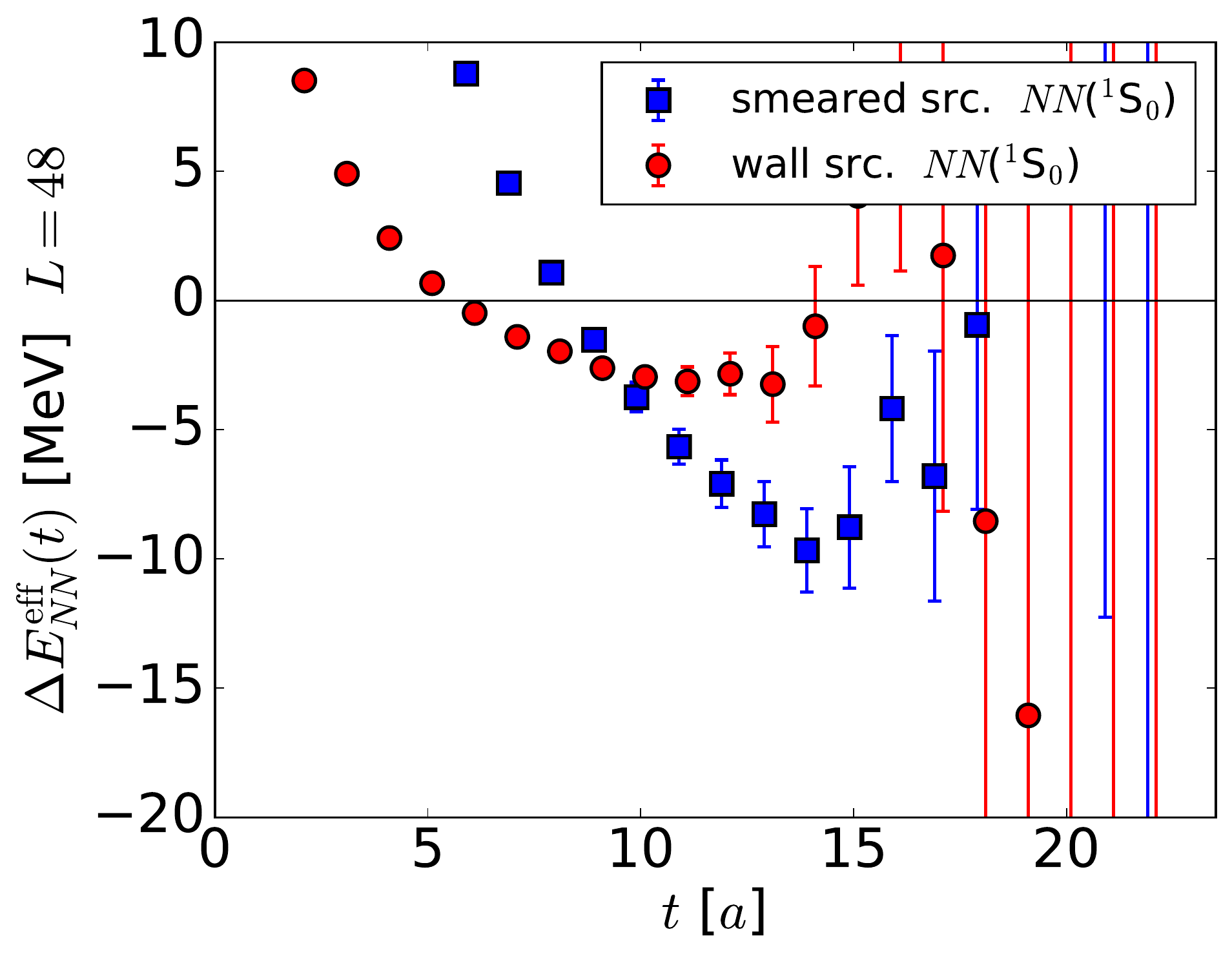}

  \includegraphics[width=0.45\textwidth]{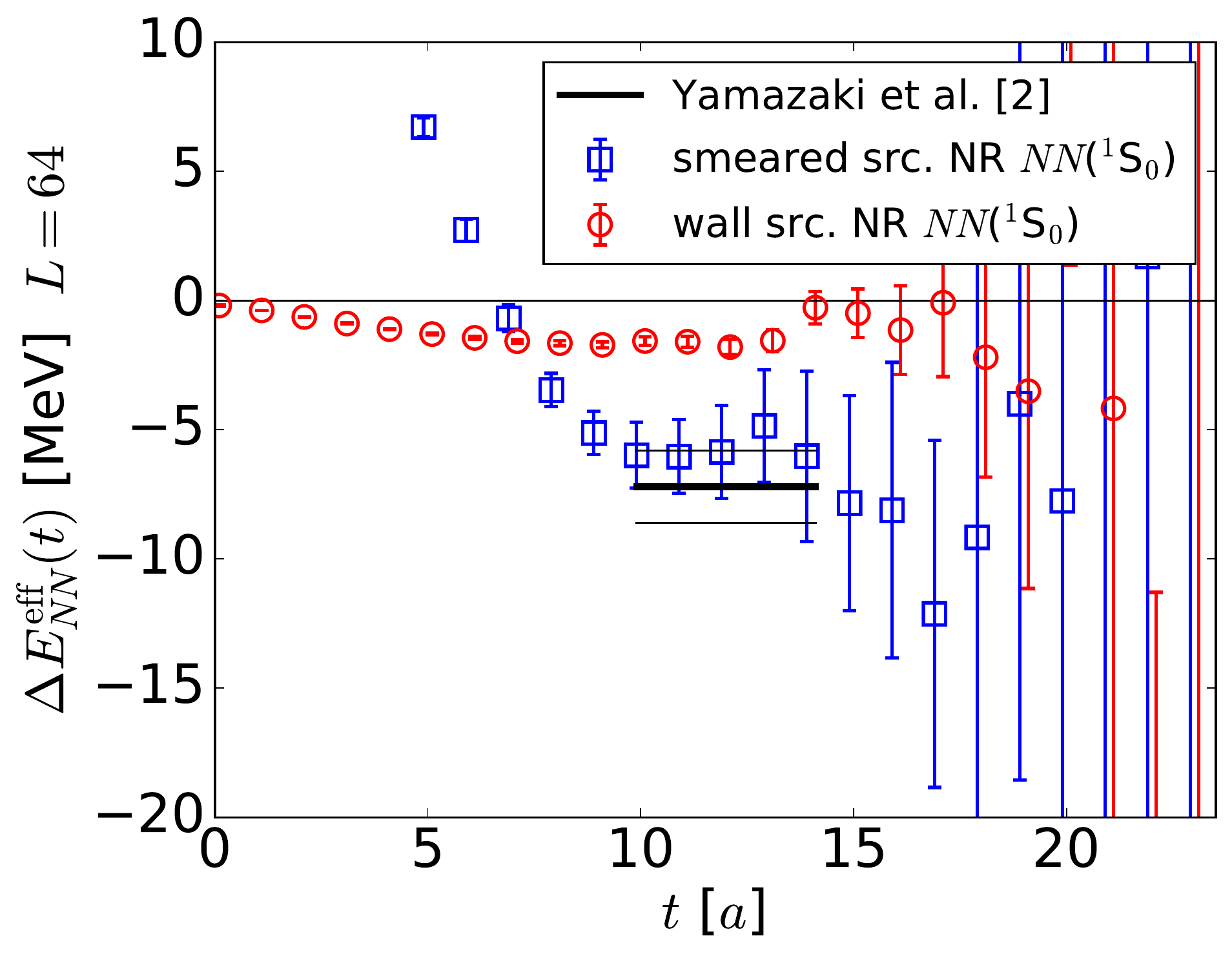}
  \includegraphics[width=0.45\textwidth]{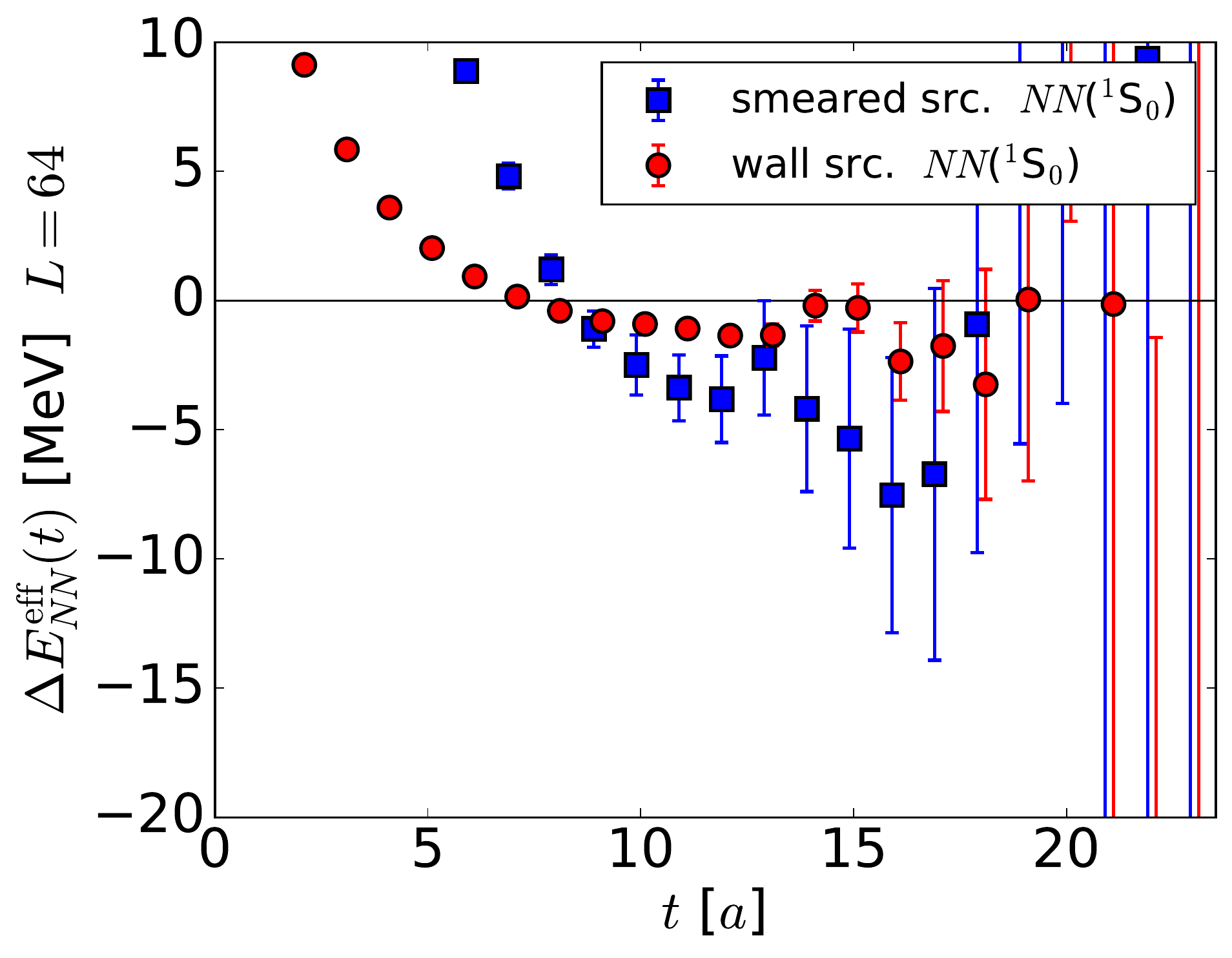}
  \caption{
  The effective energy shift $\DelEeffNN(t)$ in the $^1$S$_0$ channel for both smeared and wall sources.
  From the top to bottom, $L^3 = 32^3, 40^3, 48^3, 64^3$.
  (Left)  The results from non-relativistic operators. 
  The plateaux of Ref.~\cite{Yamazaki:2012hi} are also shown by black lines 
  (central value and 1$\sigma$ statistical errors) for comparison.
  (Right) The results from relativistic operators.
  }
   \label{fig:NN1S0}
\end{figure}

\begin{figure}[tbh]
  \centering
  \includegraphics[width=0.45\textwidth]{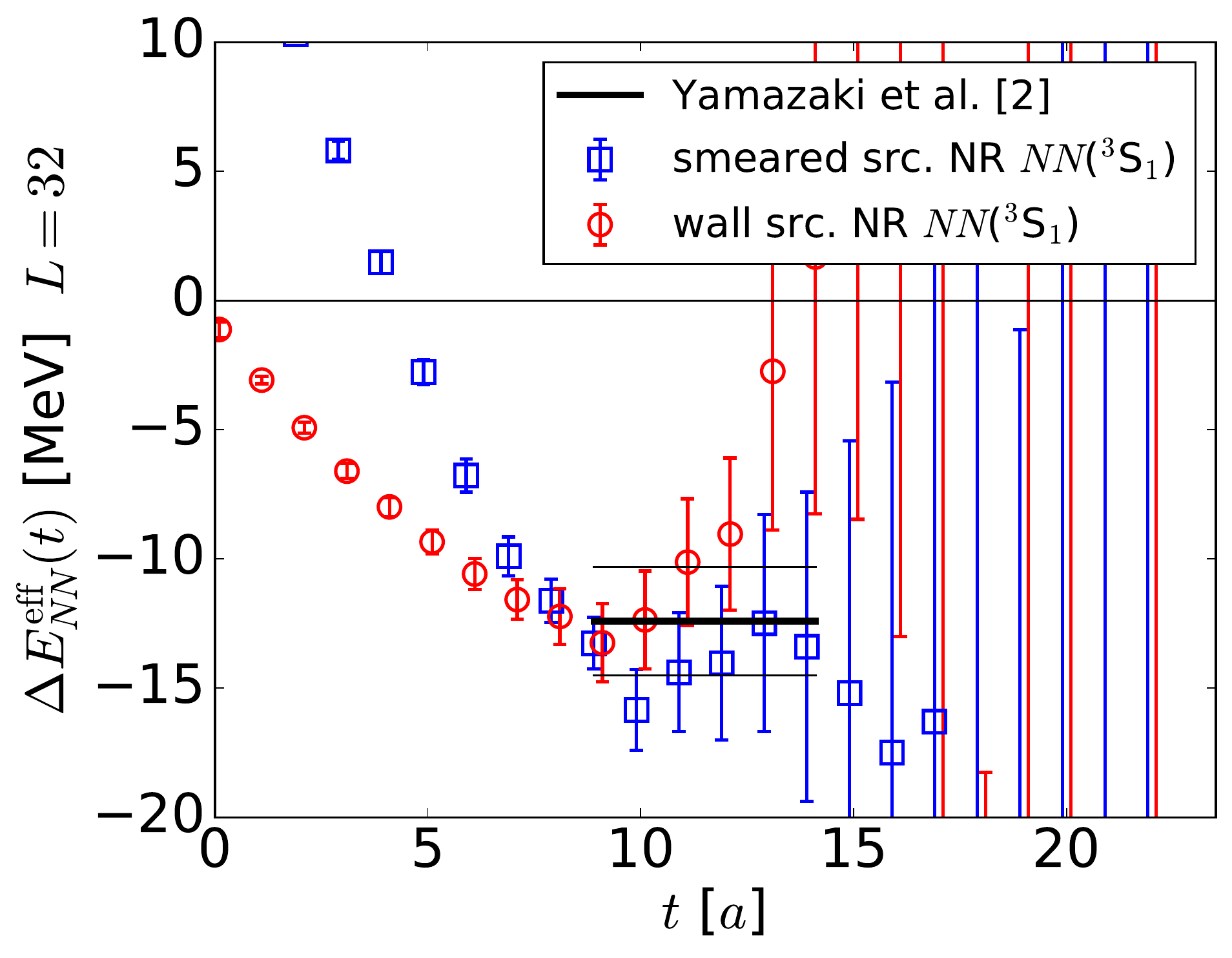}
  \includegraphics[width=0.45\textwidth]{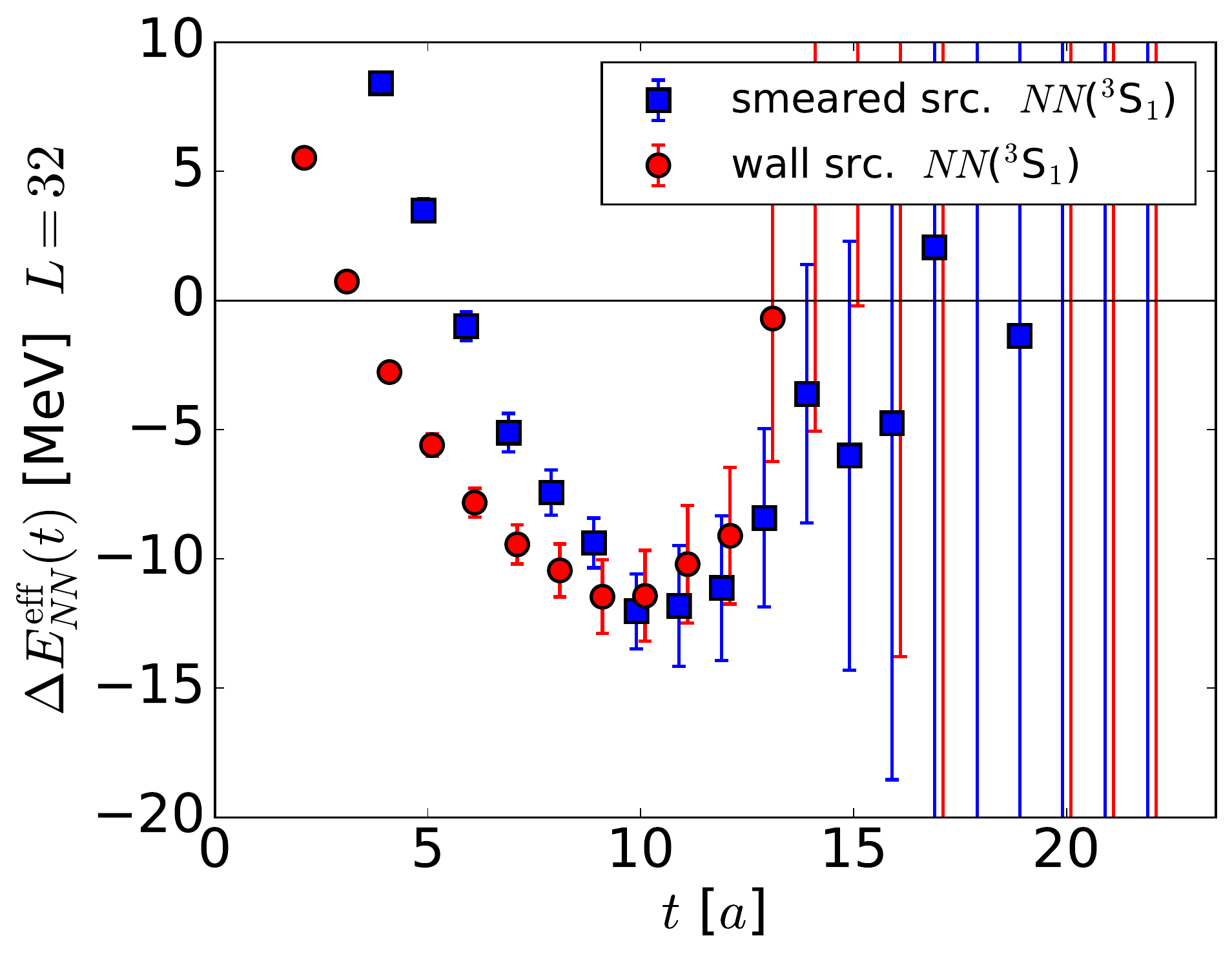}

  \includegraphics[width=0.45\textwidth]{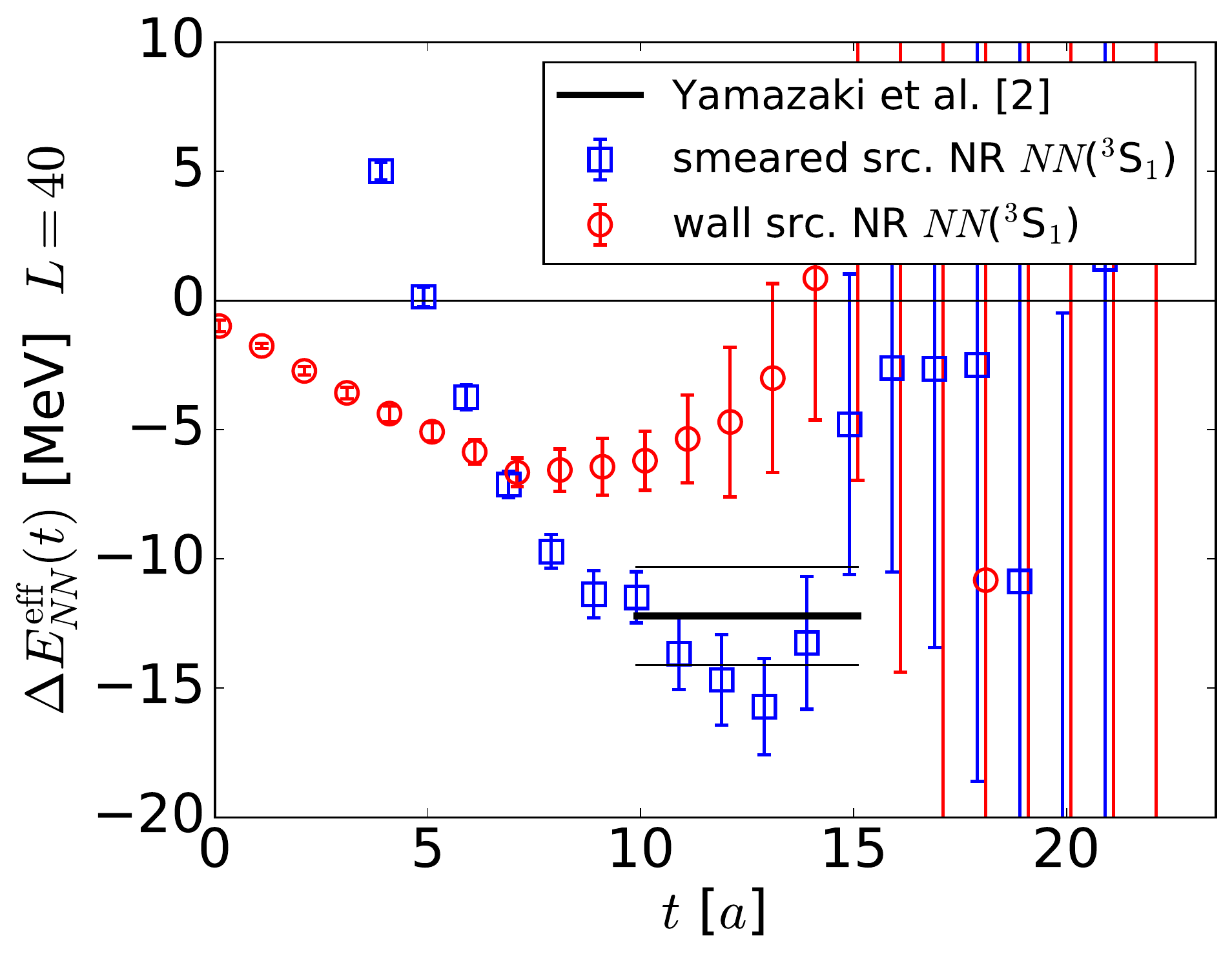}
  \includegraphics[width=0.45\textwidth]{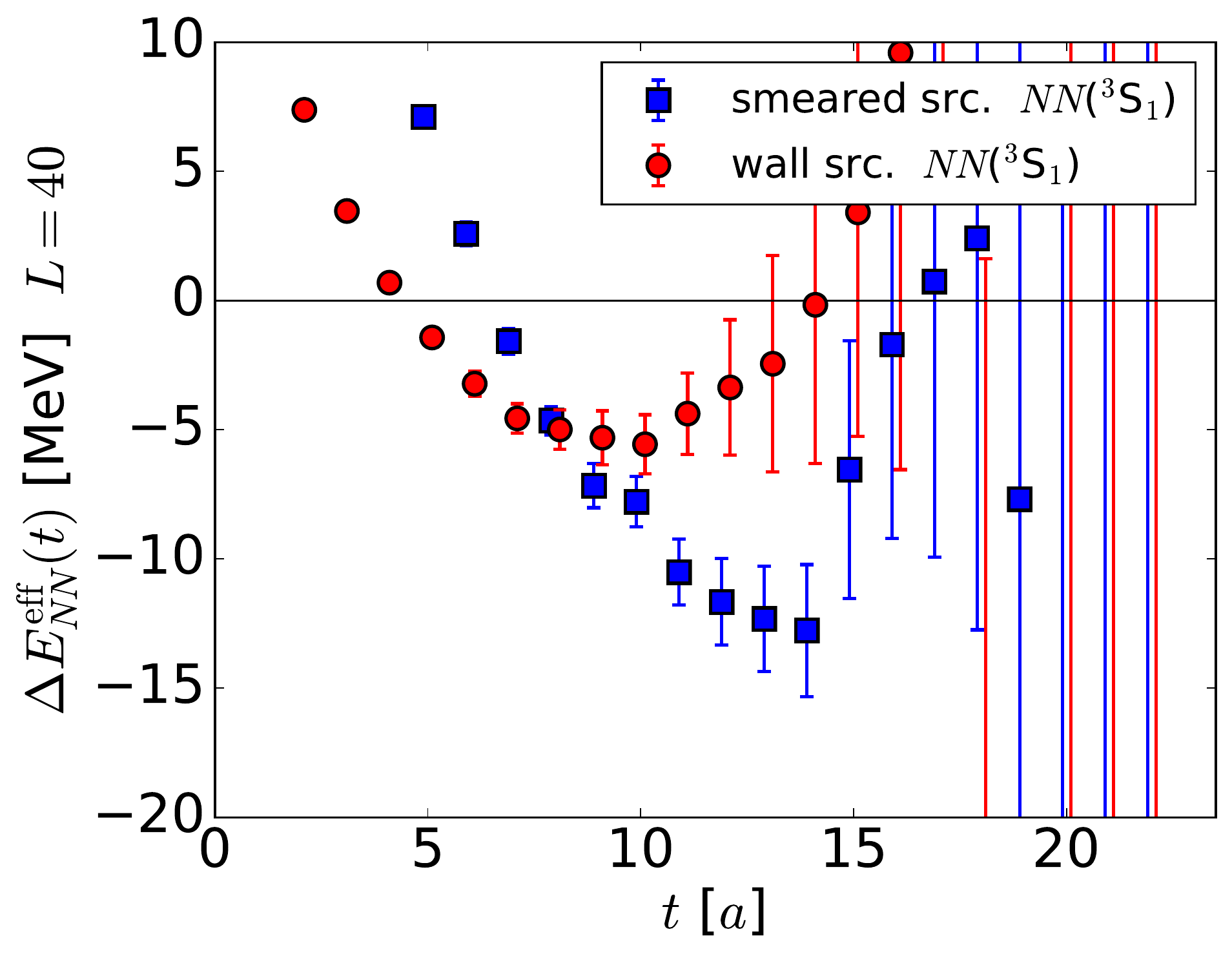}

  \includegraphics[width=0.45\textwidth]{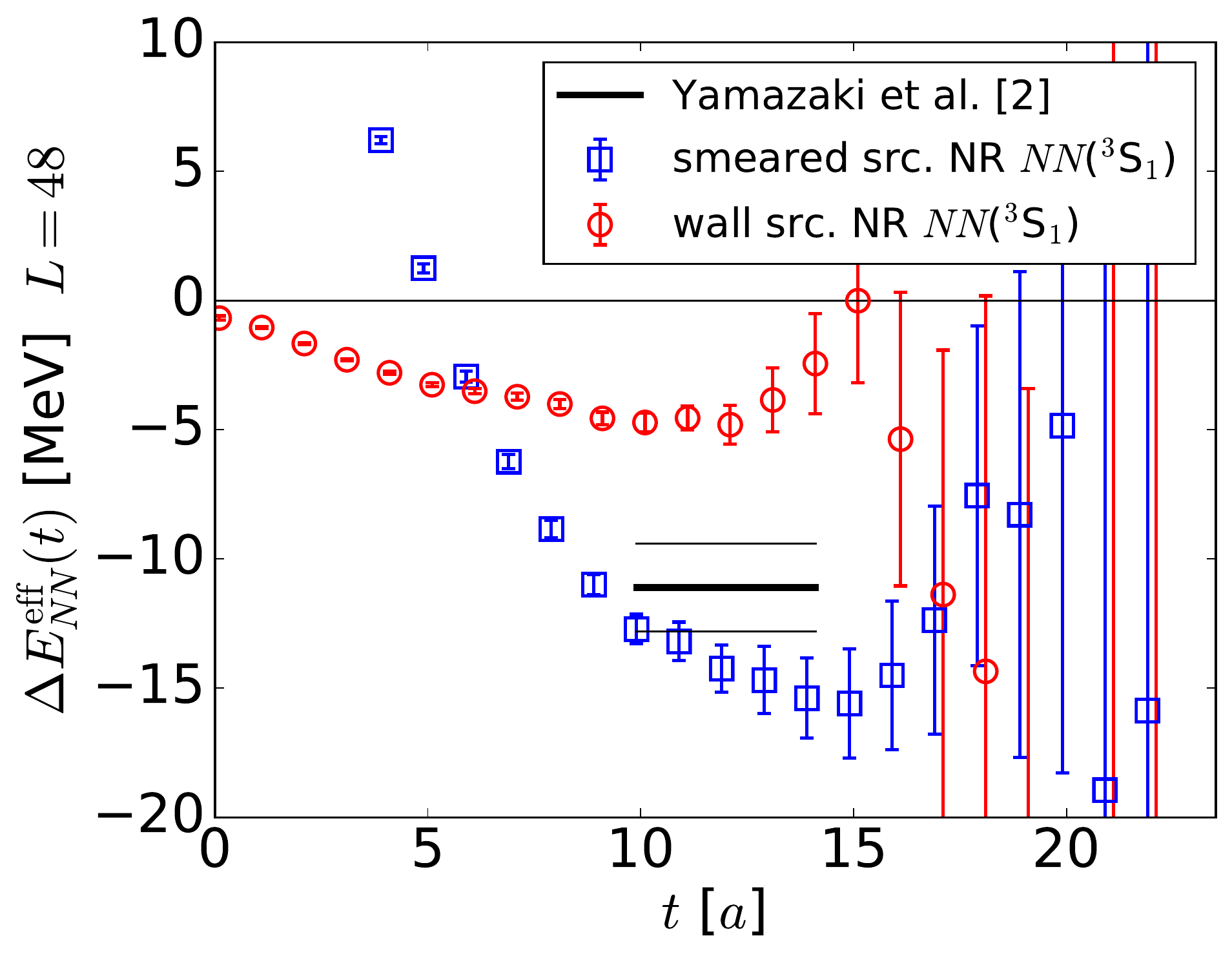}
  \includegraphics[width=0.45\textwidth]{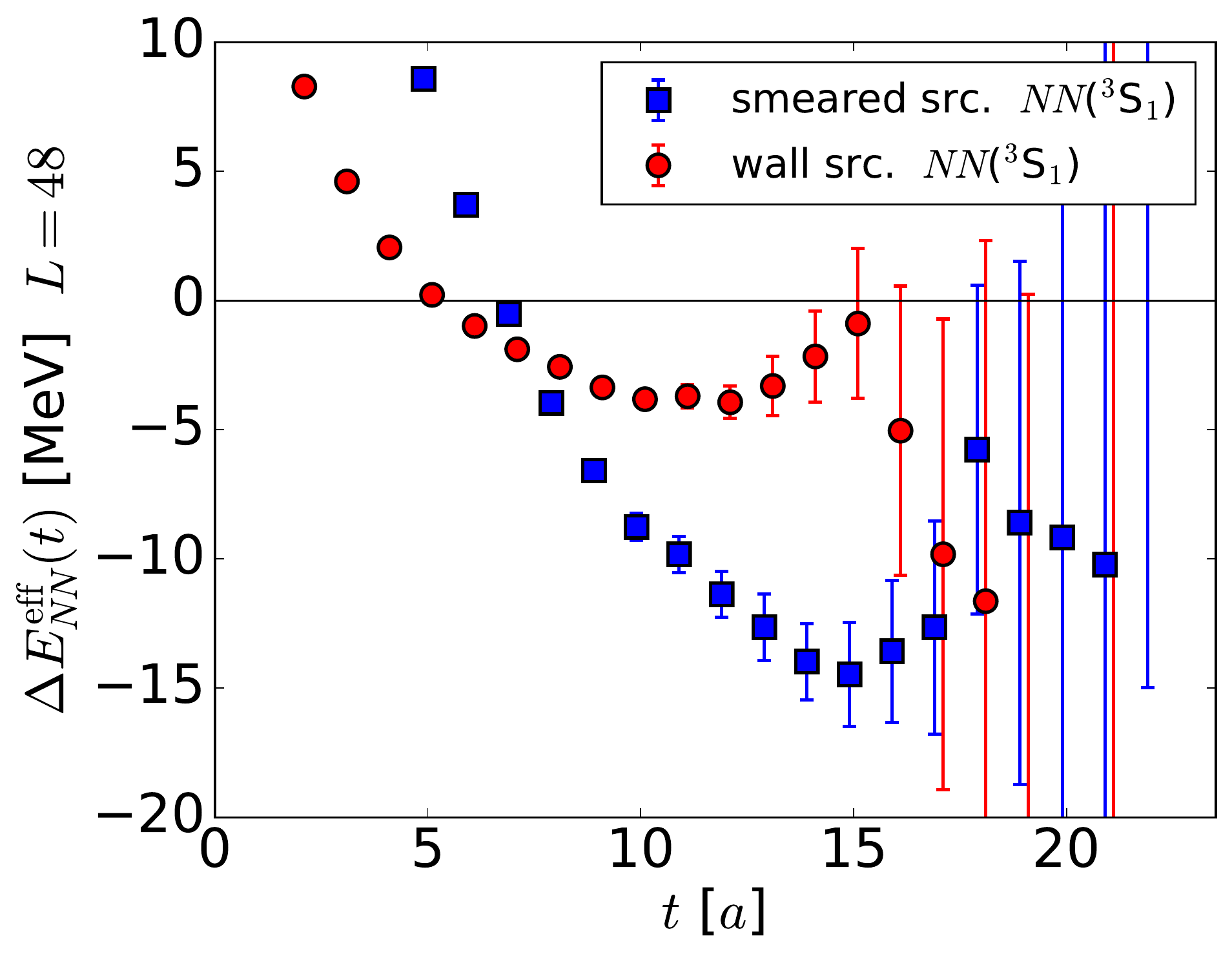}

  \includegraphics[width=0.45\textwidth]{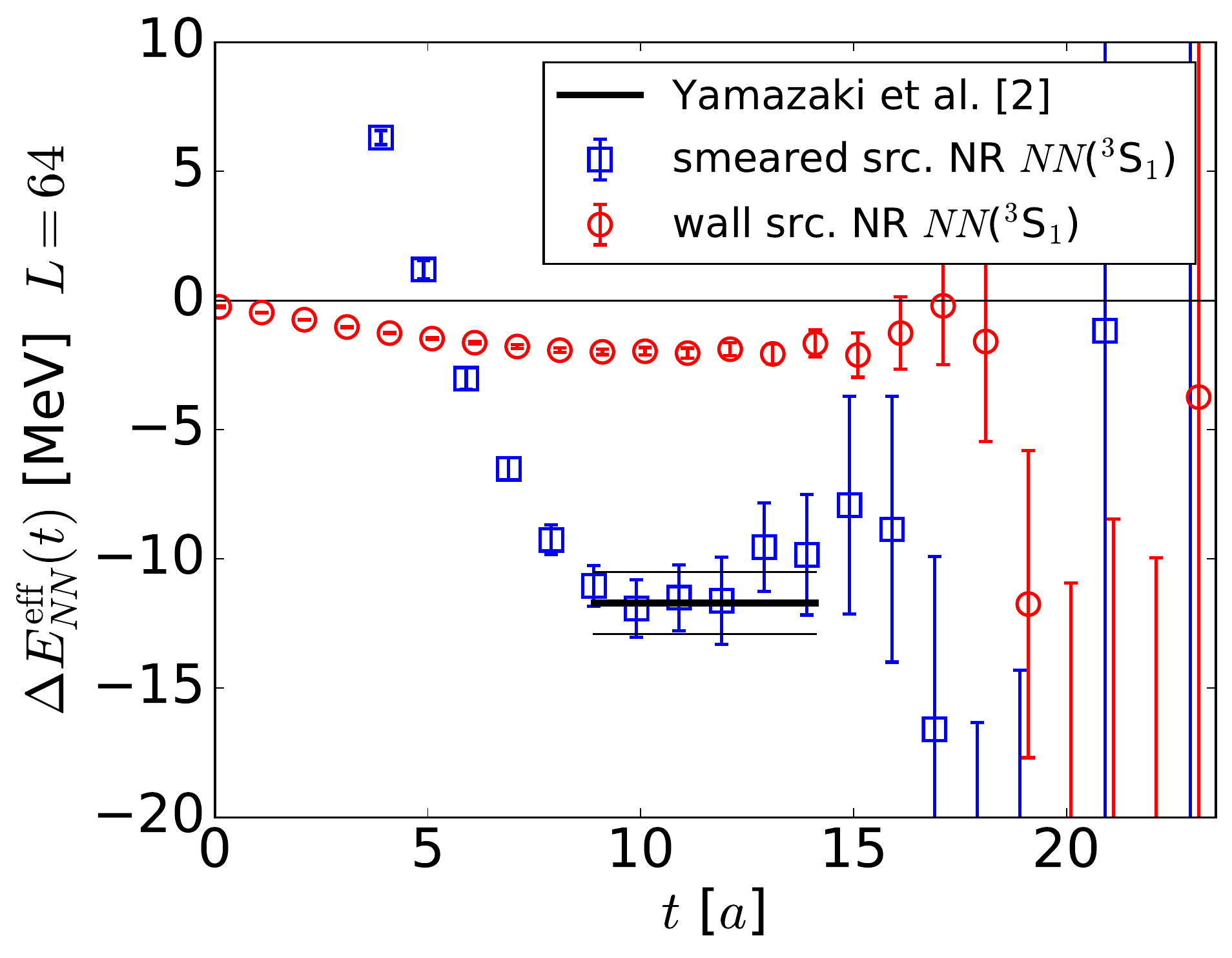}
  \includegraphics[width=0.45\textwidth]{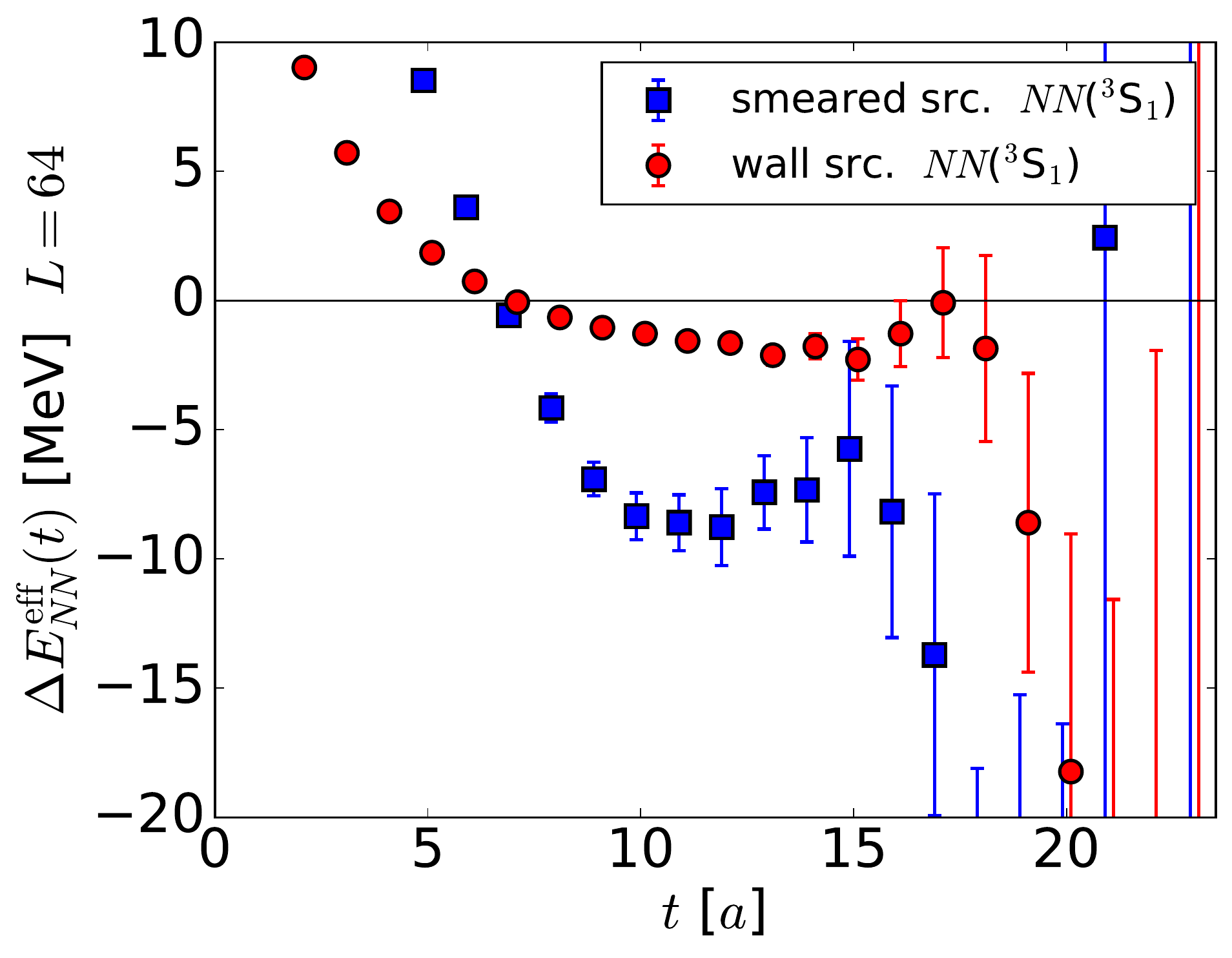}
  \caption{
  The effective energy shift $\DelEeffNN(t)$ in the $^3$S$_1$ channel for both smeared and wall sources.
  From the top to bottom, $L^3 = 32^3, 40^3, 48^3, 64^3$.
  (Left)  The results from non-relativistic operators. 
  The plateaux of Ref.~\cite{Yamazaki:2012hi} are also shown by black lines 
  (central value and 1$\sigma$ statistical errors) for comparison.
  (Right) The results from relativistic operators.
  }
   \label{fig:NN3S1}
\end{figure}

\begin{figure}[tbh]
  \centering
  \includegraphics[width=0.45\textwidth]{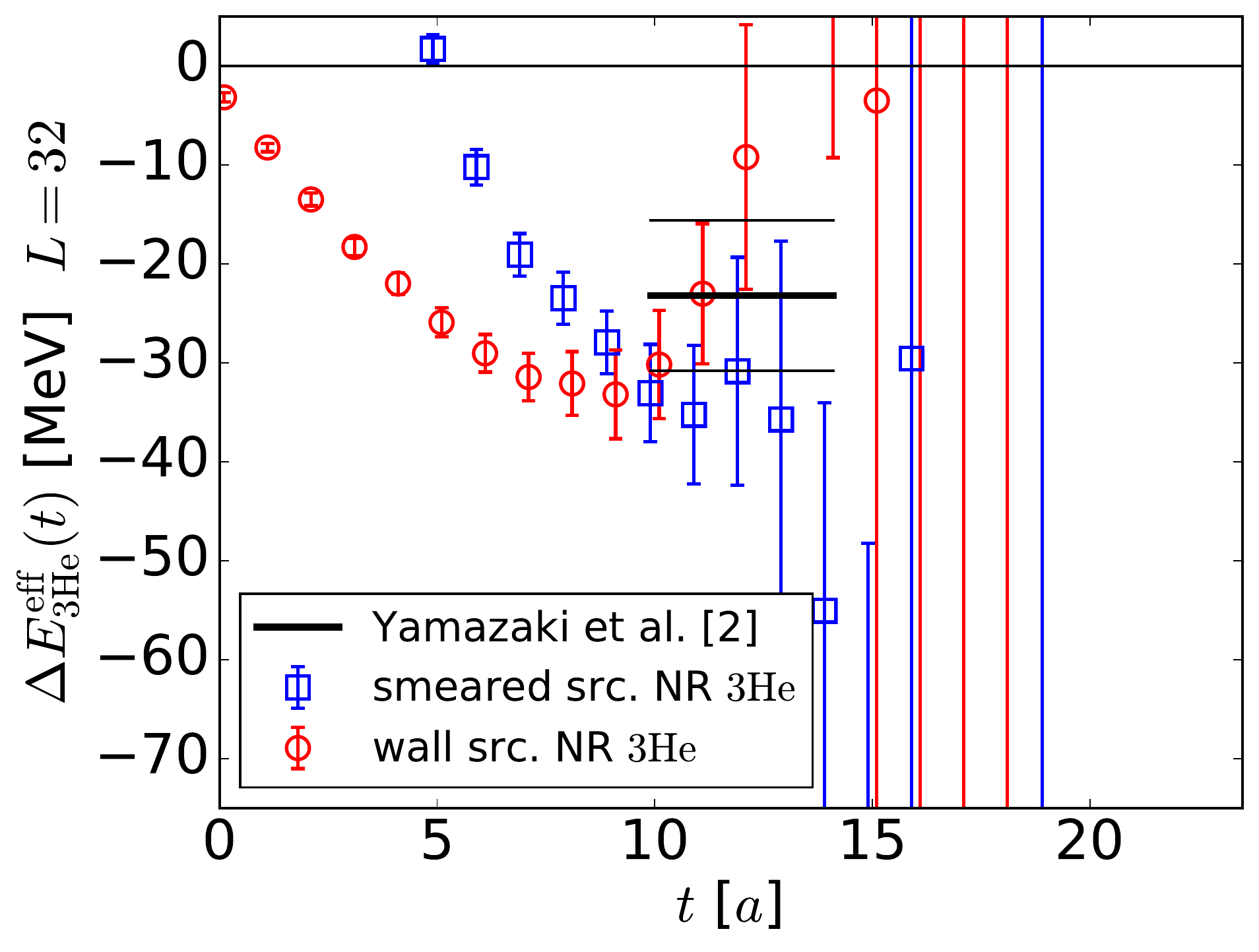}
  \includegraphics[width=0.45\textwidth]{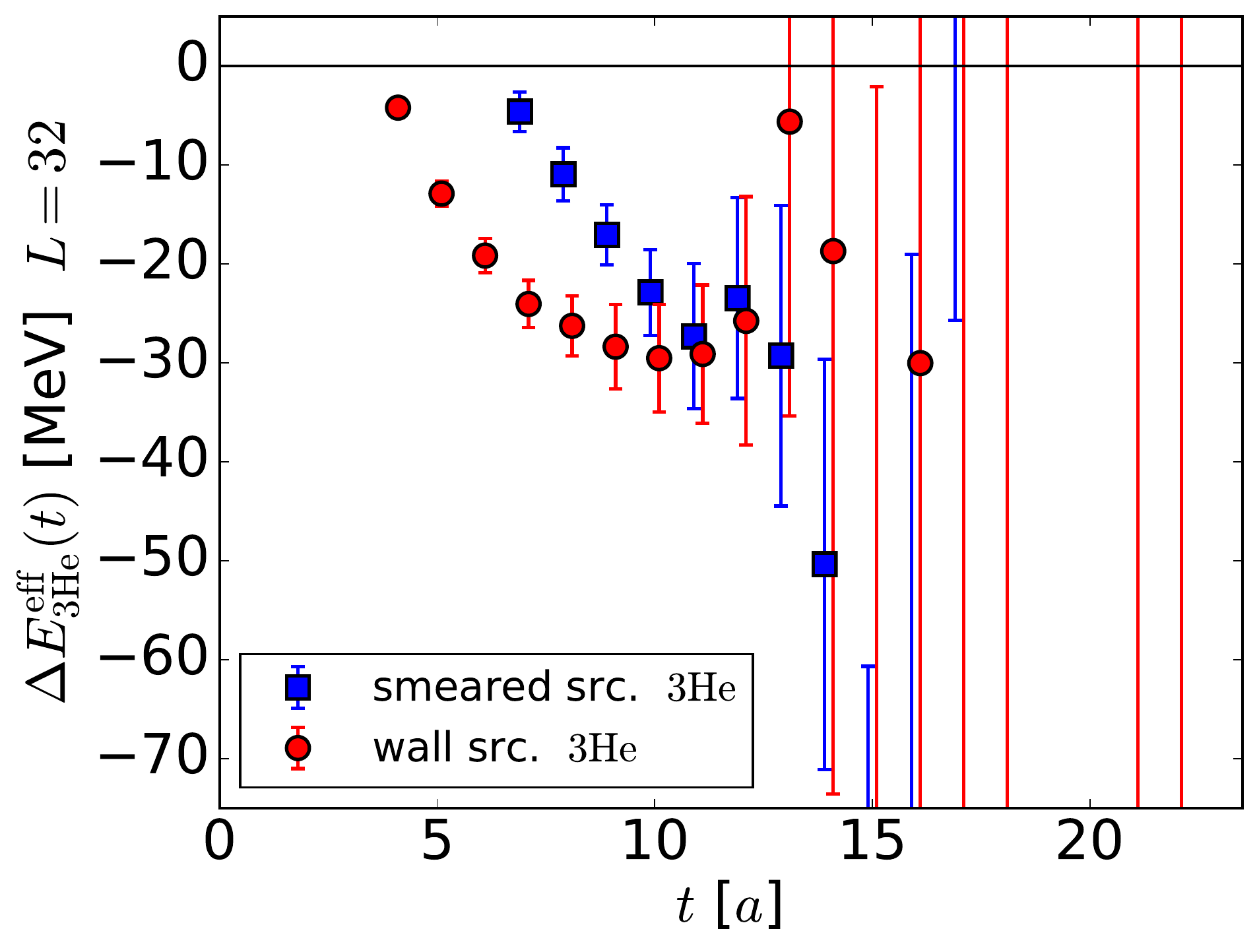}

  \includegraphics[width=0.45\textwidth]{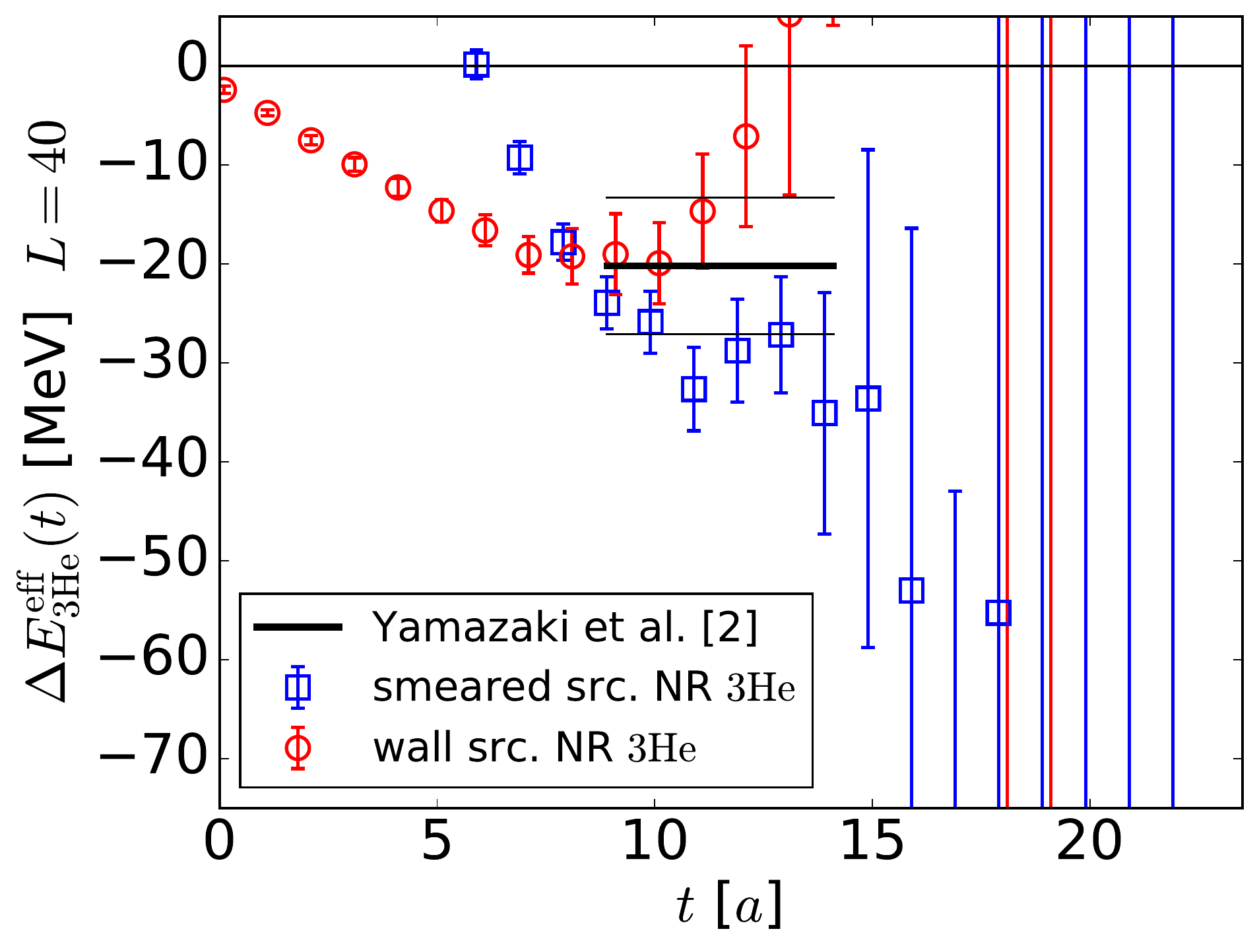}
  \includegraphics[width=0.45\textwidth]{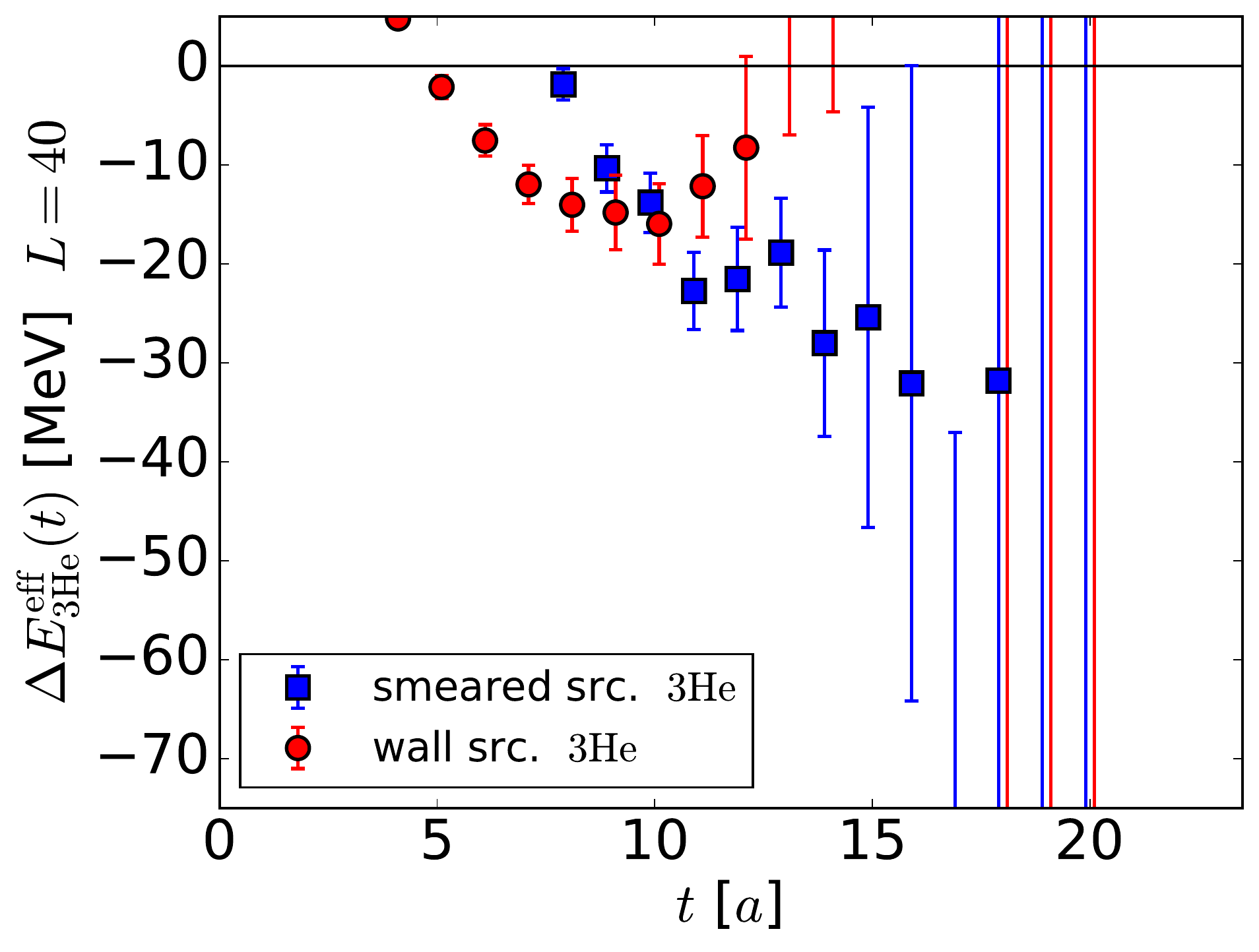}

  \includegraphics[width=0.45\textwidth]{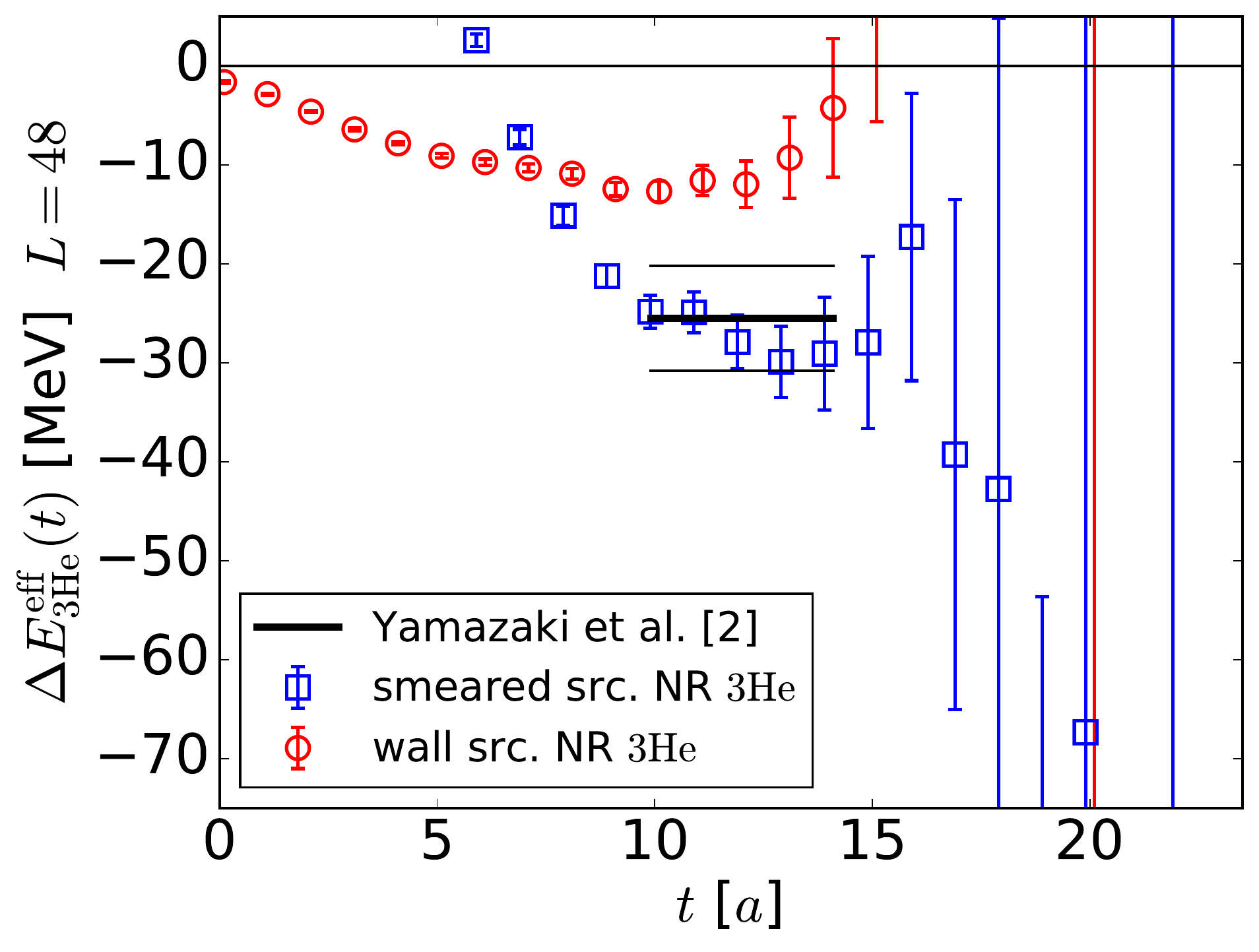}
  \includegraphics[width=0.45\textwidth]{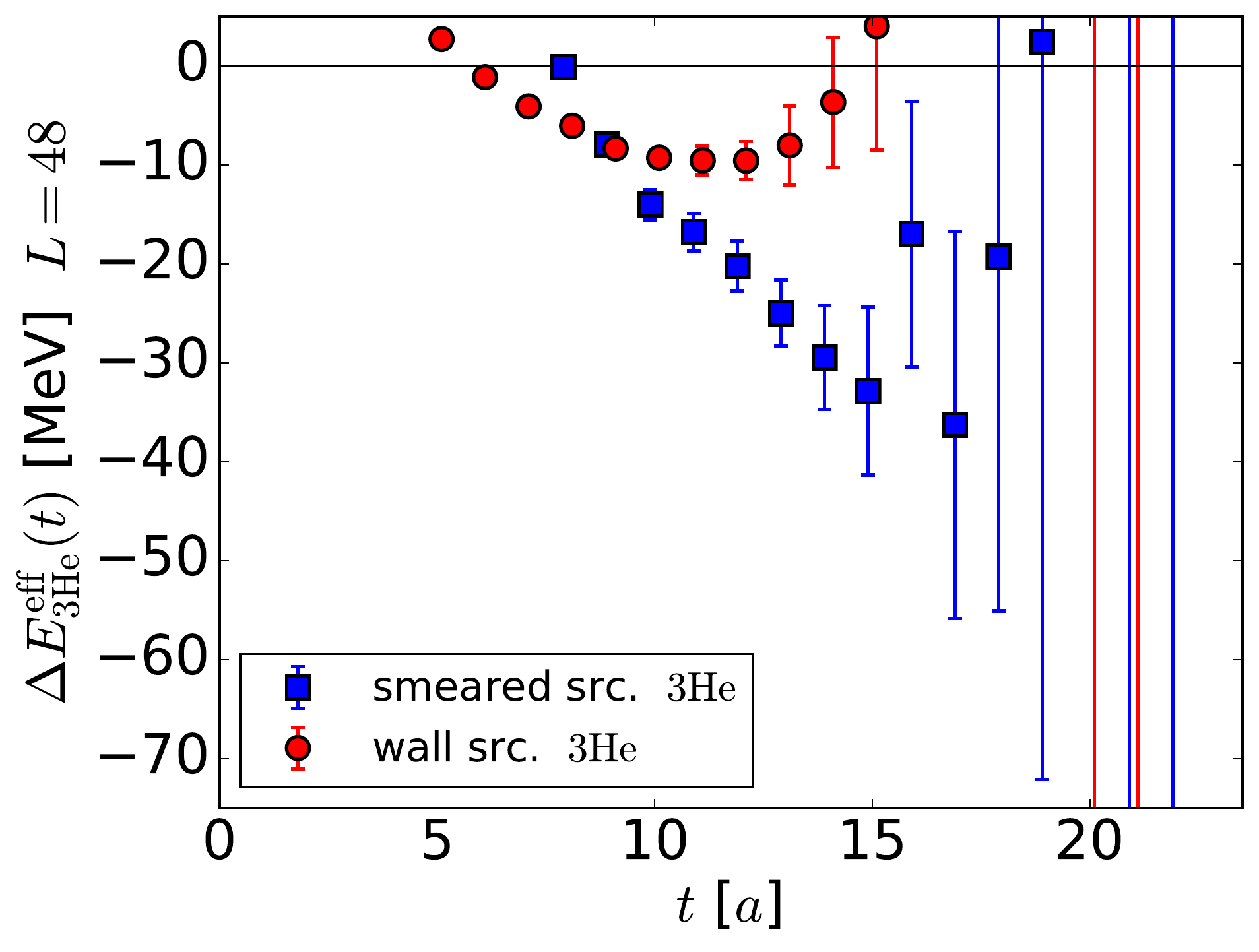}

  \includegraphics[width=0.45\textwidth]{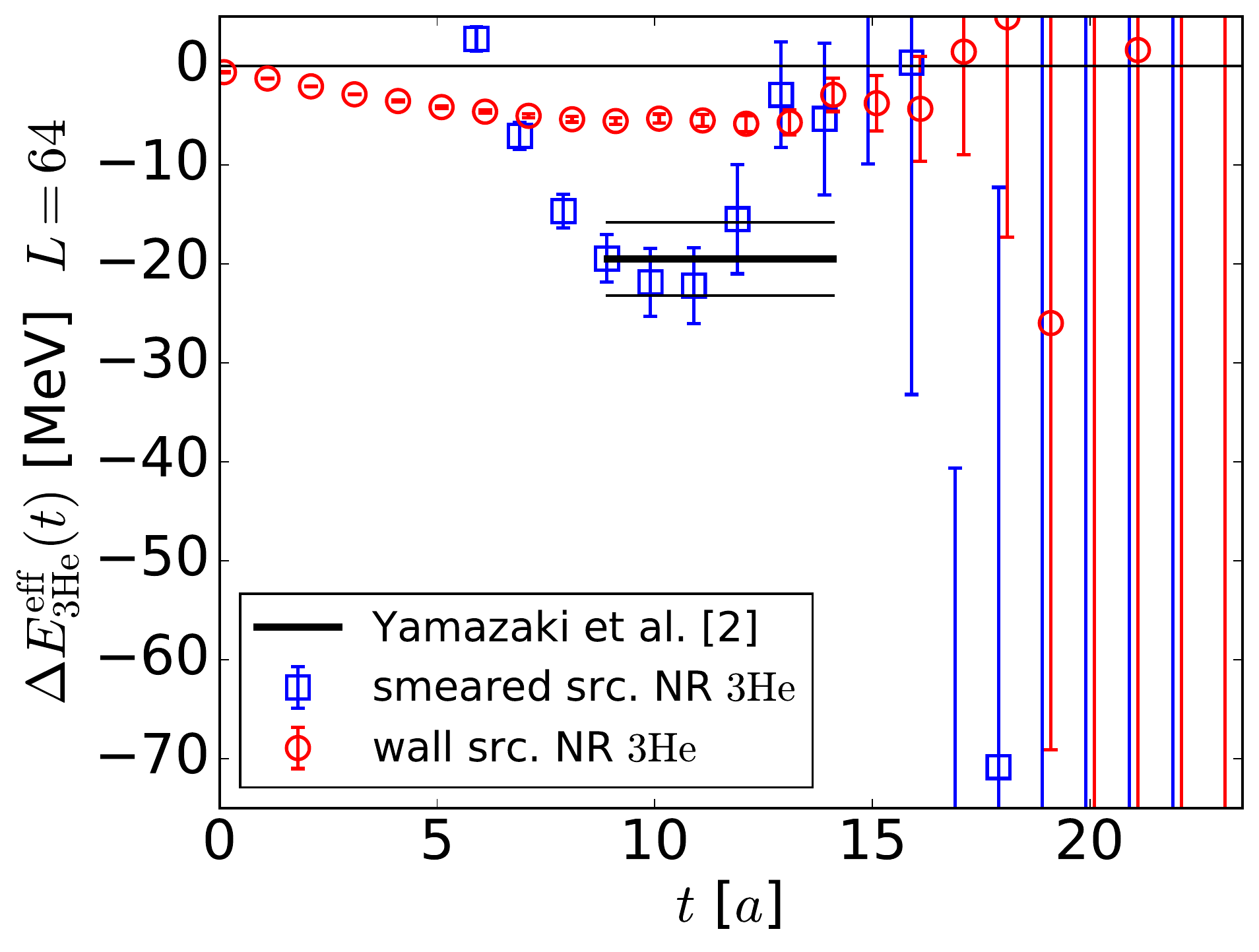}
  \includegraphics[width=0.45\textwidth]{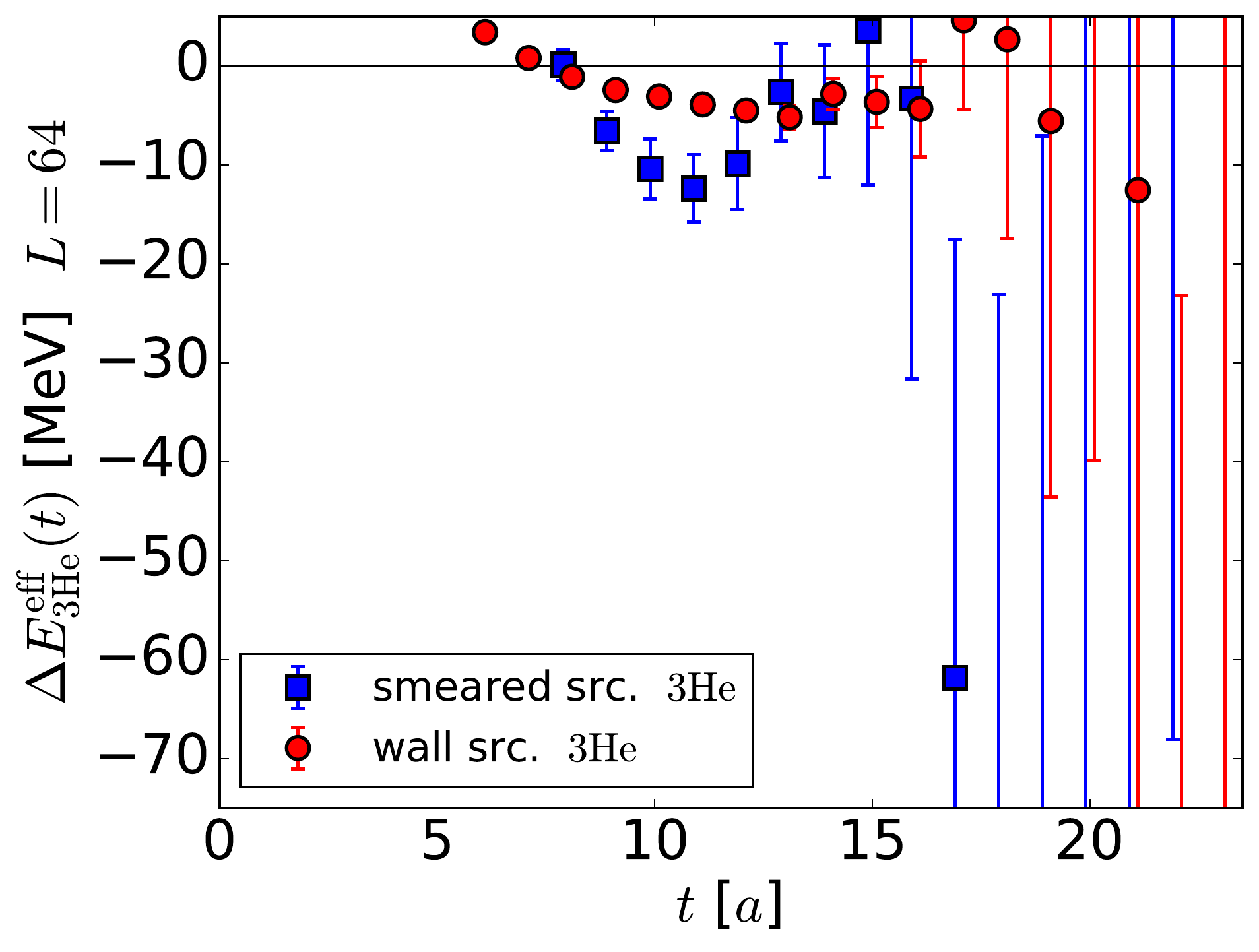}
  \caption{
  The effective energy shift $\DelEeffTri(t)$ for both smeared and wall sources.
  From the top to bottom, $L^3 = 32^3, 40^3, 48^3, 64^3$.
  (Left)  The results from non-relativistic operators. 
  The plateaux of Ref.~\cite{Yamazaki:2012hi} are also shown by black lines 
  (central value and 1$\sigma$ statistical errors) for comparison.
  (Right) The results from relativistic operators.
  }
  \label{fig:3He}
\end{figure}

\begin{figure}[tbh]
  \centering
  \includegraphics[width=0.45\textwidth]{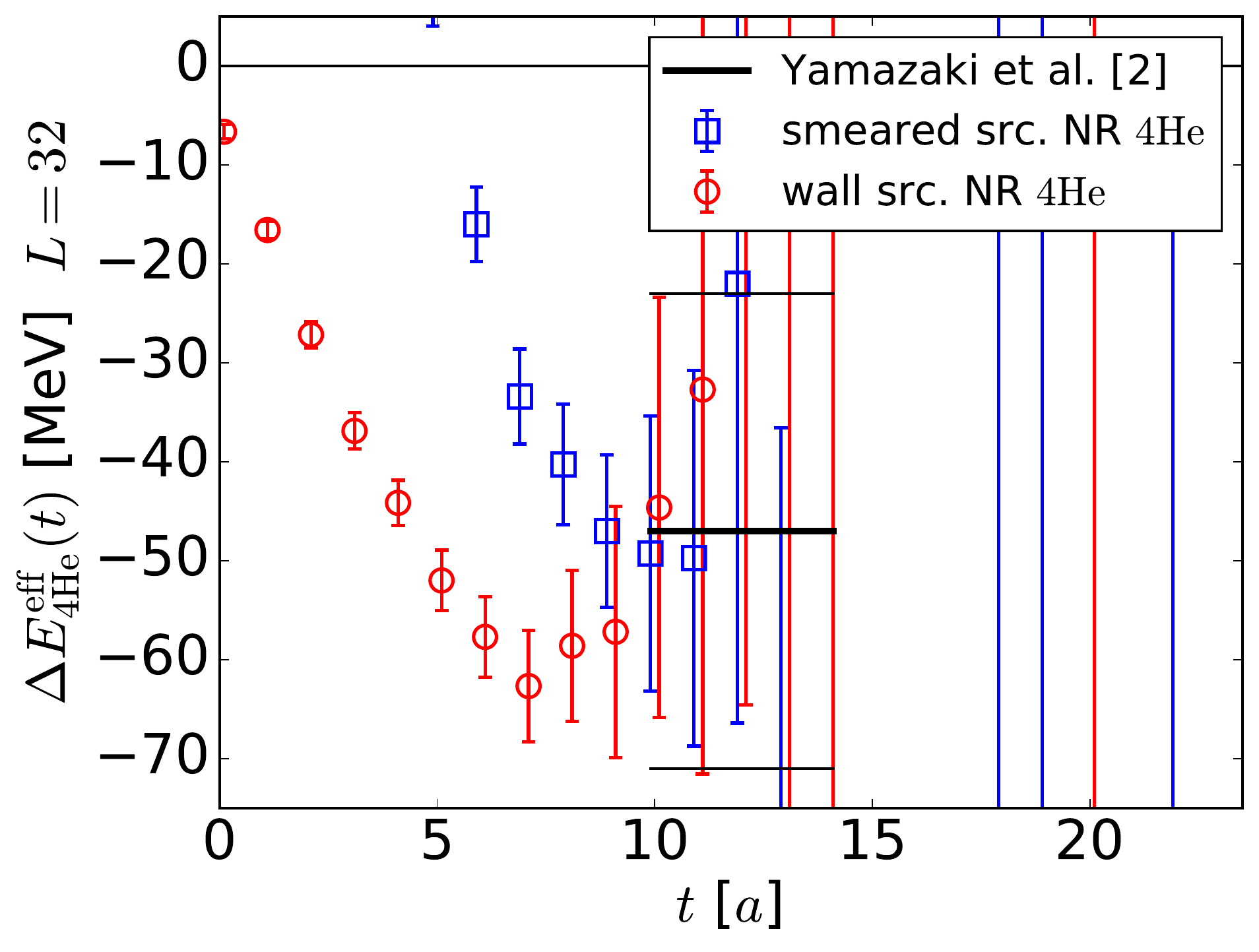}
  \includegraphics[width=0.45\textwidth]{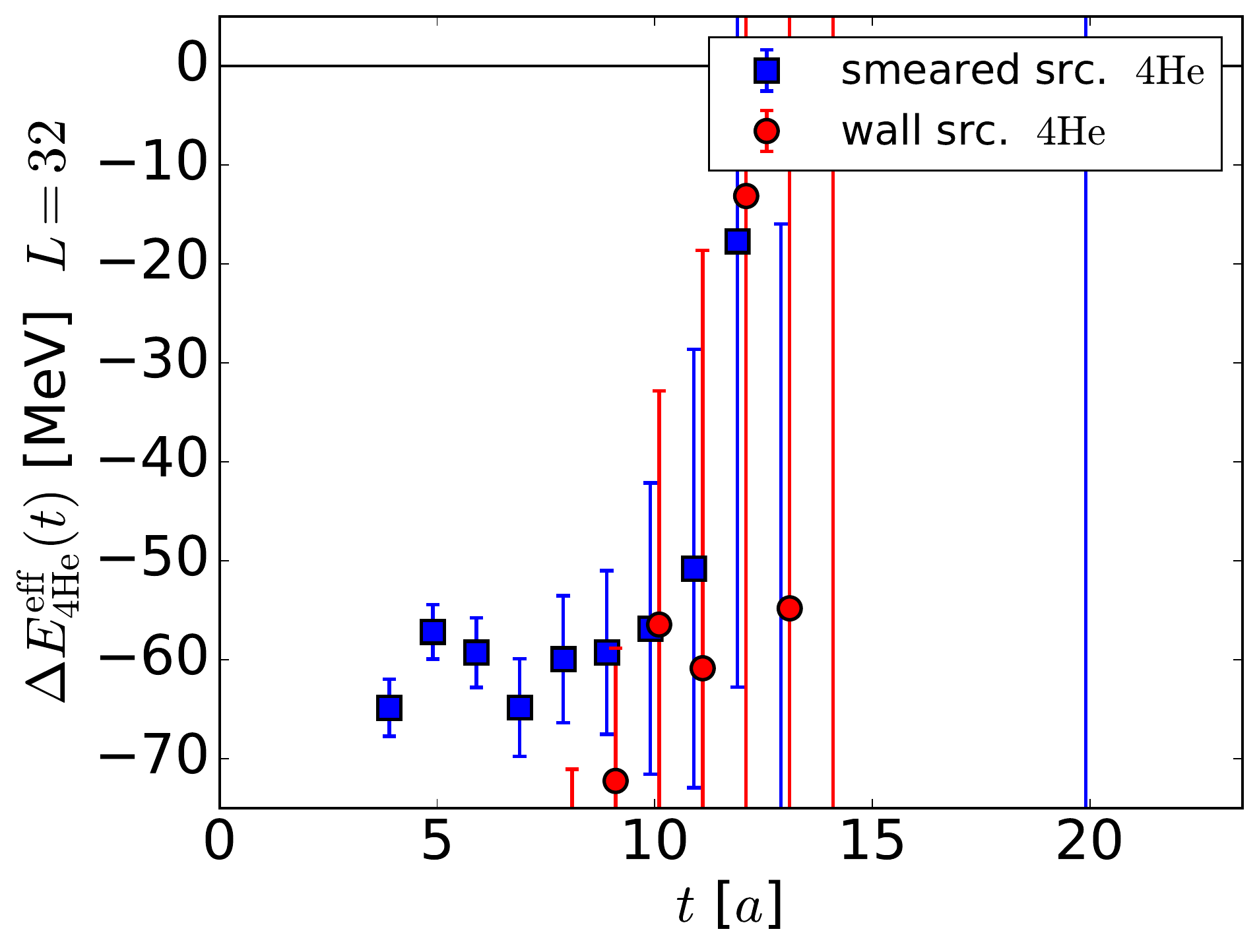}

  \includegraphics[width=0.45\textwidth]{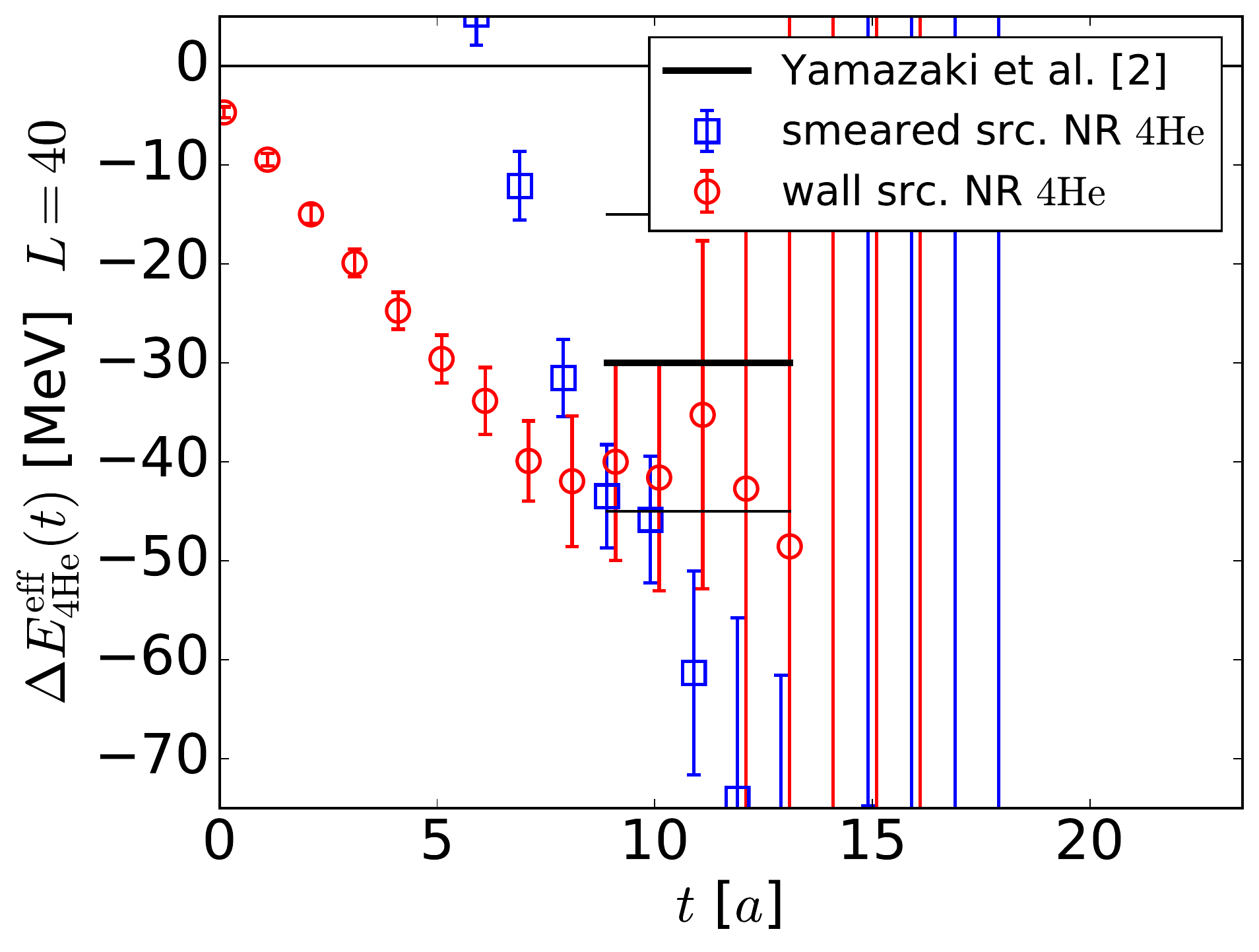}
  \includegraphics[width=0.45\textwidth]{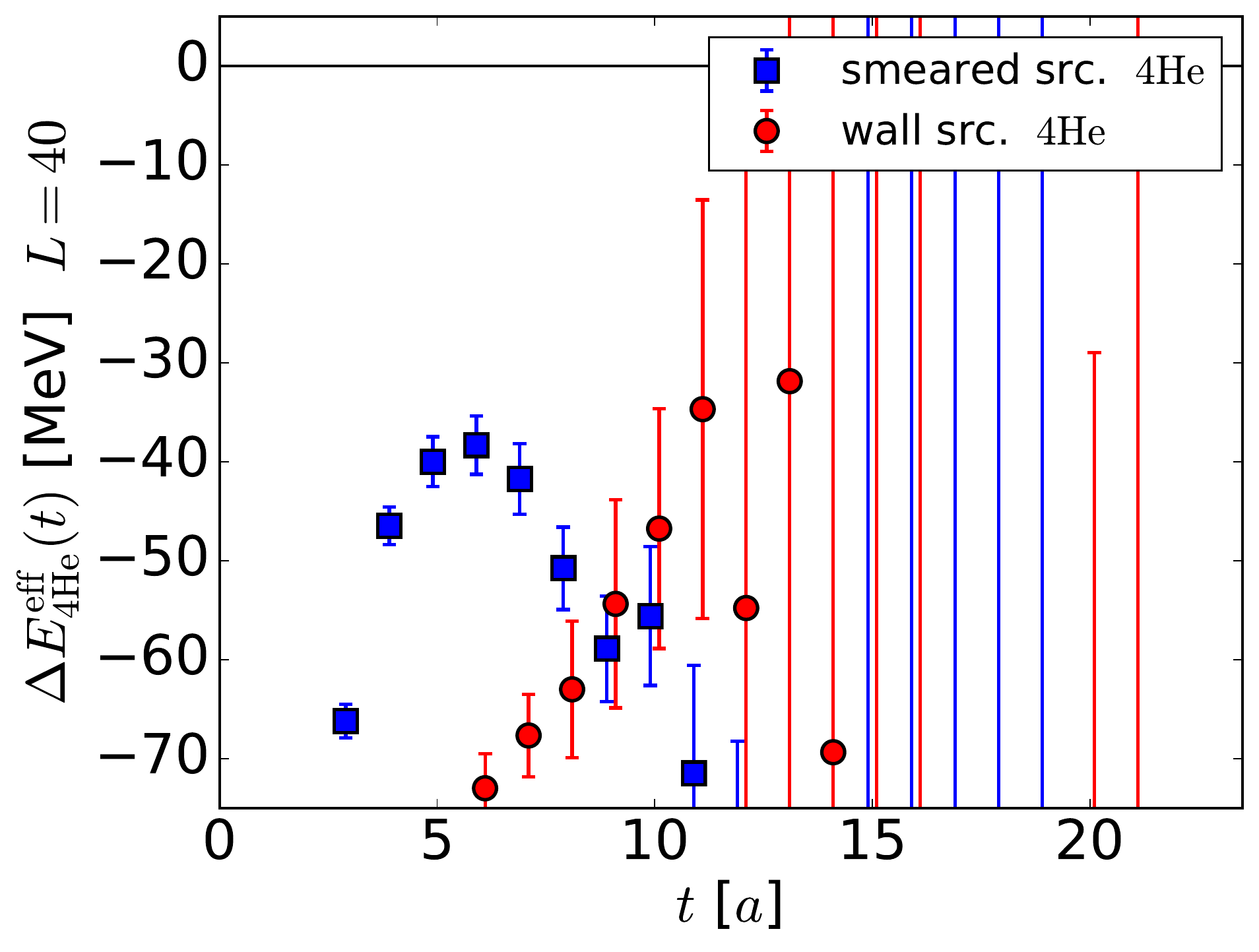}

  \includegraphics[width=0.45\textwidth]{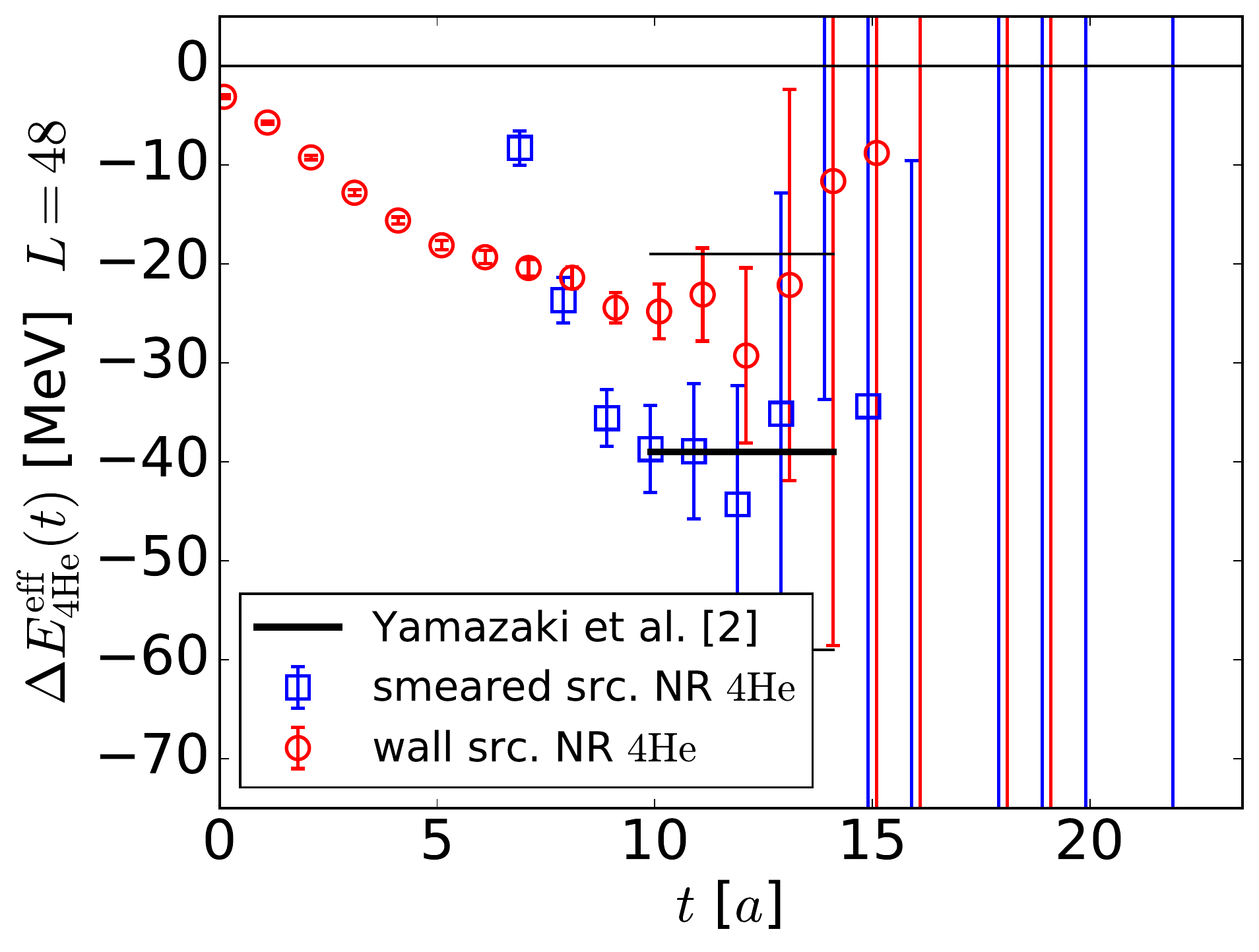}
  \includegraphics[width=0.45\textwidth]{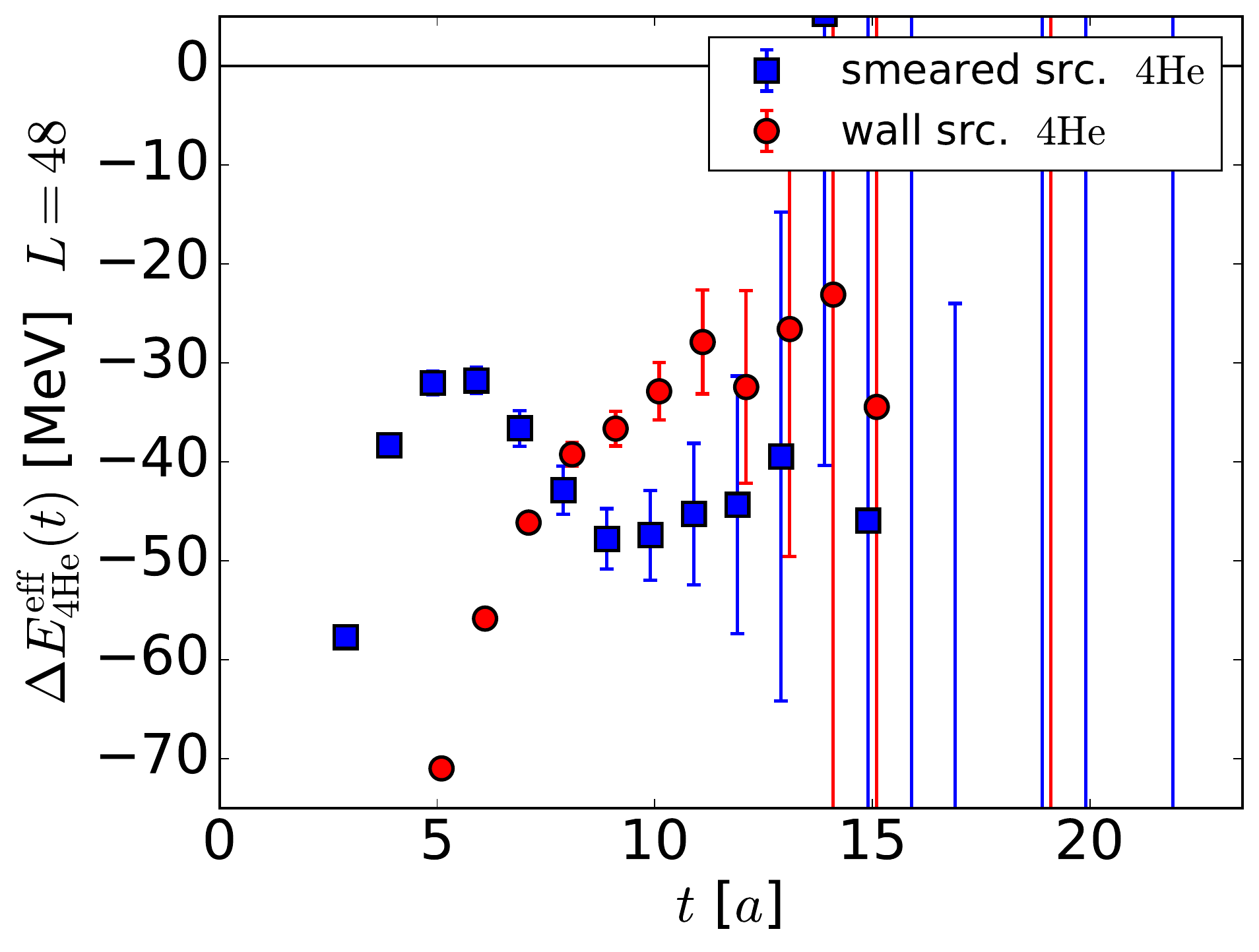}

  \includegraphics[width=0.45\textwidth]{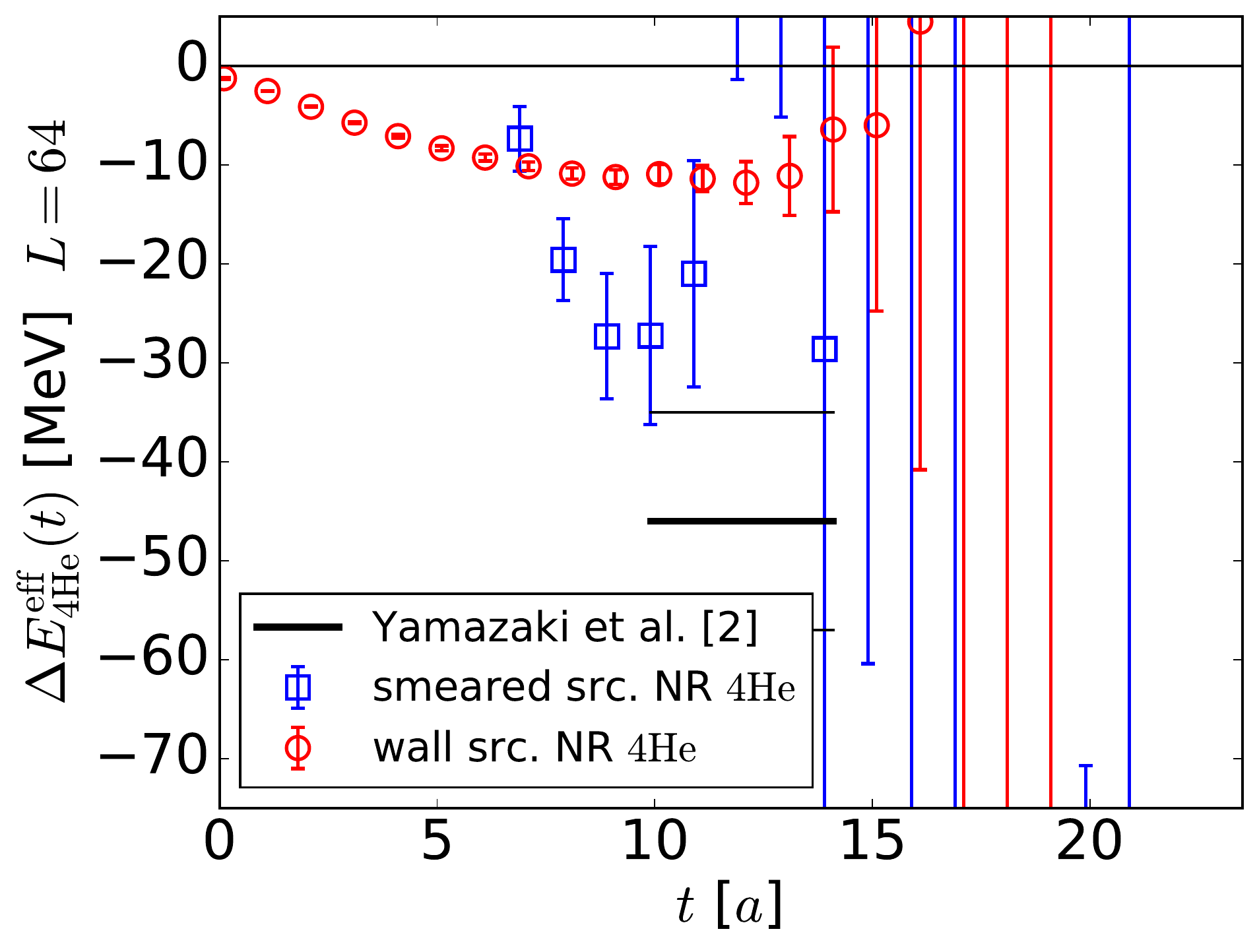}
  \includegraphics[width=0.45\textwidth]{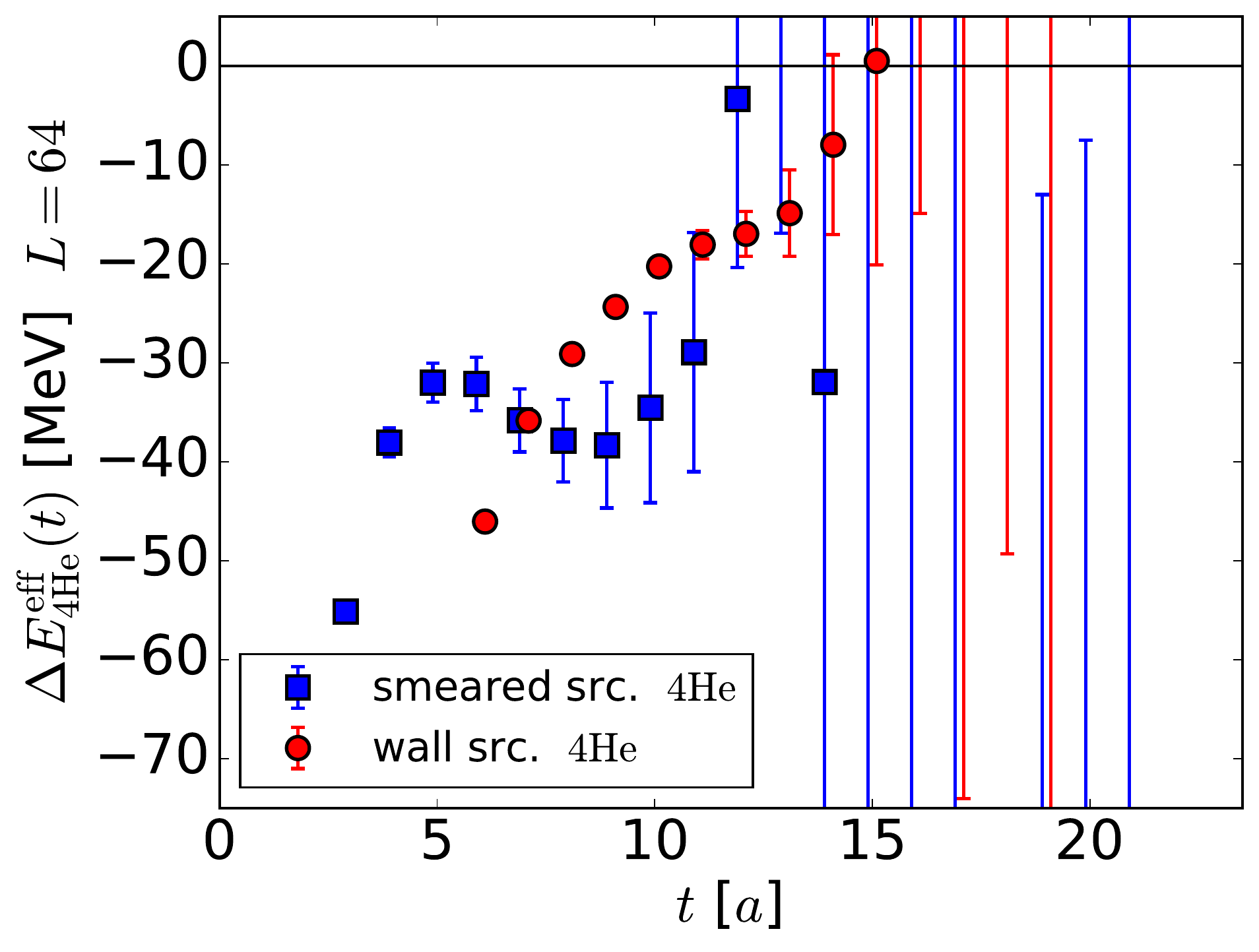}
  \caption{
  The effective energy shift $\DelEeffHe(t)$ for both smeared and wall sources.
  From the top to bottom, $L^3 = 32^3, 40^3, 48^3, 64^3$.
  (Left)  The results from non-relativistic operators. 
  The plateaux of Ref.~\cite{Yamazaki:2012hi} are also shown by black lines 
  (central value and 1$\sigma$ statistical errors) for comparison.
  (Right) The results from relativistic operators.
  }
   \label{fig:4HeNR}
\end{figure}

\end{document}